# Currents and the Energy-Momentum Tensor in Classical Field Theory: A Fresh Look at an Old Problem

Michael Forger [1] [*] and Hartmann Römer [2] [†]

[1] Departamento de Matemática Aplicada,
Instituto de Matemática e Estatística,
Universidade de São Paulo,
Caixa Postal 66281,
BR–05311-970  São Paulo, S.P., Brazil

[2] Fakultät für Physik
Albert-Ludwigs-Universität Freiburg im Breisgau
Hermann-Herder-Straße 3
D–79104  Freiburg i.Br., Germany

[1] *E-mail address:* forger@ime.usp.br
[2] *E-mail address:* hartmann.roemer@physik.uni-freiburg.de

[*]Partially supported by CNPq and FAPESP, Brazil, and by DFG, Germany
[†]Partially supported by FAPESP, Brazil

**Abstract**

We give a comprehensive review of various methods to define currents and the energy-momentum tensor in classical field theory, with emphasis on a geometric point of view. The necessity of "improving" the expressions provided by the canonical Noether procedure is addressed and given an adequate geometric framework. The main new ingredient is the explicit formulation of a principle of "ultralocality" with respect to the symmetry generators, which is shown to fix the ambiguity inherent in the procedure of improvement and guide it towards a unique answer: when combined with the appropriate splitting of the fields into sectors, it leads to the well-known expressions for the current as the variational derivative of the matter field Lagrangian with respect to the gauge field and for the energy-momentum tensor as the variational derivative of the matter field Lagrangian with respect to the metric tensor. In the second case, the procedure is shown to work even when the matter field Lagrangian depends explicitly on the curvature, thus establishing the correct relation between scale invariance, in the form of local Weyl invariance "on shell", and tracelessness of the energy-momentum tensor, required for a consistent definition of the concept of a conformal field theory.

# 1 Introduction

The energy-momentum tensor of a classical field theory combines the densities and flux densities of energy and momentum of the fields into one single object. However, the problem of giving a concise definition of this object able to provide the physically correct answer under all circumstances, for an arbitrary Lagrangian field theory on an arbitrary space-time background, has puzzled physicists for decades.

One of the most traditional approaches to the question is based on the Noether theorem, according to which a field theory with space-time translation invariance has a conserved energy-momentum tensor. Unfortunately, the so-called canonical energy-momentum tensor $\Theta^{\mu\nu}$ obtained from this procedure is in general unacceptable. In electrodynamics, for example, it is neither symmetric nor gauge invariant, and even in the simplest theory of a single scalar field where it does turn out to be symmetric and the criterion of gauge invariance is irrelevant, it has a non-vanishing trace even when the Lagrangian shows dilatation invariance.

There is a long history of attempts to cure these diseases and arrive at the physically correct energy-momentum tensor $T^{\mu\nu}$ by adding judiciously chosen "improvement" terms to $\Theta^{\mu\nu}$. A major early success in this direction was the work of Belinfante [1] and Rosenfeld [2] who, in particular, developed this strategy for Lorentz invariant field theories in flat Minkowski space-time to provide a symmetric energy-momentum tensor which, in the case of electrodynamics, is also gauge invariant and gives the physically correct expressions for the energy density and energy flux density (Poynting vector) as well as the momentum density and momentum flux density (Maxwell stress tensor) of the electromagnetic field. Later, Callan, Coleman and Jackiw [3] and Deser [4] proposed additional "improvement" terms to define a new symmetric energy-momentum tensor that, for dilatation invariant scalar field theories, is also traceless. However, all these methods of defining improved energy-momentum tensors are largely "ad hoc" procedures focussed on special models of field theory, often geared to the needs of quantum field theory and ungeometric in spirit.

In this last respect, the more recent paper of Gotay and Marsden [5], which also provides an extensive list of references witnessing the long and puzzling history of the subject, is an exception. Their approach is perhaps the first systematic attempt to tackle the problem from a truly geometric point of view.

In a geometric setting, one should consider general classical field theories on arbitrary space-time manifolds. Generically, space-time manifolds do not admit any isometries or conformal isometries at all, so there is no direct analogue of space-time translations, Lorentz transformations or dilatations, nor are there any conserved quantities in the usual sense. However, the equivalence principle and the principle of general covariance a) suggest that these rigid space-time symmetries should in a generally covariant theory be replaced by flexible space-time symmetries, which means that, in particular, the role of translations and Lorentz transformations would be taken over by space-time diffeomorphisms, and b) imply that the ordinary conservation law $\partial_\mu T^{\mu\nu} = 0$ for the energy-momentum tensor on flat space-time must be replaced by the covariant conservation law[1] $\nabla_\mu T^{\mu\nu} = 0$ for its counterpart on curved space-time. It is therefore reasonable to explore the possibility of deriving the energy-momentum tensor as a kind of covariant Noether current associated with the space-time diffeomorphism group. This is the basic idea adopted by Gotay and Marsden, worked out in detail in Ref. [5] using the modern geometric approach to general first order Lagrangian field theories, where fields are represented by sections of some fiber bundle over space-time and their first partial derivatives are represented by the first order jets of these sections, which are themselves sections of the corresponding first order jet bundle. In this framework, it is possible to derive the so-called canonical energy-momentum tensor as part of the local coordinate expression of a globally defined object with invariant geometric meaning, namely the field theoretical momentum map associated with the automorphism group of the underlying configuration bundle. In a second step, Gotay and Marsden then present an explicit construction of correction terms, leading to an improved energy-momentum tensor which is both symmetric and gauge invariant.

A first problem that arises in this approach is that, in general, diffeomorphisms of the space-time manifold $M$ do not act directly on the fields. In a geometric approach to classical field theory, fields are sections of bundles over space-time, and what really does act on such sections are automorphisms of these bundles. (If one requires certain so-called $G$-structures to be preserved, these should be $G$-automorphisms.) Given a fiber bundle $E$ over $M$, an automorphism of $E$ is a fiber preserving diffeomorphism of the total space $E$; it is called strict if it takes every fiber to itself. An automorphism of $E$ induces a diffeomorphism of the base space $M$, which is the identity on $M$ if and only if the automorphism is strict, but conversely it is not true in general that a diffeomorphism of the base space $M$ induces an automorphism of $E$. This can be restated by noting that automorphisms of $E$ form a group $\mathrm{Aut}(E)$ and strict automorphisms of $E$ form a normal subgroup $\mathrm{Aut}_s(E)$ such that

$$\mathrm{Aut}(E)/\mathrm{Aut}_s(E) \;\cong\; \mathrm{Diff}(M) \;, \qquad (1)$$

so the diffeomorphism group $\mathrm{Diff}(M)$ of $M$ is a quotient group of $\mathrm{Aut}(E)$, rather than a subgroup.

In Ref. [5], this problem is overcome by considering liftings of $\mathrm{Diff}(M)$ into $\mathrm{Aut}(E)$. What is really needed to define a geometric version of the so-called canonical energy-momentum tensor, via a field theoretic momentum map associated directly with the group of space-time diffeomorphisms, rather than the entire group of bundle automorphisms, is such a lifting only at the infinitesimal level, from vector fields $X_M$ on $M$ to projectable vector fields $X_E$ on $E$, expressing the vertical components $X^i$ of $X_E$ in terms of the components $X^\mu$ of $X_M$ and a finite number of partial derivatives thereof.

---

[1]For comments on this terminology, see Footnote 19 in Sect. 4.2.

Applying a series of partial integrations to eliminate the partial derivatives of the $X^\mu$, the authors of Ref. [5] arrive at an improved energy-momentum tensor which they claim to be the physically correct one for a general first order Lagrangian field theory.

A further and more serious problem is that, unfortunately, this claim as it stands is not correct, at least not without further qualification. The reason is that the entire analysis carried out in Ref. [5] is based on one fundamental hypothesis, namely that the Lagrangian of the theory should be invariant under a "sufficiently large" group of automorphisms of $E$. (Technically, it is sufficient to require that every compactly supported vector field on $M$ can be lifted to a vector field on $E$ whose flow leaves the Lagrangian invariant.) This guarantees that the improved energy-momentum tensor proposed in Ref. [5] is actually independent of the specific lifting employed in its construction, but according to the second Noether theorem, it also implies that this object vanishes "on shell", that is, when the fields satisfy the equations of motion. Obviously, this cannot be true for the physically correct energy-momentum tensor.

How to avoid this pitfall is of course well known and is even mentioned in Ref. [5], but this part of the story is not integrated into a coherent general picture. Rather than lumping all fields together and looking at the *total* Noether current, representing the sum of the contributions to the densities and flow densities for quantities such as charges or energy and momentum coming from *all* fields, one should split the fields into various sectors and concentrate on the exchange of these quantities between different sectors. There are two ways of performing such a split in a natural, non-artificial way.

In the first variant, one considers quantities, commonly called charges, which refer to strict bundle automorphisms, or in physical language, internal symmetries or gauge transformations. The interaction between charge carrying fields is mediated by gauge fields: these play a distinguished role as compared to all other fields, which are collectively referred to as matter fields. Correspondingly, the total Lagrangian becomes the sum of a gauge field part and a matter field part which are required to be separately invariant under the relevant symmetry group:

$$L \;=\; L_\mathrm{g} \,+\, L_\mathrm{m} \,. \qquad (2)$$

As a result, the total Noether current considered before, which indeed vanishes "on shell" (this statement is in fact nothing but the equation of motion for the gauge field – normally the Yang-Mills equation) also splits into a gauge field part and a matter field part, and it is the latter that provides the physically correct current of the theory. In other words, this current is associated with the matter field Lagrangian alone and not with the total Lagrangian obtained by adding the gauge field Lagrangian. Moreover, whenever some kind of improvement is necessary, one can completely avoid carrying out the cumbersome details of this procedure because the end result can be uniquely characterized by a certain requirement of "ultralocality" (that we shall explain in more detail later on) and turns out to be given simply by the prescription of varying the matter field Lagrangian with respect to the gauge field:

$$j^\mu = \frac{\delta L_\mathrm{m}}{\delta A_\mu} \ . \tag{3}$$

In the second variant, one considers energy and momentum, which refer to bundle automorphisms that are not strict but cover non-trivial space-time diffeomorphisms, or in physical language, space-time symmetries. The situation in this case is almost completely analogous to the previous one. The interaction between energy and momentum carrying fields is mediated by a metric tensor representing gravity: this metric tensor now plays a distinguished role as compared to all other fields, which are collectively referred to as matter fields. (Of course, both gauge fields and matter fields in the previous sense are matter fields in this sense.) Correspondingly, the total Lagrangian now becomes the sum of a gravitational part and a matter field part which are required to be separately invariant under the relevant symmetry group:

$$L = L_\mathrm{g} + L_\mathrm{m} \ . \tag{4}$$

As a result, the improved energy-momentum tensor proposed in Ref. [5], which indeed vanishes "on shell" (this statement is in fact nothing but the equation of motion for the gravitational field – normally the Einstein equation) also splits into a purely gravitational part and a matter field part, and it is the latter that provides the physically correct energy-momentum tensor of the theory. In other words, this tensor is associated with the matter field Lagrangian alone and not with the total Lagrangian obtained by adding the gravitational Lagrangian. Moreover, one can once again avoid carrying out the cumbersome details of the improvement procedure because the end result can be uniquely characterized by a certain requirement of "ultralocality" (that we shall explain in more detail later on) and turns out to be given simply by Hilbert's classical prescription of varying the matter field Lagrangian with respect to the metric tensor:

$$T^{\mu\nu} = -2 \frac{\delta L_\mathrm{m}}{\delta g_{\mu\nu}} \ . \tag{5}$$

As a corollary, this energy-momentum tensor is automatically symmetric, gauge invariant and independent of the lifting required for its construction from a Noetherian point of view.

A further important issue concerning the energy-momentum tensor is the question of whether its trace vanishes whenever we are dealing with a dilatation or scale invariant classical field theory. This requires, first of all, an appropriate definition of the concept of dilatation or scale invariance on arbitrary space-time manifolds. As is well known, this role is taken over by the notion of Weyl invariance. Weyl transformations rescale the metric tensor as well as all other fields (according to their Weyl dimension) and come in two variants: rescaling by a constant factor (global Weyl transformations) and rescaling by an arbitrary function on space-time (local Weyl transformations).

As will be shown below, local Weyl invariance "on shell" is equivalent to vanishing of the trace of the Hilbert energy-momentum tensor.² The problem that the energy-momentum tensor for certain dilatation or scale invariant scalar field theories on flat space-time has a non-vanishing trace can in the light of this result be traced back to the fact that the Lagrangian of such a theory allows many different extensions to a general space-time background, among which two play a distinguished role. The first is the standard extension obtained from the prescription of minimal coupling: this turns out to be globally but not locally Weyl invariant, not even "on shell", and that is why the trace of the "naive" energy-momentum tensor, which is the Hilbert tensor corresponding to this minimal Lagrangian, fails to vanish. The second is a non-minimal extension containing an additional term of the form $R\varphi^2$ where $\varphi$ is the scalar field and $R$ is the scalar curvature of the metric: this has the virtue of making the theory locally Weyl invariant, at least "on shell", and that is why the "new improved" energy-momentum tensor proposed in Refs [3] and [4], which is the Hilbert tensor corresponding to this non-minimal Lagrangian, has vanishing trace. Note that even though both Lagrangians have the same flat space-time limit, the corresponding Hilbert tensors are different even on flat space-time, since they are defined through variation around the flat metric.

The aim of the present work is to give a complete, coherent and geometrically motivated account of the situation concerning the energy-momentum tensor for classical field theory, with particular emphasis on the compensation of the classical trace anomaly – a subject not addressed in Ref. [5]. Reporting on an old subject with a long history, it is almost impossible to compile a reasonably complete bibliography, and we have therefore decided to concentrate on a relatively small number of key papers, apologizing in advance for our omissions. For the same reason, we shall of course not be able to avoid repeating various well-known results, but we feel that the time has come for a comprehensive and coherent account, and we hope that even an expert in the field will appreciate our approach to a certain number of fine points that in our view have not been adequately dealt with elsewhere.

The paper is organized as follows. In Section 2, we briefly review the standard Noetherian procedure for constructing the canonical total Noether current and the so-called canonical energy-momentum tensor for first order Lagrangian field theories on flat space-time, together with the procedure of "improvement" of the energy-momentum tensor à la Belinfante-Rosenfeld, but extended so as to include the correction terms required for the correct implementation of scale invariance. In Section 3, we begin with a summary of some relevant background material from geometric field theory [7–9] (Sections 3.1 and 3.2). Next, we present the construction of the canonical total Noether current from this point of view, where it appears as the local coordinate expression of a global object known as the covariant momentum map or multimomentum map, describe a framework for possible correction terms leading to an improved total Noether current and discuss Noether's first and second theorem (Section 3.3). Analysis of how the (canonical or improved) total Noether current depends on the symmetry generators then leads us to formulate, in theories with non-trivial local symmetry groups, the aforementioned "ultralocality principle", which finds an immediate application to two different types of field theories: those exhibiting gauge invariance and those exhibiting general covariance, or space-time diffeomorphism invariance (Section 3.4). However, the procedure still does not overcome the difficulty resulting from Noether's second theorem: this can only be achieved by splitting the theory into different sectors, giving a special status to gauge fields (for dealing with the physical current) and to the metric tensor (for dealing with the physical energy-momentum tensor). In order to prepare the ground for appropriately dealing with these splittings, we discuss some simple mathematical aspects of $G$-structures (including the notion of invariant fiber metrics) and of $G$-connections (Section 3.5) and then present a list of essentially all the invariant Lagrangians that are important for the construction of the field theoretical models appearing in the study of the fundamental interactions of matter (Section 3.6). In Section 4, we elaborate on the relevant splitting into a gauge field sector and a matter field sector, for currents (Section 4.1), and into a gravitational sector and a matter field sector, for the energy-momentum tensor (Section 4.2), showing that the improved expressions constructed according to the "ultralocality principle", but applied to the matter field sector alone, reproduce the standard expressions (3) and (5), respectively. The calculation of specific energy-momentum tensors is also addressed, with emphasis on two special cases that are non-trivial: scalar fields with an additional coupling to curvature ($R\varphi^2$ term) and Dirac spinor fields (Section 4.3). Finally, Section 5 exposes the relation between scale invariance, in the form of "on shell" local Weyl invariance, and tracelessness of the energy-momentum tensor, thus clarifying the question of how to decide, at the classical level (that is, without entering the question of a possible trace anomaly, for which we refer to the entertaining review [10] and the references quoted therein), whether a given field theoretical model is or is not an example of a conformal field theory.

---

²The term "on shell" in this context means that the desired invariance can be established only if one uses the equations of motion for the matter fields.

## 2 Energy-Momentum Tensors in Flat Space-Time

Consider a classical field theory on flat $n$-dimensional space-time $\mathbb{R}^n$ containing a multiplet of fields $\varphi^i$ (which we may view as lumped together into a single vector-valued field $\varphi$) and whose dynamics is supposed to be defined by a Lagrangian $L$ depending only on the point values of the $\varphi^i$ and their first order partial derivatives. This means that the equations of motion for the fields are derived from the variational principle $\delta S = 0$ for the action functional $S$ defined by

$$S[\varphi] \;=\; \int d^n x \; L(\varphi, \partial\varphi) \;. \tag{6}$$

(For simplicity, the possible explicit dependence of $L$ on the space-time coordinates is suppressed from the notation.) As is well known, these are the Euler-Lagrange

equations
$$\partial_\mu \left( \frac{\partial L}{\partial \, \partial_\mu \varphi^i} \right) - \frac{\partial L}{\partial \varphi^i} = 0 \; . \tag{7}$$

Moreover, Noether's theorem states that invariance of the action $S$ under infinitesimal transformations

$$x^\mu \;\to\; x^\mu + X_B^\mu(x) \quad , \quad \varphi^i(x) \;\to\; X_F^i(\varphi(x)) - X_B^\mu(x) \, \partial_\mu \varphi^i(x) \tag{8}$$

where the indices $B$ and $F$ stand for "base space" and "field space", respectively, leads to the conservation law

$$\partial_\mu \, \langle j_{\text{can}}^\mu, X \rangle \;=\; 0 \tag{9}$$

for the canonical total Noether current $j_{\text{can}}$ given by

$$\langle j_{\text{can}}^\mu, X \rangle \;=\; \frac{\partial L}{\partial \, \partial_\mu \varphi^i} \, X_F^i(\varphi) - \left( \frac{\partial L}{\partial \, \partial_\mu \varphi^i} \, \partial_\nu \varphi^i - \delta_\nu^\mu L \right) X_B^\nu \; . \tag{10}$$

Then the expression

$$\Theta^\mu{}_\nu \;=\; \frac{\partial L}{\partial \, \partial_\mu \varphi^i} \, \partial_\nu \varphi^i - \delta_\nu^\mu L \tag{11}$$

or

$$\Theta^{\mu\nu} \;=\; \frac{\partial L}{\partial \, \partial_\mu \varphi^i} \, \partial^\nu \varphi^i - \eta^{\mu\nu} L \tag{12}$$

provides the components of the so-called canonical energy-momentum tensor of the theory.

Different types of symmetries are distinguished by different choices for the vector fields $X_F$ and $X_B$ appearing in the (infinitesimal) transformation law (8). For example, space-time translations correspond to the choice

$$X_B^\mu(x) \;=\; a^\mu \quad , \quad X_F \;=\; 0 \tag{13}$$

where $a$ is an arbitrary constant $n$-vector. Therefore, the conservation law (9) when expressing invariance under space-time translations becomes equivalent to the conservation law

$$\partial_\mu \Theta^{\mu\nu} \;=\; 0 \tag{14}$$

for the so-called canonical energy-momentum tensor. Similarly, Lorentz transformations are described by setting

$$X_B^\mu(x) \;=\; \omega^\mu{}_\nu \, x^\nu \quad , \quad X_F^i(\varphi) \;=\; \Sigma(\omega)^i_j \, \varphi^j \;=\; \tfrac{1}{2} \, \omega_{\mu\nu} \, (\Sigma^{\mu\nu})^i_j \, \varphi^j \tag{15}$$

where $\omega$ is a constant $(n \times n)$-matrix representing an infinitesimal Lorentz transformation, which means that it satisfies the condition

$$\omega_{\mu\nu} + \omega_{\nu\mu} \;=\; 0 \tag{16}$$

but is otherwise arbitrary, whereas $\Sigma$ stands for the representation of the Lie algebra of infinitesimal Lorentz transformations on the space of field values. (This representation, which is assumed to be given, may be irreducible or reducible, so that in the general case, we encounter various multiplets of fields of different spin.) Therefore, the conservation law (9) when expressing Lorentz invariance becomes equivalent to the conservation law

$$\partial_\mu \Theta^{\mu\kappa\lambda} = 0 \tag{17}$$

for the so-called canonical moment tensor, defined by

$$\Theta^{\mu\kappa\lambda} = x^\kappa \Theta^{\mu\lambda} - x^\lambda \Theta^{\mu\kappa} + \Sigma^{\mu\kappa\lambda} \tag{18}$$

where

$$\Sigma^{\mu\kappa\lambda} = \pi_i^\mu \, (\Sigma^{\kappa\lambda})^i_j \, \varphi^j \tag{19}$$

is called its internal part or spin part. Finally, dilatations or scale transformations are represented by

$$X_B^\mu(x) = x^\mu \quad , \quad X_F^i(\varphi) = -d_\varphi \, \varphi^i \tag{20}$$

where $d_\varphi$ is a real number called the scaling dimension of the field $\varphi$. (For simplicity, we assume here that all field components have the same scaling dimension: the general case containing various multiplets of fields of different scaling dimension is handled similarly.) Therefore, the conservation law (9) when expressing scale invariance becomes equivalent to the conservation law

$$\partial_\mu \Theta^\mu = 0 \tag{21}$$

for the so-called canonical dilatation or scaling current, defined by

$$\Theta^\mu = x_\nu \Theta^{\mu\nu} + \Sigma^\mu \tag{22}$$

where

$$\Sigma^\mu = d_\varphi \, \pi_i^\mu \, \varphi^i \tag{23}$$

is called its internal part.

Of course, it is well known that the so-called canonical energy-momentum tensor suffers from a number of serious problems.

- In theories containing gauge fields, it fails to be gauge invariant: this already happens in the simplest case, namely electrodynamics.

- In general, it fails to be symmetric.

- In general, it fails to be traceless even in theories which are manifestly scale invariant.

Similar statements hold for the so-called canonical moment tensor and scaling current.

These defects can be illustrated by looking at two simple examples. One of these is free electrodynamics, where
$$L = -\tfrac{1}{4} F^{\mu\nu} F_{\mu\nu} \tag{24}$$
with
$$F_{\mu\nu} = \partial_\mu A_\nu - \partial_\nu A_\mu \tag{25}$$
and
$$\Theta^{\mu\nu} = \tfrac{1}{4} \eta^{\mu\nu} F^{\kappa\lambda} F_{\kappa\lambda} - F^\mu{}_\kappa \, \partial^\nu A^\kappa \tag{26}$$
which is not gauge invariant, not symmetric and not traceless, even in four space-time dimensions where scale invariance of the action is achieved by setting $d_A = 1$. The other example is the theory of a real scalar field, where
$$L = \tfrac{1}{2} \partial^\mu \varphi \, \partial_\mu \varphi - U(\varphi) \tag{27}$$
and
$$\Theta^{\mu\nu} = \partial^\mu \varphi \, \partial^\nu \varphi - \tfrac{1}{2} \eta^{\mu\nu} \partial^\kappa \varphi \, \partial_\kappa \varphi + \eta^{\mu\nu} U(\varphi) \tag{28}$$
which is not traceless even though scale invariance of the action may be achieved by setting $d_\varphi = \tfrac{1}{2}(n-2)$ and choosing the potential $U$ to vanish when $n=2$ and to be given by
$$U(\varphi) \sim (\varphi^2)^{\frac{n}{n-2}} \tag{29}$$
when $n>2$: of course, the only solutions for which $U$ is a polynomial in $\varphi$ are $\varphi^6$ for $n=3$, $\varphi^4$ for $n=4$ (the famous massless $\varphi^4$ theory in 4 dimensions) and $\varphi^3$ for $n=6$.

For physically meaningful quantities such as the components of the energy-momentum tensor, representing the densities and flux densities of energy and momentum, such defects are of course unacceptable. After all, these are physically observable quantities and must therefore be gauge invariant. Similarly, asymmetric energy-momentum tensors lead to unacceptable predictions regarding the local torques exerted by the fields of the theory on all other forms of matter; a lucid discussion of this aspect can be found in Chapter 5.7 of Ref. [6]. Moreover, the requirement of symmetry of the energy-momentum tensor is indispensable if energy and momentum are to serve as the source of gravity, as required by the principles of general relativity.

On the other hand, it is worth noting that, even in flat space-time, a symmetric conserved energy-momentum tensor, which we shall generally denote by $T^{\mu\nu}$, has many pleasant properties. One of these is that it allows to define a conserved moment tensor without the need for an internal or spin part:
$$T^{\mu\kappa\lambda} = x^\kappa T^{\mu\lambda} - x^\lambda T^{\mu\kappa} \; . \tag{30}$$
Indeed, if we suppose that
$$T^{\nu\mu} = T^{\mu\nu} \; , \tag{31}$$
as well as
$$\partial_\mu T^{\mu\nu} = 0 \; , \tag{32}$$

then this definition of the moment tensor immediately implies that

$$\partial_\mu T^{\mu\kappa\lambda} = 0 \ . \tag{33}$$

Similarly, a traceless symmetric conserved energy-momentum tensor can be used to define a conserved dilatation or scaling current without the need for an internal part:

$$T^\mu = x_\nu T^{\mu\nu} \ . \tag{34}$$

Indeed, if in addition to eqs (31) and (32), we also suppose that

$$T^\mu_\mu = 0 \ , \tag{35}$$

then this definition of the dilatation or scaling current immediately implies that

$$\partial_\mu T^\mu = 0 \ . \tag{36}$$

As an additional benefit, we can infer from eqs (31), (32) and (35) that the conformal current, defined by

$$C^{\mu\nu} = \left(2x^\nu x_\kappa - x^2 \delta^\nu_\kappa\right) T^{\mu\kappa} \ , \tag{37}$$

is also conserved:

$$\partial_\mu C^{\mu\nu} = 0 \ . \tag{38}$$

This means that conformal invariance becomes a consequence of combining Poincaré invariance and scale invariance.

The strategy for constructing a symmetric and – in the presence of scale invariance – traceless energy-momentum tensor is always the same. The so-called canonical energy-momentum tensor is "improved" by adding a term whose divergence vanishes identically, for reasons of symmetry, that is

$$T^{\mu\nu} = \Theta^{\mu\nu} + \partial_\kappa f^{\kappa\mu\nu} \tag{39}$$

where $f$ is supposed to be antisymmetric in its first two indices

$$f^{\kappa\mu\nu} + f^{\mu\kappa\nu} = 0 \tag{40}$$

in order to guarantee that the vanishing of $\partial_\mu T^{\mu\nu}$ becomes equivalent to that of $\partial_\mu \Theta^{\mu\nu}$. The problem is to determine $f^{\kappa\mu\nu}$ in such a way that $T^{\mu\nu}$ becomes symmetric and – in the presence of scale invariance – traceless. Further restrictions arise from the postulate that $f^{\kappa\mu\nu}$ should be a pointwise defined function of the fields and their partial derivatives (up to some finite order). For example, in free electrodynamics, the substitution (39) with $f^{\kappa\mu\nu} = -F^{\kappa\mu}A^\nu$ leads to

$$T^{\mu\nu} = \tfrac{1}{4}\eta^{\mu\nu} F^{\kappa\lambda}F_{\kappa\lambda} - F^\mu_{\ \kappa}F^{\kappa\nu} \tag{41}$$

where we have omitted a term that vanishes "on shell", that is, as a result of the equations of motion of the theory, which in this case are the free Maxwell equations $\partial_\mu F^{\mu\nu} = 0$.

For a more systematic treatment of the method of improving the so-called canonical energy-momentum tensor, it is convenient to replace the Ansatz (39) by the equivalent Ansatz

$$T^{\mu\nu} = \Theta^{\mu\nu} + \tfrac{1}{2} \partial_\kappa \left(\tau^{\kappa\mu\nu} + \tau^{\mu\nu\kappa} - \tau^{\nu\kappa\mu}\right) \tag{42}$$

where $\tau$ is supposed to be antisymmetric in its last two indices

$$\tau^{\kappa\mu\nu} + \tau^{\kappa\nu\mu} = 0 \tag{43}$$

in order to guarantee that

$$f^{\kappa\mu\nu} = \tfrac{1}{2} \left(\tau^{\kappa\mu\nu} + \tau^{\mu\nu\kappa} - \tau^{\nu\kappa\mu}\right) \tag{44}$$

be antisymmetric in its first two indices. Conversely, $\tau$ can be expressed in terms of $f$:

$$\tau^{\kappa\mu\nu} = f^{\kappa\mu\nu} + f^{\nu\kappa\mu} . \tag{45}$$

Now from eqs (42) and (43) we obtain

$$T^{\mu\nu} - T^{\nu\mu} = \Theta^{\mu\nu} - \Theta^{\nu\mu} + \partial_\kappa \tau^{\kappa\mu\nu}$$

whereas eqs (14) and (18) yield

$$\partial_\kappa \Theta^{\kappa\mu\nu} = \Theta^{\mu\nu} - \Theta^{\nu\mu} + \partial_\kappa \Sigma^{\kappa\mu\nu} ,$$

so we get

$$T^{\mu\nu} - T^{\nu\mu} = \partial_\kappa \left(\Theta^{\kappa\mu\nu} - \Sigma^{\kappa\mu\nu} + \tau^{\kappa\mu\nu}\right) . \tag{46}$$

Similarly, from eqs (42) and (43) we obtain

$$T^\mu_\mu = \Theta^\mu_\mu + \eta_{\mu\nu} \partial_\kappa \tau^{\mu\nu\kappa}$$

whereas eqs (14) and (18) yield

$$\partial_\mu \Theta^\mu = \Theta^\mu_\mu + \partial_\mu \Sigma^\mu ,$$

so we get

$$T^\mu_\mu = \partial_\kappa \left(\Theta^\kappa - \Sigma^\kappa + \eta_{\mu\nu} \tau^{\mu\nu\kappa}\right) . \tag{47}$$

These relations can be used to draw the following conclusions:

1. In a Poincaré invariant theory, where $\Theta^{\mu\nu}$ and $\Theta^{\mu\kappa\lambda}$ are conserved (see eqs (14) and (17)), symmetry of $T^{\mu\nu}$ can be achieved by setting

   $$\tau^{\kappa\mu\nu} = \Sigma^{\kappa\mu\nu} . \tag{48}$$

   Then the improved moment tensor of eq. (30) is given by the following expression:

   $$T^{\mu\kappa\lambda} = \Theta^{\mu\kappa\lambda} + \tfrac{1}{2} \partial_\nu \left(x^\kappa \left(\Sigma^{\nu\mu\lambda} + \Sigma^{\mu\lambda\nu} - \Sigma^{\lambda\nu\mu}\right) - x^\lambda \left(\Sigma^{\nu\mu\kappa} + \Sigma^{\mu\kappa\nu} - \Sigma^{\kappa\nu\mu}\right)\right) . \tag{49}$$

2. In a Poincaré and scale invariant theory, where $\Theta^{\mu\nu}$, $\Theta^{\mu\kappa\lambda}$ and $\Theta^\mu$ are conserved (see eqs (14), (17) and (21)), symmetry and tracelessness of $T^{\mu\nu}$ can be achieved by setting

$$\tau^{\kappa\mu\nu} = \Sigma^{\kappa\mu\nu} + \frac{1}{n-1}\left(\eta^{\kappa\mu}\partial^\nu f - \eta^{\kappa\nu}\partial^\mu f\right), \tag{50}$$

provided $f$ satisfies the differential equation

$$\Box f = \partial_\kappa\left(\Sigma^\kappa - \eta_{\mu\nu}\Sigma^{\mu\nu\kappa}\right). \tag{51}$$

Then the improved moment tensor of eq. (30) and the improved dilatation or scaling current of eq. (34) are given by the following more complicated expressions:

$$T^{\mu\kappa\lambda} = \Theta^{\mu\kappa\lambda} + \tfrac{1}{2}\partial_\nu\left(x^\kappa\left(\Sigma^{\nu\mu\lambda} + \Sigma^{\mu\lambda\nu} - \Sigma^{\lambda\nu\mu}\right) - x^\lambda\left(\Sigma^{\nu\mu\kappa} + \Sigma^{\mu\kappa\nu} - \Sigma^{\kappa\nu\mu}\right)\right)$$
$$+ \frac{1}{n-1}\left(\left(x^\kappa\eta^{\mu\lambda} - x^\lambda\eta^{\mu\kappa}\right)\Box f - \left(x^\kappa\partial^\lambda - x^\lambda\partial^\kappa\right)\partial^\mu f\right). \tag{52}$$

$$T^\mu = \Theta^\mu - \Sigma^\mu + \eta_{\kappa\lambda}\Sigma^{\kappa\lambda\mu} + \tfrac{1}{2}\partial_\kappa\left(x_\nu\left(\Sigma^{\kappa\mu\nu} + \Sigma^{\mu\nu\kappa} - \Sigma^{\nu\kappa\mu}\right)\right)$$
$$+ \frac{1}{n-1}\left(x^\mu\Box f - x_\nu\partial^\nu\partial^\mu f\right). \tag{53}$$

As an example, let us look at the scale invariant scalar field theory given by the Lagrangian (27) with the potential (29), where the scaling dimension of the field $\varphi$ has the value $d_\varphi = \tfrac{1}{2}(n-2)$ and the so-called canonical energy-momentum tensor (28) is symmetric but not traceless. In this case,

$$\Sigma^{\mu\kappa\lambda} = 0 \quad, \quad \Sigma^\mu = \tfrac{1}{2}(n-2)\,\varphi\,\partial^\mu\varphi,$$

so that eq. (51) reduces to

$$\Box f = \tfrac{1}{2}(n-2)\,\partial_\kappa\left(\varphi\,\partial^\kappa\varphi\right) = \tfrac{1}{4}(n-2)\,\Box\left(\varphi^2\right),$$

with the obvious solution

$$f = \tfrac{1}{4}(n-2)\,\varphi^2.$$

Thus the improved energy-momentum tensor is

$$T^{\mu\nu} = \partial^\mu\varphi\,\partial^\nu\varphi - \tfrac{1}{2}\eta^{\mu\nu}\partial^\kappa\varphi\,\partial_\kappa\varphi + \eta^{\mu\nu}U(\varphi) - \frac{n-2}{4(n-1)}\left(\partial^\mu\partial^\nu - \eta^{\mu\nu}\Box\right)\varphi^2, \tag{54}$$

in agreement with the result obtained in Refs [3] and [4].

Of course, the substitution $\Theta^{\mu\nu} \to T^{\mu\nu}$ performed in eqs (39) and (42) is not just a formal step. After all, the components of the energy-momentum tensor, which are the densities and flux densities of energy and momentum, are observable quantities, and experiment must decide what are the correct expressions. In particular, the expression given in eq. (41) for the electromagnetic field has passed all experimental tests. The same applies to the components of the moment tensor, which are the density and flux density of angular momentum, and, to a somewhat lesser degree, to the dilatation or scaling current. It is also clear that the physically correct expression must be free of diseases such as asymmetry or lack of gauge invariance, which means that the so-called canonical energy-momentum tensor is really an unphysical object resulting from a formal construction: it is neither canonical nor does it represent the physical energy or momentum density of anything.[3] From this point of view, one may even be tempted to call the entire strategy of "improvement" into question, at least as long as it remains essentially an "ad hoc" procedure for getting the right result from the wrong one. What is really needed is a general method that allows to derive the physically correct energy-momentum tensor of any field theory, on flat as well as curved space-time, either by means of a general guideline for directing the strategy of "improvement" towards a definite and unique result or else from scratch and without any intermediate steps. This is the problem that we shall address in the remainder of this paper.

# 3 Geometric Formulation

## 3.1 General Considerations

In a geometric setting, classical fields are sections of fiber bundles over space-time, which we assume to be an $n$-dimensional manifold $M$. Simple examples show that these bundles cannot in general be expected to be trivial and that even when they are topologically trivial, they do not carry any distinguished trivialization. It is also important to note that these bundles do not in general carry any additional structure, except when one restricts oneself to special types of fields.

- *Vector bundles* arise naturally in theories with *linear matter fields* and also in general relativity: the metric tensor is an example.

- *Affine bundles* can be employed to incorporate *gauge fields*, since connections in a principal $G$-bundle $P$ over space-time $M$ can be viewed as sections of the *connection bundle* of $P$ – an affine bundle $CP$ over $M$ constructed from $P$.

- *General fiber bundles* are used to handle *nonlinear matter fields*, in particular those corresponding to maps from space-time $M$ to some target manifold $Q$: a standard example are the nonlinear sigma models.

In what follows, we shall therefore suppose that the fields of the theory under study can be represented as the sections of some given fiber bundle $E$ over $M$, with projection $\pi : E \to M$ and typical fiber $Q$, usually referred to as the *configuration bundle* of the theory; another frequently used term is *field bundle*.[4] The standard procedure here is to gather *all* the various bundles carrying the various fields that appear in a given theory into one big "total" configuration bundle. However, the resulting picture tends to obscure the special role played by interaction mediating fields such as the metric tensor and Yang-Mills fields. In fact, physically realistic models of field theory are always made up of various sectors containing different types of fields that interact among themselves and with each other. Such sectors can be defined by assuming the "total" configuration bundle $E$ over $M$ to be the *fiber product* over $M$ of various "partial" configuration bundles $E^{(1)}, \ldots, E^{(r)}$ over $M$,

$$E = E^{(1)} \times_M \ldots \times_M E^{(r)}, \tag{55}$$

since a section $\varphi$ of this fiber product corresponds simply to a multiplet $(\varphi^{(1)}, \ldots, \varphi^{(r)})$ formed by sections $\varphi^{(k)}$ of $E^{(k)}$ ($k = 1, \ldots, r$). As we shall see in the next section, such splittings play a central role for the correct understanding of concepts such as currents and the energy-momentum tensor.

---

[3]This is why we insist on carrying along the prefix "so-called".

In this general framework, symmetries are always represented by bundle automorphisms and their generators are represented by projectable vector fields. Briefly, these concepts are defined as follows. First, a *bundle automorphism* of $E$ is a fiber preserving diffeomorphism $\phi_E : E \to E$. This means, of course, that $\phi_E$ induces a diffeomorphism $\phi_M : M \to M$ such that the diagram

$$\begin{array}{ccc} E & \xrightarrow{\phi_E} & E \\ \pi \downarrow & & \downarrow \pi \\ M & \xrightarrow{\phi_M} & M \end{array} \tag{56}$$

commutes; then $\phi_E$ is said to *cover* $\phi_M$. Moreover, a bundle automorphism is said to be *strict* if it covers the identity on $M$. The set $\mathrm{Aut}(E)$ of all bundle automorphisms of $E$ is a group, called the *automorphism group* of $E$, and the set $\mathrm{Aut}_s(E)$ of all strict bundle automorphisms of $E$ is a normal subgroup of $\mathrm{Aut}(E)$ such that

$$\mathrm{Aut}(E)/\mathrm{Aut}_s(E) \cong \mathrm{Diff}(M). \tag{57}$$

We also define the *support* of a diffeomorphism $\phi_M$ of $M$ to be the closure of the set of points in $M$ on which it does not act as the identity:

$$\mathrm{supp}\, \phi_M = \overline{\{x \in M / \phi_M(x) \neq x\}}. \tag{58}$$

---

[4] We prefer not to adopt this term because according to standard terminology, derivatives of fields are again fields, whereas derivatives of sections of $E$ are no longer sections of $E$ but of $JE$, the first order jet bundle of $E$; see below.

Similarly, the *base support* of an automorphism $\phi_E$ of $E$ is defined as the closure of the set of points in $M$ for which it does not act as the identity on the respective fiber:
$$\operatorname{supp}\phi_E \;=\; \overline{\{x \in M / \phi_E(e) \neq e \text{ for some } e \in E_x\}} \;. \tag{59}$$
Next, a vector field $X_E$ on $E$ is said to be *projectable* if
$$T_{e_1}\pi(X_E(e_1)) \;=\; T_{e_2}\pi(X_E(e_2)) \qquad \text{for } e_1, e_2 \in E \text{ with } \pi(e_1) = \pi(e_2) \tag{60}$$
where $T\pi : TE \to TM$ is the tangent map to the projection $\pi : E \to M$. This means, of course, that $X_E$ induces a vector field $X_M$ on $M$ to which it is $\pi$-related:
$$X_M(x) \;=\; T_e\pi(X_E(e)) \qquad \text{for } x \in M \text{ and } e \in E \text{ with } \pi(e) = x \;. \tag{61}$$
Then $X_E$ is also said to *cover* $X_M$. Moreover, a projectable vector field on $E$ is said to be *vertical* if it covers the zero vector field on $M$. The set $\mathfrak{X}_P(E)$ of all projectable vector fields on $E$ is a Lie algebra and the set $\mathfrak{X}_V(E)$ of all vertical vector fields on $E$ is an ideal in $\mathfrak{X}_P(E)$ such that
$$\mathfrak{X}_P(E)/\mathfrak{X}_V(E) \;\cong\; \mathfrak{X}(M) \;. \tag{62}$$
Of course, a vector field on $E$ is projectable or vertical if and only if its flow consists of (local) bundle automorphisms or (local) strict bundle automorphisms, respectively, so formally $\mathfrak{X}_P(E)$ is the Lie algebra of $\operatorname{Aut}(E)$ and $\mathfrak{X}_V(E)$ is the Lie algebra of $\operatorname{Aut}_s(E)$; we also refer to projectable vector fields as *infinitesimal bundle automorphisms* and to vertical vector fields as *infinitesimal strict bundle automorphisms*. An alternative interpretation that we shall often adopt without further mention is to think of a bundle automorphism as being the pair $(\phi_M, \phi_E)$, simply denoted by $\phi$, and similarly of an infinitesimal bundle automorphism as being the pair $(X_M, X_E)$, simply denoted by $X$. In this sense, a bundle automorphism $\phi$ acts on a section $\varphi$ of $E$ according to
$$(\phi \cdot \varphi)(x) \;=\; \phi_E \cdot \bigl(\varphi(\phi_M^{-1}(x))\bigr) \qquad \text{for } x \in M \;. \tag{63}$$
By differentiation, this leads to the following formula for the variation $\delta_X \varphi$ of a section $\varphi$ of $E$ under an infinitesimal bundle automorphism $X$, noting that this variation, as any other one, must be a section of the vector bundle $\varphi^*(VE)$ over $M$ obtained from the vertical bundle $VE$ of $E$ by pull-back via $\varphi$:
$$(\delta_X \varphi)(x) \;=\; X_E(\varphi(x)) \,-\, T_x\varphi \cdot X_M(x) \qquad \text{for } x \in M \;. \tag{64}$$
In this context, it should be noted that since $\operatorname{Diff}(M)$ is a quotient group but not a subgroup of $\operatorname{Aut}(E)$, there is "a priori" no natural way of letting a diffeomorphism of $M$ act on a section of $E$: this can only be defined by giving a prescription for lifting diffeomorphisms of $M$ to automorphisms of $E$. In general, such a lifting procedure does not exist globally, although for many purposes it is sufficient to define it infinitesimally, that is, as a lifting from vector fields on $M$ to projectable vector fields on $E$, but even when it does exist, its definition often requires additional input data: this happens, for example, when $E$ is a vector bundle associated to some principal $G$-bundle $P$ over $M$ describing internal symmetries. If, on the other hand, $E$ is one of the tensor bundles $T_s^r M$ of $M$, a natural and globally defined lifting procedure does exist: it consists in taking $\phi_E$ to be the corresponding tensor power $T_s^r \phi_M$ of the tangent map $T\phi_M$ to $\phi_M$.

## 3.2 First Order Lagrangian Formalism

In the by now standard geometric first order Lagrangian formalism of classical field theory (see, e.g., Refs [7,8] but note that we shall follow the notation employed in Ref. [9]), one starts out from a configuration bundle $E$ over $M$ as above and introduces its first order jet bundle $JE$, which can be defined as follows. Given a point $e$ in $E$ with base point $x = \pi(e)$ in $M$, the fiber $J_e E$ of $JE$ at $e$ consists of all linear maps from the tangent space $T_x M$ of the base space $M$ at $x$ to the tangent space $T_e E$ of the total space $E$ at $e$ whose composition with the tangent map $T_e \pi : T_e E \to T_x M$ to the projection $\pi : E \to M$ gives the identity on $T_x M$:

$$J_e E \;=\; \{\, u_e \in L(T_x M, T_e E) \,/\, T_e \pi \circ u_e = \mathrm{id}_{T_x M} \,\} \; . \tag{65}$$

Obviously, $J_e E$ is an affine subspace of the vector space $L(T_x M, T_e E)$ of all linear maps from $T_x M$ to $T_e E$ and hence $JE$, the disjoint union of all the spaces $J_e E$ as $e$ varies over $E$, is an affine bundle over $E$ with respect to the *target projection* $\tau_{JE} : JE \to E$ that takes all of $J_e E$ to the point $e$. Of course, it is also a fiber bundle over $M$ with respect to the *source projection* $\sigma_{JE} : JE \to M$ defined by $\sigma_{JE} = \pi \circ \tau_{JE}$, but without any additional structure. A section $\varphi$ of $E$ over $M$ gives rise to a section $j\varphi$ of $JE$ over $M$ called its *jet prolongation* and defined by taking $(j\varphi)(x) \in J_{\varphi(x)} E$ to be the tangent map $T_x \varphi$ to $\varphi$ at $x$; it will also be denoted by $(\varphi, \partial \varphi)$ to suggest that its value at any point of $M$ incorporates all the information contained in the value of $\varphi$ and of its first order (partial) derivatives at that point. Indeed, in adapted local coordinates $(x^\mu, q^i)$ for $E$ derived from local coordinates $x^\mu$ for $M$, local coordinates $q^i$ for $Q$ and a local trivialization of $E$ over $M$, as well as the induced local coordinates $(x^\mu, q^i, q^i_\mu)$ for $JE$, a section $\varphi$ fixes the $q^i$ to be given functions of $x$, $q^i = \varphi^i(x)$, whereas its jet prolongation $j\varphi = (\varphi, \partial \varphi)$ fixes the $q^i_\mu$ to be their partial derivatives, $q^i_\mu = \partial_\mu \varphi^i(x)$. Moreover, given a function $\mathcal{L}$ on $JE$, we also use the abbreviations

$$\frac{\partial \mathcal{L}}{\partial \varphi^i} \;=\; \frac{\partial \mathcal{L}}{\partial q^i}(\varphi, \partial \varphi) \quad , \quad \frac{\partial \mathcal{L}}{\partial\, \partial_\mu \varphi^i} \;=\; \frac{\partial \mathcal{L}}{\partial q^i_\mu}(\varphi, \partial \varphi) \tag{66}$$

to denote the pull-back of the partial derivatives of $\mathcal{L}$ with respect to the field variables by the jet prolongation $j\varphi = (\varphi, \partial \varphi)$ of $\varphi$; this is intended to make closer contact with the notation used in the previous section, which is the standard one employed by physicists. Similarly, we sometimes even abbreviate $\mathcal{L}(\varphi, \partial \varphi)$ to $\mathcal{L}$, by abuse of notation.

The analysis of symmetries in this framework relies on the observation that every bundle automorphism $\phi_E$ of $E$ over $M$ has a *jet prolongation* to an affine bundle automorphism $\phi_{JE}$ of $JE$ over $E$ covering $\phi_E$, i.e., such that the diagram

$$
\begin{array}{ccc}
JE & \xrightarrow{\phi_{JE}} & JE \\
\tau_{JE} \downarrow & & \downarrow \tau_{JE} \\
E & \xrightarrow{\phi_E} & E \\
\pi \downarrow & & \downarrow \pi \\
M & \xrightarrow{\phi_M} & M
\end{array}
\qquad (67)
$$

commutes: it can be defined explicitly by

$$\phi_{JE}(u_e) \;=\; T_e \phi_E \circ u_e \circ (T_x \phi_M)^{-1} \qquad \text{for } u_e \in J_e E \;. \tag{68}$$

Similarly, every projectable vector field $X_E$ on $E$ has a *jet prolongation* to a projectable vector field $X_{JE}$ on $JE$ covering $X_E$: it can be defined explicitly by applying eq. (68) to (local) one-parameter groups of (local) bundle automorphisms and differentiating with respect to the flow parameter. The result is most conveniently expressed in adapted local coordinates as above, where

$$ X_M \;=\; X^\mu \frac{\partial}{\partial x^\mu} \;, \tag{69}$$

and

$$ X_E \;=\; X^\mu \frac{\partial}{\partial x^\mu} \;+\; X^i \frac{\partial}{\partial q^i} \;, \tag{70}$$

so that

$$ X_{JE} \;=\; X^\mu \frac{\partial}{\partial x^\mu} \;+\; X^i \frac{\partial}{\partial q^i} \;+\; \left( \frac{\partial X^i}{\partial x^\mu} + q^j_\mu \frac{\partial X^i}{\partial q^j} - q^i_\nu \frac{\partial X^\nu}{\partial x^\mu} \right) \frac{\partial}{\partial q^i_\mu} \;. \tag{71}$$

Note that the coefficient functions $X^\mu$ in eq. (70) depend only on the base coordinates $x^\rho$ but not on the fiber coordinates $q^r$ of $E$, expressing the fact that $X_E$ is projectable and covers $X_M$. Similarly, the coefficient functions $X^\mu$ and $X^i$ in eq. (71) depend only on the coordinates $x^\rho$ and $q^r$ but not on the fiber coordinates $q^r_\rho$ of $JE$, expressing the fact that $X_{JE}$ is projectable and covers $X_E$; moreover, the affine dependence of the remaining coefficients of $X_{JE}$ on the fiber coordinates $q^r_\rho$ of $JE$ reflects the fact that the (local) bundle automorphisms generated by this vector field are affine.

Extending the alternative interpretation of bundle automorphisms and infinitesimal bundle automorphisms mentioned at the end of the previous subsection, we shall also think of the former as triples $(\phi_M, \phi_E, \phi_{JE})$, again simply denoted by $\phi$, and similarly of the latter as triples $(X_M, X_E, X_{JE})$, again simply denoted by $X$. In this sense, a
bundle automorphism $\phi$ acts on the 1-jet $j\varphi = (\varphi, \partial\varphi)$ of a section $\varphi$ of $E$ according to

$$(\phi \cdot j\varphi)(x) = \phi_{JE} \cdot \left(j\varphi(\phi_M^{-1}(x))\right) \qquad \text{for } x \in M \ . \tag{72}$$

As before, this leads to the following formula for the variation $\delta_X j\varphi = (\delta_X \varphi, \delta_X \partial\varphi)$
of the 1-jet $j\varphi = (\varphi, \partial\varphi)$ of a section $\varphi$ of $E$ under an infinitesimal bundle automorphism $X$:

$$(\delta_X j\varphi)(x) = X_{JE}(j\varphi(x)) - T_x j\varphi \cdot X_M(x) \qquad \text{for } x \in M \ . \tag{73}$$

In adapted local coordinates as above, eq. (64) assumes the form

$$\delta_X \varphi^i = X^i(\varphi) - \partial_\nu \varphi^i X^\nu \ , \tag{74}$$

while eq. (73) becomes

$$\delta_X \partial_\mu \varphi^i = \frac{\partial X^i}{\partial x^\mu}(\varphi) + \partial_\mu \varphi^j \frac{\partial X^i}{\partial q^j}(\varphi) - \partial_\nu \varphi^i \frac{\partial X^\nu}{\partial x^\mu} - \partial_\nu \partial_\mu \varphi^i X^\nu \ . \tag{75}$$

It is amusing to note that this implies the useful identity

$$\delta_X \partial_\mu \varphi^i = \partial_\mu \delta_X \varphi^i \ . \tag{76}$$

The geometric first order Lagrangian formalism of classical field theory starts from
the assumption that the dynamics of the theory is fixed by prescribing a *Lagrangian* $\hat{\mathcal{L}}$
which, formally, is a map

$$\hat{\mathcal{L}} : JE \longrightarrow \pi^*(\bigwedge^n T^*M) \tag{77}$$

of fiber bundles over $E$. Assuming, as always, that $M$ is orientable,[5] this guarantees
that $\hat{\mathcal{L}}$ can be integrated to define the *action* or, more precisely, the action over any
compact subset $K$ of $M$, which is the functional on sections $\varphi$ of $E$ defined by

$$S_K[\varphi] = \int_K \hat{\mathcal{L}}(\varphi, \partial\varphi) \ . \tag{78}$$

In adapted local coordinates as above, with $K$ contained in the domain of definition of
the space-time coordinates $x^\mu$, we write

$$\hat{\mathcal{L}} = \mathcal{L} \, d^n x \tag{79}$$

where by abuse of language, $\mathcal{L}$ is also called the *Lagrangian*, to obtain

$$S_K[\varphi] = \int_K d^n x \, \mathcal{L}(\varphi, \partial\varphi) \ . \tag{80}$$

---

[5] If $M$ is not orientable, we replace it by its twofold orientation cover.

(For simplicity, the possible explicit dependence of $\mathcal{L}$ on the space-time coordinates is suppressed from the notation.) As in the special case discussed in the previous section, the equations of motion for the field $\varphi$ are derived from the requirement that, for any compact subset $K$ of $M$, the action $S_K[\varphi]$ should be stationary under variations $\delta\varphi$ of the field $\varphi$ that vanish on the boundary of $K$. In adapted local coordinates as above, with $K$ contained in the domain of definition of the space-time coordinates $x^\mu$, this stationarity condition can be evaluated by performing an explicit partial integration to compute the following expression for the variation of the action induced by an arbitrary variation $\delta\varphi$ of $\varphi$:

$$\begin{aligned}
\delta S_K[\varphi] &= \int_K d^n x \left( \frac{\partial \mathcal{L}}{\partial q^i}(\varphi, \partial\varphi)\, \delta\varphi^i + \frac{\partial \mathcal{L}}{\partial q^i_\mu}(\varphi, \partial\varphi)\, \delta\,\partial_\mu \varphi^i \right) \\
&= \int_K d^n x \left( \frac{\partial \mathcal{L}}{\partial q^i}(\varphi, \partial\varphi) - \partial_\mu \left( \frac{\partial \mathcal{L}}{\partial q^i_\mu}(\varphi, \partial\varphi) \right) \right) \delta\varphi^i \qquad (81) \\
&\quad + \int_K d^n x\; \partial_\mu \left( \frac{\partial \mathcal{L}}{\partial q^i_\mu}(\varphi, \partial\varphi)\, \delta\varphi^i \right) .
\end{aligned}$$

Indeed, since the second term vanishes due to the boundary condition imposed on $\delta\varphi$ and since, except for this boundary condition, $\delta\varphi$ is arbitrary, we arrive at the standard Euler-Lagrange equation which, taking into account the abbreviation (66), is the same as in the simple ungeometric situation discussed in the previous section:

$$\frac{\partial \mathcal{L}}{\partial \varphi^i} - \partial_\mu \left( \frac{\partial \mathcal{L}}{\partial\, \partial_\mu \varphi^i} \right) = 0 . \qquad (82)$$

Globally, note that a variation $\delta\varphi$ of $\varphi$ is a section of the vector bundle $\varphi^*(VE)$ over $M$ obtained from the vertical bundle $VE$ of $E$ by pull-back via $\varphi$, and what has just been proved is that the induced variation (or functional derivative) of the action can be written in the special form

$$\delta S_K[\varphi] = \int_K \frac{\delta \hat{\mathcal{L}}}{\delta \varphi}[\varphi] \cdot \delta\varphi \qquad (83)$$

or

$$\delta S_K[\varphi] = \int_K d^n x\, \frac{\delta \mathcal{L}}{\delta \varphi}[\varphi] \cdot \delta\varphi = \int_K d^n x\, \frac{\delta \mathcal{L}}{\delta \varphi^i}[\varphi]\, \delta\varphi^i \qquad (84)$$

where $\delta\hat{\mathcal{L}}/\delta\varphi$ evaluated at the section $\varphi$ is an $n$-form with values in $\varphi^*(V^*E)$ and (locally) $\delta\mathcal{L}/\delta\varphi$ evaluated at the section $\varphi$ is a section of $\varphi^*(V^*E)$, the vector bundle over $M$ dual to $\varphi^*(VE)$, called the variational derivative or Euler-Lagrange derivative of $\hat{\mathcal{L}}$ and of $\mathcal{L}$, respectively, at $\varphi$.[6] Explicitly,

$$\frac{\delta \mathcal{L}}{\delta \varphi^i} = \frac{\partial \mathcal{L}}{\partial \varphi^i} - \partial_\mu \left( \frac{\partial \mathcal{L}}{\partial\, \partial_\mu \varphi^i} \right) . \qquad (85)$$

---

[6]What is special about eq. (84), say, is that it shows the functional derivative of the action, viewed as a linear functional on the space of smooth sections of $\varphi^*(VE)$, to be a regular functional, with a smooth integral kernel given by eq. (85), and not just a distribution.

## 3.3 Noether's Theorems

Our goal in this subsection will be to formulate, within the first order Lagrangian formalism outlined in the previous subsection, the well-known relation between (continuous) symmetries and conservation laws established by Noether's theorems. We begin by defining, for a given Lagrangian $\hat{\mathcal{L}}$ as in eq. (77) and a given bundle automorphism $\phi$, the transform of $\hat{\mathcal{L}}$ under $\phi$ to be the Lagrangian

$$\hat{\mathcal{L}}^\phi = \phi_M^* \circ \hat{\mathcal{L}} \circ \phi_{JE} . \tag{86}$$

Then $\hat{\mathcal{L}}$ is said to be *invariant* under $\phi$ and $\phi$ is said to be a *symmetry* of $\hat{\mathcal{L}}$ if

$$\hat{\mathcal{L}}^\phi = \hat{\mathcal{L}} . \tag{87}$$

Similarly, for a given Lagrangian $\hat{\mathcal{L}}$ as in eq. (77) and a given infinitesimal bundle automorphism $X$, $\hat{\mathcal{L}}$ is said to be *invariant* under $X$ and $X$ is said to be an *infinitesimal symmetry* of $\hat{\mathcal{L}}$ if

$$L_X \hat{\mathcal{L}} = 0 \tag{88}$$

where $L_X$ denotes the Lie derivative along $X$, applied to functions on $JE$ with values in $n$-forms on $M$, which can be obtained as an appropriate combination of the Lie derivative of functions on $JE$ along $X_{JE}$ with the Lie derivative of $n$-forms on $M$ along $X_M$. Of course, $\hat{\mathcal{L}}$ is invariant under an infinitesimal bundle automorphism $X$ if and only if it is invariant under the (local) one-parameter group of (local) bundle automorphisms generated by $X$. It is also clear that a symmetry $\phi$ of $\hat{\mathcal{L}}$ leaves the action invariant, in the sense that for any section $\varphi$ of $E$ over $M$ and any compact subset $K$ of $M$,

$$S_K[\phi \cdot \varphi] = S_{\phi_M(K)}[\varphi] . \tag{89}$$

In particular, it transforms solutions of the equations of motion into solutions of the equations of motion. Similarly, an infinitesimal symmetry $X$ of $\hat{\mathcal{L}}$ leaves the action unchanged, in the sense that for any section $\varphi$ of $E$ over $M$ and any compact subset $K$ of $M$,

$$\delta_X S_K[\varphi] = \int_K L_X \hat{\mathcal{L}}(\varphi, \partial\varphi) = 0 . \tag{90}$$

In adapted local coordinates as above, with $K$ contained in the domain of definition of the space-time coordinates $x^\mu$, this infinitesimal invariance condition can be evaluated by computing the following explicit expression for the variation of the action induced by a variation $\delta_X \varphi$ of $\varphi$ generated by an infinitesimal bundle automorphism $X$:

$$\delta_X S_K[\varphi] = \int_K d^n x \left( \partial_\mu \Big( \mathcal{L}(\varphi, \partial\varphi) X^\mu \Big) + \delta_X^F \mathcal{L}(\varphi, \partial\varphi) \right) . \tag{91}$$

Here, the first term comes from the variation of the integration domain (representing the Lie derivative of $n$-forms on $M$ along $X_M$ mentioned above), while the second denotes the part of the variation of $\mathcal{L}$ due to the variation of the fields alone (representing the Lie derivative of functions on $JE$ along $X_{JE}$ mentioned above, pulled back to $M$):

$$\delta_X^F \mathcal{L}(\varphi, \partial \varphi) \;=\; \frac{\partial \mathcal{L}}{\partial q^i}(\varphi, \partial \varphi)\, \delta_X \varphi^i \;+\; \frac{\partial \mathcal{L}}{\partial q^i_\mu}(\varphi, \partial \varphi)\, \delta_X \partial_\mu \varphi^i \;. \tag{92}$$

Thus using eq. (76), integrating by parts and inserting the abbreviations (66) and (85), we obtain

$$\delta_X S_K[\varphi] \;=\; \int_K d^n x \left( \frac{\delta \mathcal{L}}{\delta \varphi^i}\, \delta_X \varphi^i \;+\; \partial_\mu \langle j^\mu_{\text{can}}, X \rangle (\varphi, \partial \varphi) \right) , \tag{93}$$

where the expression

$$\langle j^\mu_{\text{can}}, X \rangle (\varphi, \partial \varphi) \;=\; \mathcal{L}\, X^\mu \;+\; \frac{\partial \mathcal{L}}{\partial\, \partial_\mu \varphi^i}\, \delta_X \varphi^i \;. \tag{94}$$

defines what we shall call the *canonical total Noether current*. Explicitly, we may use eq. (74) to write it in the form

$$\langle j^\mu_{\text{can}}, X \rangle (\varphi, \partial \varphi) \;=\; \frac{\partial \mathcal{L}}{\partial\, \partial_\mu \varphi^i}\, X^i(\varphi) \;-\; \left( \frac{\partial \mathcal{L}}{\partial\, \partial_\mu \varphi^i}\, \partial_\nu \varphi^i \;-\; \delta^\mu_\nu\, \mathcal{L} \right) X^\nu \;, \tag{95}$$

in complete agreement with eq. (10). (The term "total" is meant to indicate that $j_{\text{can}}$ encompasses both a "current type piece" in the original sense of the word "current", referring to internal symmetries, and an "energy-momentum tensor type piece", referring to space-time symmetries.) From eq. (93), we can then immediately derive the *first Noether theorem*, which states that if $X$ is an infinitesimal symmetry of $\hat{\mathcal{L}}$, then the canonical total Noether current is conserved "on shell" (i.e., provided $\varphi$ is a solution of the equations of motion):

$$\partial_\mu \langle j^\mu_{\text{can}}, X \rangle (\varphi, \partial \varphi) \;=\; 0 \;. \tag{96}$$

It also states that the same is true for the *improved total Noether current*, defined by adding the curl of an (as yet unspecified) *correction term*:

$$j^\mu_{\text{imp}}(X; \varphi, \partial \varphi) \;=\; \langle j^\mu_{\text{can}}, X \rangle (\varphi, \partial \varphi) \;+\; \partial_\nu j^{\mu\nu}_{\text{cor}}(X; \varphi, \partial \varphi) \;. \tag{97}$$

Indeed, supposing the correction term to be antisymmetric in its indices, we see that eq. (96) implies

$$\partial_\mu j^\mu_{\text{imp}}(X; \varphi, \partial \varphi) \;=\; 0 \;. \tag{98}$$

In the language of differential forms, the (canonical or improved) total Noether current is to be considered as an $(n-1)$-form and the correction term as an $(n-2)$-form on $M$, with respective local coordinate representations

$$\langle j_{\text{can}}, X \rangle (\varphi, \partial \varphi) \;=\; \langle j^\mu_{\text{can}}, X \rangle (\varphi, \partial \varphi)\, d^n x_\mu \;, \tag{99}$$

$$j_{\text{imp}}(X;\varphi,\partial\varphi) = j_{\text{imp}}^{\mu}(X;\varphi,\partial\varphi)\, d^n x_{\mu}\,, \tag{100}$$

and

$$j_{\text{cor}}(X;\varphi,\partial\varphi) = \tfrac{1}{2}\, j_{\text{cor}}^{\mu\nu}(X;\varphi,\partial\varphi)\, d^n x_{\mu\nu}\,, \tag{101}$$

where

$$d^n x_{\mu} = i_{\partial_{\mu}} d^n x\,, \quad d^n x_{\mu\nu} = i_{\partial_{\nu}} i_{\partial_{\mu}} d^n x\,, \tag{102}$$

so that eq. (97) becomes

$$j_{\text{imp}}(X;\varphi,\partial\varphi) = \langle j_{\text{can}}, X\rangle(\varphi,\partial\varphi) + d\, j_{\text{cor}}(X;\varphi,\partial\varphi)\,. \tag{103}$$

The first Noether theorem now states that if $X$ is an infinitesimal symmetry of $\hat{\mathcal{L}}$, then the forms $\langle j_{\text{can}}, X\rangle(\varphi,\partial\varphi)$ and $j_{\text{imp}}(X;\varphi,\partial\varphi)$ are closed "on shell" (i.e., provided $\varphi$ is a solution of the equations of motion):

$$d\, \langle j_{\text{can}}, X\rangle(\varphi,\partial\varphi) = 0\,, \tag{104}$$

$$d\, j_{\text{imp}}(X;\varphi,\partial\varphi) = 0\,. \tag{105}$$

Note that, in the second case, this statement is completely independent of the choice of the correction term.

Further insight into the nature of the various types of total Noether currents, as well as of the allowed correction terms, can be gained from studying their functional dependence on the infinitesimal bundle automorphism $X$ and also on the solution $\varphi$. The latter is simpler and will therefore be dealt with first. Here, the basic assumption (or, in the case of the canonical total Noether current, the basic fact, to be proved below) is that all these "field dependent" differential forms on $M$ are obtained from corresponding "field independent" differential forms on $JE$ by pull-back with the jet prolongation $(\varphi,\partial\varphi)$ of $\varphi$. Thus, as already suggested by the notation, $\langle j_{\text{can}}, X\rangle(\varphi,\partial\varphi)$ is the pull-back of an $(n-1)$-form $\langle j_{\text{can}}, X\rangle$ on $JE$ representing the *field independent canonical total Noether current* and similarly $j_{\text{imp}}(X;\varphi,\partial\varphi)$ is the pull-back of an $(n-1)$-form $j_{\text{imp}}(X)$ on $JE$ representing the *field independent improved total Noether current*, whereas $j_{\text{cor}}(X;\varphi,\partial\varphi)$ is the pull-back of an $(n-2)$-form $j_{\text{cor}}(X)$ on $JE$ representing a *field independent correction term*; the relation (103) is then guaranteed to hold, for any choice of solution (or even field configuration) $\varphi$, if we require that

$$j_{\text{imp}}(X) = \langle j_{\text{can}}, X\rangle + d\, j_{\text{cor}}(X)\,. \tag{106}$$

Note that this condition is sufficient but is by no means necessary: what is really needed is only eq. (103) which states that eq. (106) holds modulo forms that vanish "on shell".

In order to substantiate this picture, let us pause to explain the global significance of the canonical total Noether current, which in this context can be viewed as expressing the covariant momentum map of multisymplectic field theory [7–9]. Namely, we have

$$\langle j_{\text{can}}, X\rangle = i_{X_{JE}} \theta_{\hat{\mathcal{L}}} \tag{107}$$

where $\theta_{\hat{\mathcal{L}}}$ is the multicanonical form on $JE$ obtained from the multicanonical form $\theta$ on the affine dual of $JE$ by pull-back via the covariant Legendre transformation induced by $\hat{\mathcal{L}}$. Indeed, in adapted local coordinates, $\theta_{\hat{\mathcal{L}}}$ is given by

$$\theta_{\hat{\mathcal{L}}} \;=\; \frac{\partial \mathcal{L}}{\partial q_\mu^i} \, dq^i \wedge d^n x_\mu \;+\; \left( \mathcal{L} - q_\mu^i \, \frac{\partial \mathcal{L}}{\partial q_\mu^i} \right) d^n x \;. \qquad (108)$$

Therefore, according to eq. (71),

$$\begin{aligned}
\langle j_{\mathrm{can}} , X \rangle \;=\; & \left( \frac{\partial \mathcal{L}}{\partial q_\mu^i} \, X^i + \left( \mathcal{L} - q_\nu^j \, \frac{\partial \mathcal{L}}{\partial q_\nu^j} \right) X^\mu \right) d^n x_\mu \\
& - \, \frac{1}{2} \left( \frac{\partial \mathcal{L}}{\partial q_\mu^i} \, X^\nu - \frac{\partial \mathcal{L}}{\partial q_\nu^i} \, X^\mu \right) dq^i \wedge d^n x_{\mu\nu} \;.
\end{aligned} \qquad (109)$$

Pulling back with $(\varphi, \partial \varphi)$ and using the second of the relations

$$dx^\kappa \wedge d^n x_\mu \;=\; \delta^\kappa_\mu \, d^n x \;, \qquad (110)$$

$$dx^\kappa \wedge d^n x_{\mu\nu} \;=\; \delta^\kappa_\nu \, d^n x_\mu - \delta^\kappa_\mu \, d^n x_\nu \;, \qquad (111)$$

we see that two of the five terms cancel out and the remaining three reproduce eq. (95). The statement of the first Noether theorem can also be checked in this purely differential formulation by first writing out the infinitesimal invariance condition (88) in terms of the (local) function $\mathcal{L}$ on $JE$ given by eq. (79), which – apart from an additional term of the form $\mathcal{L} \, \partial_\mu X^\mu$ due to the fact that $\mathcal{L}$ is the coefficient of an $n$-form rather than a function – follows directly from eq. (71):

$$\frac{\partial (\mathcal{L} X^\mu)}{\partial x^\mu} + \frac{\partial \mathcal{L}}{\partial q^i} \, X^i + \frac{\partial \mathcal{L}}{\partial q_\mu^i} \left( \frac{\partial X^i}{\partial x^\mu} + q_\mu^j \, \frac{\partial X^i}{\partial q^j} - q_\nu^i \, \frac{\partial X^\nu}{\partial x^\mu} \right) \;=\; 0 \;. \qquad (112)$$

Pulling back with $(\varphi, \partial \varphi)$ and inserting the resulting expression into that obtained by explicitly taking the divergence of eq. (95) gives

$$d \, \langle j_{\mathrm{can}} , X \rangle (\varphi, \partial \varphi) \;=\; \left( \partial_\mu \left( \frac{\partial \mathcal{L}}{\partial \, \partial_\mu \varphi^i} \right) - \frac{\partial \mathcal{L}}{\partial \varphi^i} \right) \left( X^i(\varphi) - \partial_\nu \varphi^i \, X^\nu \right) d^n x \;, \qquad (113)$$

which vanishes due to the equations of motion (82).

Another important feature of the canonical total Noether current is expressed through the *second Noether theorem*, which states that if $X$ is an infinitesimal local symmetry of $\hat{\mathcal{L}}$, then this current is not only conserved but actually vanishes "on shell". To prove this statement and appreciate its consequences, we need a precise definition, at least at the infinitesimal level, of the concept of a local symmetry. The basic idea is that an infinitesimal symmetry of a Lagrangian should be regarded as local if it remains an infinitesimal symmetry even when modified through multiplication by an arbitrary space-time dependent weight factor – a factor that can be viewed as modifying, in a space-time dependent manner, the "speed" of the one-parameter group it generates. The mathematical implementation of this intuitive idea, however, is far from trivial: it hinges on a proper definition of what one means by the symmetry group of the Lagrangian and, in particular, is sensitive to the distinction between the abstract symmetry group itself and its concrete action on the fields of the theory; we shall return to this question and discuss examples in the next subsection. But the symmetry group pertinent to the canonical total Noether current is simply the group of symmetries of the Lagrangian as defined at the beginning of this subsection, which is a subgroup of the group of all bundle automorphisms of $E$, with the natural action on fields given by pull-back of sections, so we may in this specific context define an infinitesimal bundle automorphism $X$ to be an *infinitesimal local symmetry* of $\hat{\mathcal{L}}$ if, for any function $f$ on $M$, the rescaled infinitesimal bundle automorphism $fX$ is an infinitesimal symmetry of $\hat{\mathcal{L}}$.[7] Obviously, the infinitesimal local symmetries of $\hat{\mathcal{L}}$ in this sense form a module over the ring $\mathfrak{F}(M)$ of functions on $M$ and also a Lie subalgebra of the Lie algebra of all infinitesimal symmetries of $\hat{\mathcal{L}}$ which in turn is a Lie subalgebra of the Lie algebra $\mathfrak{X}_P(E)$ of all projectable vector fields on $E$: this can be seen directly from the defining condition (88), using elementary properties of the Lie derivative on differential forms. (Indeed, observe that the Lie derivative of a function on $E$ which is the pull-back of a function $f$ on $M$ along a vector field $X_E$ on $E$ which is projectable to a vector field $X_M$ on $M$ is again the pull-back of a function on $M$, namely the Lie derivative of $f$ along $X_M$. Thus for any two infinitesimal local symmetries $X_E$ and $Y_E$ of $\hat{\mathcal{L}}$ and any function $f$ on $M$, the expression $f[X_E, Y_E] = [fX_E, Y_E] + (Y_E \cdot f) X_E$ is an infinitesimal symmetry of $\hat{\mathcal{L}}$ because $fX_E$ and $(Y_E \cdot f) X_E = (Y_M \cdot f) X_E$ both are, showing that the Lie bracket $[X_E, Y_E]$ is again an infinitesimal local symmetry of $\hat{\mathcal{L}}$.) With this definition, it is easy to see that the canonical total Noether current vanishes "on shell" as soon as $X$ is an infinitesimal local symmetry of $\hat{\mathcal{L}}$, because given a solution $\varphi$ of the equations of motion and an infinitesimal bundle automorphism $X$ such that, for any function $f$ on $M$, $fX$ is an infinitesimal symmetry of $\hat{\mathcal{L}}$, we may apply eq. (104) with $X$ and with $fX$ to get

$$0 = d \langle j_{\text{can}}, fX \rangle(\varphi, \partial\varphi) - f d \langle j_{\text{can}}, X \rangle(\varphi, \partial\varphi) = df \wedge \langle j_{\text{can}}, X \rangle(\varphi, \partial\varphi),$$

where in the second equation, we have also used the fact that $\langle j_{\text{can}}, X \rangle$ and its pull-back $\langle j_{\text{can}}, X \rangle(\varphi, \partial\varphi)$ are $\mathfrak{F}(M)$-linear in $X$ (that is, in $X_E$) – a property which is not obvious from the global formula (107) but can for instance be read off from the local coordinate representations (109) or (95), in which the derivatives of the coefficient functions $X^\mu$ and $X^i$ that are present in the formula (71) for $X_{JE}$ have dropped out. Now the claim follows from the fact that a differential form whose exterior product with every 1-form is zero must vanish.

The argument just given makes it clear that applying the same kind of reasoning to the improved total Noether current (or its constituents) will require a more careful analysis of its functional dependence on the infinitesimal bundle automorphism $X$, a dependence that has not been specified so far since no assumption has been made on the nature of the correction term: the $(n-2)$-form $j_{\text{cor}}(X)$ on $JE$ is still completely arbitrary. In order to fix this term and arrive at a unique expression for the improved total Noether current which provides the correct expressions for both the physical current and the physical energy-momentum tensor, something more is needed, something that lies beyond the realm of a purely Noetherian approach. This is the subject that we shall turn to next.

---

[7]We identify functions $f$ on $M$ with their pull-back to $E$, defined simply by composition with the projection $\pi$.

## 3.4 Improvement and the Ultralocality Principle

Within the formalism developed in the previous subsections, symmetries in first order Lagrangian field theories can be described by fixing a symmetry group, which is (at least formally) a Lie group $\hat{G}$, together with an action of $\hat{G}$ on the configuration bundle $E$ by bundle automorphisms, or at the infinitesimal level, by fixing a symmetry algebra, which is a Lie algebra $\hat{\mathfrak{g}}$, together with a representation (up to sign) of $\hat{\mathfrak{g}}$ by projectable vector fields on $E$, that is, a Lie algebra homomorphism[8]

$$\hat{\mathfrak{g}} \longrightarrow \mathfrak{X}_P(E) \,. \tag{114}$$

Typically, $\hat{G}$ and $\hat{\mathfrak{g}}$ will be finite-dimensional when we are dealing with global symmetries but will be infinite-dimensional as soon as we are dealing with local symmetries. Moreover, in this case, there is an additional structure: just like $\mathfrak{X}_P(E)$, $\hat{\mathfrak{g}}$ is not only a Lie algebra but also a module over the ring $\mathfrak{F}(M)$ of functions on space-time – normally the module of sections of some vector bundle over $M$. However, the linear map (114) will normally *not* be a homomorphism of $\mathfrak{F}(M)$-modules. As a result, its composition with the (field independent) canonical total Noether current, which can itself be viewed as a linear map

$$\begin{array}{rcl} j_{\text{can}} : \ \mathfrak{X}_P(E) & \longrightarrow & \Omega^{n-1}(JE) \\ X & \longmapsto & \langle j_{\text{can}}, X \rangle \end{array} \tag{115}$$

and actually, as argued above, as a homomorphism of $\mathfrak{F}(M)$-modules, will normally *not* be a homomorphism of $\mathfrak{F}(M)$-modules either. *Improvement is the strategy designed to cure this defect.* More explicitly, the (field independent) improved total Noether current, viewed as a linear map

$$\begin{array}{rcl} j_{\text{imp}} : \ \mathfrak{X}_P(E) & \longrightarrow & \Omega^{n-1}(JE) \\ X & \longmapsto & j_{\text{imp}}(X) \end{array} \tag{116}$$

is fixed by imposing the following principle (the terminology has been adapted from the theory of integrable systems [11]):

---

[8]Strictly speaking, the infinitesimal version of a (left) group action is a Lie algebra antihomomorphism, but this can be easily corrected by an adequate choice of sign.

**Ultralocality Principle**: *The correction term in eq. (106) must be chosen in such a way that the composition of the linear map (114) with the linear map (116) becomes a homomorphism of $\mathfrak{F}(M)$-modules.*

The content of this principle is best understood by considering two special cases of fundamental physical importance; they will be dealt with separately but in parallel in order to emphasize the differences as well as the similarities.

1. Field theories with gauge invariance:

   The expression "gauge invariance" is usually employed to indicate the presence of a local internal symmetry group $\hat{G}$ acting on the configuration bundle of the theory by strict bundle automorphisms. The mathematical formulation of a standard gauge theory whose structure group is a (finite-dimensional) Lie group $G$ with corresponding (finite-dimensional) Lie algebra $\mathfrak{g}$, say, starts with the introduction of a principal $G$-bundle $P$ over $M$ to which all other bundles appearing in the theory are associated. For example, the configuration bundle $E$ is associated to $P$ by means of a given action of $G$ on its typical fiber $Q$. Similarly, the group $\hat{G}$ is defined to be the group of $G$-equivariant strict automorphisms of $P$, which is isomorphic to the space $\Gamma(P \times_G G)$ of sections of the associated bundle $P \times_G G$ obtained by letting $G$ act on $G$ itself by conjugation, and correspondingly, the Lie algebra $\hat{\mathfrak{g}}$ is defined to be the Lie algebra of $G$-equivariant vertical vector fields on $P$, which is isomorphic to the space $\Gamma(P \times_G \mathfrak{g})$ of sections of the associated bundle $P \times_G \mathfrak{g}$ obtained by letting $G$ act on $\mathfrak{g}$ via the adjoint representation; of course, $\Gamma(P \times_G G)$ is (at least formally) a Lie group because $P \times_G G$ is a bundle of Lie groups, and correspondingly, $\Gamma(P \times_G \mathfrak{g})$ is a Lie algebra because $P \times_G \mathfrak{g}$ is a bundle of Lie algebras. (Variants are obtained by considering instead appropriate "sufficiently large" subgroups and corresponding subalgebras, such as the group $\Gamma_c(P \times_G G)$ of sections of $P \times_G G$ with compact support ($\equiv 1$ outside some compact subset of $M$) and the Lie algebra $\Gamma_c(P \times_G \mathfrak{g})$ of sections of $P \times_G \mathfrak{g}$ with compact support ($\equiv 0$ outside some compact subset of $M$), but the considerations and arguments to be presented in what follows are easily adapted to cover such situations; this will be left to the reader without further mention in order not to overload the notation.) In addition, $\Gamma(P \times_G \mathfrak{g})$ is obviously an $\mathfrak{F}(M)$-module: its elements are the common infinitesimal local symmetries for all gauge invariant Lagrangians. This provides one realization of the general structure advocated in eq. (114) above: the explicit action of the symmetry generators on the fields is coded into a Lie algebra homomorphism

   $$\begin{array}{rcl} \Gamma(P \times_G \mathfrak{g}) & \longrightarrow & \mathfrak{X}_P(E) \\ \xi & \longmapsto & \xi_E \end{array}. \qquad (117)$$

   The map (117) is of course $\mathbb{R}$-linear but not necessarily $\mathfrak{F}(M)$-linear. In general, it is only required to be *local*, which means that if $\xi$ vanishes in some region of space-time, that is, in a certain open subset $U$ of $M$, then $\xi_E$ vanishes in that same region, or rather in the open subset $\pi^{-1}(U)$ of $E$. According to Peetre's theorem, this can be recast into the condition that, for every point $x$ in $M$, the value of $\xi_E$ at any point in the fiber $E_x$ of $E$ over $x$ should depend only on the value of $\xi$ and its partial derivatives up to a certain fixed order at $x$. Explicitly, in adapted local coordinates as before, such a map (117) amounts to a procedure for defining the vertical components $\xi_E^i$ of $\xi_E$ in terms of the components $\xi^a$ of $\xi$ by a formula of the type

$$\xi_E^i \;=\; C_a^i \, \xi^a \;+\; C_a^{i,\rho} \, \partial_\rho \xi^a \;+\; \ldots \;+\; C_a^{i,\rho_1\ldots\rho_k} \, \partial_{\rho_1} \ldots \partial_{\rho_k} \xi^a \qquad (118)$$

where the coefficients $C_a^i, C_a^{i,\rho}, \ldots, C_a^{i,\rho_1\ldots\rho_k}$ are (local) functions on $E$, whereas the horizontal components $\xi_E^\mu$ of $\xi_E$ vanish. From this point of view, requiring the map (117) to be $\mathfrak{F}(M)$-linear, or equivalently, to be *ultralocal*, in the sense that, for any point $x$ in $M$, the value of $\xi_E$ at any point in the fiber $E_x$ of $E$ over $x$ should depend only on the value of $\xi$ at $x$ but not on that of its partial derivatives, is a much stronger condition, satisfied in many models but not in all, depending on the specific nature of the fields appearing in the theory; examples will be given later. If it is satisfied, we can simply compose the linear map (117) with the linear map (115) and are done; otherwise, this composition needs improvement. In general, composing the linear map (117) with the linear map (116) gives a linear map

$$\begin{array}{rcl} \Gamma(P \times_G \mathfrak{g}) & \longrightarrow & \Omega^{n-1}(JE) \\ \xi & \longmapsto & \langle j, \xi \rangle \;=\; j_{\mathrm{imp}}(\xi_E) \end{array} \qquad (119)$$

that is claimed to be the improved current of the theory, provided the correction term used in the improvement has been chosen according to the ultralocality principle which states that the correct choice is the one that guarantees the map (119) to be $\mathfrak{F}(M)$-linear. Note that this condition has to a certain extent already been incorporated into the notation used in eq. (119), which suggests that the dependence of the improved current on the symmetry generators is given by a simple algebraic pairing and can thus be absorbed into a definition of $j$ as an $(n-1)$-form on $JE$ with coefficients in the vector bundle $P \times_G \mathfrak{g}^*$ dual to $P \times_G \mathfrak{g}$ (or rather its pull-back from $M$ to $JE$). Conversely, the ultralocality principle is an immediate consequence of this property.

2. Field theories with general covariance, or space-time diffeomorphism invariance:

   The expressions "general covariance" or "space-time diffeomorphism invariance" are usually employed to indicate the presence of a "sufficiently large" local space-time symmetry group $\hat{G}$ acting on the configuration bundle of the theory by non-strict bundle automorphisms. In a generally covariant field theory, the group $\hat{G}$ is the diffeomorphism group $\mathrm{Diff}(M)$ of $M$, and correspondingly, the Lie algebra $\hat{\mathfrak{g}}$ is the Lie algebra $\mathfrak{X}(M)$ of vector fields on $M$. (Variants are obtained by considering instead appropriate "sufficiently large" subgroups and corresponding subalgebras, such as the group $\mathrm{Diff}_c(M)$ of diffeomorphisms of $M$ with compact support and the Lie algebra $\mathfrak{X}_c(M)$ of vector fields on $M$ with compact support, but the considerations and arguments to be presented in what follows are easily adapted to cover such situations; this will be left to the reader without further mention in order not to overload the notation.) In addition, $\mathfrak{X}(M)$ is obviously an $\mathfrak{F}(M)$-module: its elements are the common infinitesimal local symmetries for all generally covariant Lagrangians.[9] This provides another realization of the general structure advocated in eq. (114) above: the explicit action of the symmetry generators on the fields is coded into a Lie algebra homomorphism

$$\begin{array}{rcl} \mathfrak{X}(M) & \longrightarrow & \mathfrak{X}_P(E) \\ X_M & \longmapsto & X_E \end{array}, \qquad (120)$$

which is a splitting of the exact sequence of Lie algebras

$$\{0\} \longrightarrow \mathfrak{X}_V(E) \longrightarrow \mathfrak{X}_P(E) \longrightarrow \mathfrak{X}(M) \longrightarrow \{0\} \qquad (121)$$

(see eq. (62)). Again, the map (120) – which is precisely the infinitesimal version of the *lifting* of $\mathrm{Diff}(M)$ into $\mathrm{Aut}(E)$ mentioned in the introduction and used in Ref. [5] to derive explicit expressions for the energy-momentum tensor – is of course $\mathbb{R}$-linear but not necessarily $\mathfrak{F}(M)$-linear.[10] In general, it is only required to be *local*, which means that if $X_M$ vanishes in some region of space-time, that is, in a certain open subset $U$ of $M$, then $X_E$ vanishes in that same region, or rather in the open subset $\pi^{-1}(U)$ of $E$. According to Peetre's theorem, this can be recast into the condition that, for every point $x$ in $M$, the value of $X_E$ at any point in the fiber $E_x$ of $E$ over $x$ should depend only on the value of $X_M$ and its partial derivatives up to a certain fixed order at $x$. Explicitly, in adapted local coordinates as before, such a map (120) amounts to a procedure for defining the vertical components $X^i$ of $X_E$ in terms of the components $X^\mu$ of $X_M$ by a formula of the type [5]

$$X^i \;=\; C^i_\mu X^\mu \,+\, C^{i,\rho}_\mu \, \partial_\rho X^\mu \,+\, \ldots \,+\, C^{i,\rho_1\ldots\rho_k}_\mu \, \partial_{\rho_1}\ldots\partial_{\rho_k} X^\mu \qquad (122)$$

where the coefficients $C^i_\mu, C^{i,\rho}_\mu, \ldots, C^{i,\rho_1\ldots\rho_k}_\mu$ are (local) functions on $E$, whereas the splitting condition imposed above fixes the horizontal components $X^\mu$ of $X_E$ to be equal to those of $X_M$. From this point of view, requiring the map (120) to be $\mathfrak{F}(M)$-linear, or equivalently, to be *ultralocal*, in the sense that, for any point $x$ in $M$, the value of $X_E$ at any point in the fiber $E_x$ of $E$ over $x$ should depend only on the value of $X_M$ at $x$ but not on that of its partial derivatives, is a much stronger condition, satisfied in some models but by far not in all, depending on the specific nature of the fields appearing in the theory; examples will be given later.

---

[9] A more detailed justification of this statement will be given later.

[10] In general, the existence of such a lifting can only be guaranteed for vector fields $X_M$ with compact support contained in a sufficiently small open subset of $M$, for example a coordinate patch, but for the purpose of constructing an improved energy-momentum tensor, this is enough.

If it is satisfied, we can simply compose the linear map (120) with the linear map (115) and are done; otherwise, this composition needs improvement. In general, composing the linear map (120) with the linear map (116) gives a linear map

$$\begin{aligned} \mathfrak{X}(M) &\longrightarrow \Omega^{n-1}(JE) \\ X_M &\longmapsto \langle T, X_M \rangle = j_{\mathrm{imp}}(X_E) \end{aligned} \qquad (123)$$

that is claimed to be the improved energy-momentum tensor of the theory, provided the correction term used in the improvement has been chosen according to the ultralocality principle which states that the correct choice is the one that guarantees the map (123) to be $\mathfrak{F}(M)$-linear. Note that this condition has to a certain extent already been incorporated into the notation used in eq. (123), which suggests that the dependence of the improved energy-momentum tensor on the symmetry generators is given by a simple algebraic pairing and can thus be absorbed into a definition of $T$ as an $(n-1)$-form on $JE$ with coefficients in the cotangent bundle $T^*M$ of $M$ (or rather its pull-back from $M$ to $JE$). Conversely, the ultralocality principle is an immediate consequence of this property.

If one is not interested in the specific form of the correction terms, the result of the entire procedure can be reformulated as a direct definition of the improved current $j$ and the improved energy-momentum tensor $T$ in integral form, as follows.

**Definition 3.1** *In first order Lagrangian field theories with gauge invariance, the (field independent)* **improved current** *$j$ is an $(n-1)$-form on $JE$ with coefficients in the vector bundle $P \times_G \mathfrak{g}^*$ with the property that, for any smooth hypersurface $\Sigma$ in $M$, any compact subset $K$ of $M$ with smooth boundary $\partial K$ intersecting $\Sigma$ in a smooth $(n-2)$-dimensional submanifold, any infinitesimal gauge transformation $\xi$ with support contained in $K$ and any section $\varphi$ of $E$ with jet prolongation $(\varphi, \partial \varphi)$, we have*

$$\int_{K \cap \Sigma} \langle j_{\mathrm{can}}, \xi_E \rangle (\varphi, \partial \varphi) = \int_{K \cap \Sigma} \langle j, \xi \rangle (\varphi, \partial \varphi) ,$$

*or more explicitly, using local coordinate notation*

$$\int_{K \cap \Sigma} d\sigma_\mu(x) \, \langle j^\mu_{\mathrm{can}}, \xi_E \rangle (\varphi, \partial \varphi)(x) = \int_{K \cap \Sigma} d\sigma_\mu(x) \, j^\mu_a(\varphi, \partial \varphi)(x) \, \xi^a(x) . \qquad (124)$$

**Definition 3.2** *In generally covariant first order Lagrangian field theories, the (field independent)* **improved energy-momentum tensor** *$T$ is an $(n-1)$-form on $JE$ with coefficients in the cotangent bundle $T^*M$ of $M$ with the property that, for any smooth hypersurface $\Sigma$ in $M$, any compact subset $K$ of $M$ with smooth boundary $\partial K$ intersecting $\Sigma$ in a smooth $(n-2)$-dimensional submanifold, any vector field $X_M$ on $M$ with support contained in $K$ and any section $\varphi$ of $E$ with jet prolongation $(\varphi, \partial \varphi)$, we have*

$$\int_{K \cap \Sigma} \langle j_{\mathrm{can}}, X_E \rangle (\varphi, \partial \varphi) = \int_{K \cap \Sigma} \langle T, X_M \rangle (\varphi, \partial \varphi) ,$$

*or more explicitly, using local coordinate notation*

$$\int_{K \cap \Sigma} d\sigma_\mu(x) \, \langle j^\mu_{\text{can}}, X_E \rangle(\varphi, \partial\varphi)(x) = \int_{K \cap \Sigma} d\sigma_\mu(x) \, T^\mu{}_\nu(\varphi, \partial\varphi)(x) \, X_M^\nu(x) \,. \quad (125)$$

The proof of these assertions relies on Stokes' theorem, which implies that the contributions to the rhs of eqs (124) and (125) coming from the correction terms can be converted into integrals over the boundary $\partial K \cap \Sigma$ which vanish since $\xi$ and $X_M$ are supposed to have support inside $K$.

Of course, this characterization of candidates for the current and the improved energy tensor is not new and has appeared many times before in the literature, with widely varying terminology; see, e.g., [5, Theorem 1.1]. It should also be pointed out that the definitions above do not by themselves guarantee existence of $j$ and/or $T$; this will be a consequence of other arguments to be presented in the next section. However, eqs (124) and (125), which can be regarded as defining $j$ and $T$ as quantities measuring the *linear response* of the fields in the theory to gauge transformations and to space-time diffeomorphisms, respectively, are by themselves sufficient to guarantee uniqueness of $j$ and $T$; in particular, $T$ is independent of the choice of the lifting map (120) used in its definition.

There is now only one fundamental problem that remains to be settled. This problem becomes apparent when we notice that the ultralocality principle hidden in eqs (124) and (125) guarantees that the second Noether theorem continues to hold for the improved expressions: the fact that the improvement terms have been chosen precisely so as to render the assignments (119) and (123) $\mathfrak{F}(M)$-linear implies that the gauge field current $j$ and the energy-momentum tensor $T$ are not only conserved but actually vanish "on shell". (The proof is the same as the one given for the canonical total Noether current at the end of the previous subsection.) Obviously, this simply cannot be true for the physical current or the physical energy-momentum tensor: these do *not* vanish just because all fields in the theory satisfy their adequate equations of motion!

Regarding our original problem of giving a general definition of the energy-momentum tensor for an arbitrary Lagrangian field theory, we thus find that the Noetherian approach, even when combined with a procedure of posterior improvement, has led us into a dilemma. On one side, we have field theories with invariance under "small" global symmetry groups, such as the special relativistic field theories studied in the previous section which are invariant under space-time translations, Lorentz transformations and, in some cases, also under dilatations. As stated before, improvement is in this case essentially an "ad hoc" procedure: what is missing is a simple physical criterion for directing it towards a definite and unique answer. At the opposite extreme, we have field theories with "large" local symmetry groups such as the ones studied in Ref. [5] where this problem can be overcome but the result is useless in practice because it vanishes "on shell". Intuitively, this can be understood by observing that if there are too many conserved currents in the theory, obtained from each other by arbitrarily redistributing local weight factors, then these currents must in fact all be zero. Of course, this is well known and corresponds to the fact that local symmetries induce constraints that restrict the phase space of the theory to the zero level of the momentum map. In particular, we conclude that in the presence of large local symmetry groups, the Noetherian construction alone is not able to provide a physically meaningful energy-momentum tensor.

In order to find a way out of this impasse, let us return to the starting point of our geometric analysis and recall that the formalism presented in this section was based on the idea that all the different bundles carrying different types of fields have been lumped together into a single big fiber bundle $E$ over $M$. If this is done, then in an entirely geometric field theory with no external data and no artificial internal ingredients (such as an explicit symmetry breaking potential), Einstein's principle of general covariance (or coordinate invariance) combined with the principle of gauge invariance, suitably reformulated in global language, will impose invariance of the total Lagrangian $\hat{\mathcal{L}}$ under the entire group $\mathrm{Aut}(E)$ of automorphisms of $E$, provided one requires *all* fields of the theory to transform appropriately. As we have seen, it then follows from the second Noether theorem that the canonical total Noether current introduced above vanishes identically "on shell", the same being true for the improved current and the improved energy-momentum tensor. But as has already been mentioned once before, physically realistic models of field theory are always made up of various sectors containing different types of fields that interact among themselves and with each other. Moreover, currents and the energy-momentum tensor will describe the mutual exchange of charges and of energy-momentum between the various sectors, that is, the flow or transfer of these quantities between the fields in different sectors. From this point of view, it becomes clear that if one lumps all fields together into a single total field $\varphi$, that total field will by definition have no partner for the exchange of charges or energy-momentum, and this is why the improved current and the improved energy-momentum tensor as defined above vanish identically "on shell". A physically meaningful definition of non-trivial conserved quantities requires a subdivision of $E$ and $\varphi$ into various sectors and a physical discussion of the distinguished role played by certain kinds of fields, especially the ones that mediate the four fundamental interactions, namely gauge fields (for the electromagnetic, weak and strong interactions) and, of course, the metric tensor (for gravity).

In order to prepare the ground for such an analysis, let us first introduce some important notions that allow to distinguish between different types of fields that appear in any given sector.

First, one may distinguish between fields that enter the Lagrangian together with their first order partial derivatives (type A) and fields that enter the Lagrangian without derivatives (type B). In other words, for fields of type A, which constitute the generic case, $\hat{\mathcal{L}}$ depends on their entire jet, whereas for fields of type B, $\hat{\mathcal{L}}$ depends only on point values of the fields. Of course, there is one other physically relevant case, namely general relativity, where the Lagrangian is constructed from curvature invariants and hence contains the metric tensor and its partial derivatives up to second order. In certain special cases, this situation can still be incorporated into the first order formalism by adopting the Palatini formulation, based on introducing, in addition to the metric tensor and as an "a priori" independent new field, a linear connection on space-time, in such a way that vanishing of the torsion and of the covariant derivative of the metric tensor (which uniquely characterize the Levi-Civita connection) become part of the equations of motion; see the extensive discussion in Chapter 21.2 of Ref. [6]. However, this procedure fails when curvature terms also appear in the matter field Lagrangian – a situation that we shall encounter in the next two sections.

More important is the distinction between external and dynamical fields. External fields represent the influence of the outside world on the system under consideration but cannot be influenced by anything happening inside the system; in particular, their space-time dependence is fixed and prescribed from outside, so they act like external parameters in a dynamical system. (Quite often, but not always, external fields are of type B; examples will be given soon.) Dynamical fields represent the physics of the system itself, and determining their space-time dependence is, both physically and mathematically, the main goal of field theory. The laws governing this space-time dependence are of course the equations of motion, so we may say briefly that dynamical fields satisfy equations of motion while external fields don't (or to be more precise, may or may not do so). Moreover, Lagrangian field theory supposes that the equations of motion are derived from a principle of stationary action, so the previous statement can be viewed as resulting from the prescription that in this derivation, the dynamical fields are varied while the external fields are kept fixed. The situation may be different for variations induced by symmetry transformations since these will sometimes not leave all external fields fixed. If they do, then it may be convenient to replace a totally symmetric formulation of the theory by one with manifest invariance only under restricted bundle automorphisms, i.e., bundle automorphisms that leave all the external fields fixed. In some cases, this is achieved by choosing special coordinates in which the external fields assume a particularly simple form given, for instance, by constant functions: an important example is a manifestly Lorentz invariant, rather than generally coordinate invariant, formulation of field theory on flat space-time, where the flat space-time metric plays the role of the external field. More generally, the most important class of external fields invariant under all admissible variations is provided by invariant fiber metrics, which in turn are a special case of $G$-structures – a concept to be discussed next.

## 3.5 $G$-structures and $G$-connections

In the general framework outlined in the previous subsections, it is difficult to construct invariant Lagrangians directly. This is due to the fact that the full automorphism group of the configuration bundle, which at this stage is the only natural candidate for a symmetry group, is simply too large: it must somehow be reduced in order to arrive at subgroups that do admit interesting invariants. (In a similar spirit, we may observe that the first order jet bundle of a fiber bundle is not a vector bundle but only an affine bundle and that it is difficult to construct invariants under affine transformations, since the translation part generates large orbits.) Inspection of concrete models reveals that, as a rule, such symmetry reductions are achieved by requiring invariance of certain additional structures that appear naturally, but often tacitly, in the construction of the pertinent Lagrangian: this leads us into the mathematical theory of $G$-structures.

Given a Lie group $G$, a fiber bundle $E$ over $M$, with bundle projection $\pi : E \to M$ and typical fiber $Q$, is said to carry a *G-structure* and simply called a *G-bundle* if there is an action
$$\begin{aligned} G \times Q &\longrightarrow Q \\ (g,q) &\longmapsto g \cdot q \end{aligned} \tag{126}$$
of $G$ on $Q$ such that one can find a bundle atlas for which the transition functions $f^E_{\alpha\beta}$ between any two bundle charts in the atlas can be written in terms of $G$-valued functions $g_{\alpha\beta}$ on the overlap of the respective domains according to
$$f^E_{\alpha\beta}(x,q) = (x, g_{\alpha\beta}(x) \cdot q) . \tag{127}$$

Of course, if we regard the diffeomorphism group $\mathrm{Diff}(M)$ as a (formal) Lie group, then every fiber bundle with typical fiber $Q$ carries a canonical (formal) $\mathrm{Diff}(Q)$-structure. Usually, however, $G$ is much smaller, and typically it can be assumed to be a finite-dimensional Lie group; it is in this situation that the concept of a $G$-bundle becomes useful. As an example, assuming $Q$ to be an $N$-dimensional vector space and $G$ to be the general linear group $\mathrm{GL}(N)$ leads us to regard $N$-dimensional vector bundles as $\mathrm{GL}(N)$-bundles. Similarly, supposing $Q$ to be an $N$-dimensional real/complex vector space equipped with a given scalar product and $G$ to be the orthogonal group $\mathrm{O}(N)$ / unitary group $\mathrm{U}(N)$ leads us to regard $N$-dimensional real/complex vector bundles equipped with a fixed Riemannian/Hermitean fiber metric as $\mathrm{O}(N)$-bundles / $\mathrm{U}(N)$-bundles. Examples of nonlinear nature also appear, namely in the so-called nonlinear sigma models where, typically, $Q$ is a homogeneous space for $G$: $Q = G/H$.

The modern approach to $G$-bundles consists in introducing a distinguished class of $G$-bundles called *principal G-bundles* and to derive all others from these as *associated bundles*. Briefly, given a $G$-bundle $E$ over $M$ with bundle projection $\pi : E \to M$ and typical fiber $Q$ as above, the idea is to define a principal $G$-bundle $P$ over $M$ with bundle projection $\rho : P \to M$ and typical fiber $G$ whose transition functions can be written in terms of the same $G$-valued functions $g_{\alpha\beta}$ as before, according to
$$f^P_{\alpha\beta}(x,g) = (x, g_{\alpha\beta}(x)g) . \tag{128}$$

Making use of the fact that left multiplication on $G$ commutes with right multiplication on $G$ (which is nothing but the associativity law $g_1(gg_2) = (g_1g)g_2$ for the group multiplication in $G$), this can be seen to guarantee the possibility to introduce a globally defined free right action

$$\begin{aligned} P \times G &\longrightarrow P \\ (p,g) &\longmapsto p \cdot g \end{aligned} \qquad (129)$$

of $G$ on $P$ whose orbits are precisely the fibers of $P$. Moroever, taking the Cartesian product $P \times Q$ of $P$ and $Q$, on which $G$ can be made to act from the right, say, according to $(p,q) \cdot g = (p \cdot g, g^{-1} \cdot q)$, and taking the quotient space with respect to this combined action, one constructs a $G$-bundle $P \times_G Q$ over $M$ known as the $G$-bundle associated to $P$ (with respect to the action of $G$ on $Q$), which turns out to be naturally isomorphic to the original $G$-bundle $E$ over $M$.

Among the advantages offered by the approach based on the introduction of an appropriate principal $G$-bundle, we may quote the possibility to give relatively simple definitions of various important concepts such as $G$-bundle automorphisms, $G$-invariant fiber metrics and $G$-connections. For example, a $G$-bundle automorphism is, roughly speaking, a bundle automorphism that leaves the given $G$-structure invariant. This can be made more precise by first defining an automorphism of a principal $G$-bundle $P$ over $M$ to be an automorphism $\phi_P$ of $P$ as a fiber bundle over $M$ that is equivariant, i.e., commutes with the right action of $G$ on $P$:

$$\phi_P(p \cdot g) \;=\; \phi_P(p) \cdot g \;. \qquad (130)$$

It is easy to see that such an equivariant automorphism $\phi_P$ induces an automorphism $\phi_{P \times_G Q}$ of any associated $G$-bundle $P \times_G Q$. Explicitly,

$$\phi_{P \times_G Q}([p,q]) \;=\; [\phi(p), q] \qquad (131)$$

where $[p,q] \in P \times_G Q$ denotes the equivalence class of $(p,q) \in P \times Q$. Conversely, a bundle automorphism $\phi_E$ of a $G$-bundle $E$ is called a *G-bundle automorphism* if it can be obtained in this way. For example, when $E$ is an $N$-dimensional vector bundle and $G = \mathrm{GL}(N)$, $P$ will be the corresponding linear frame bundle of $E$ and a bundle automorphism of $E$ will be a $G$-bundle automorphism if and only if it is fiberwise linear, that is, a vector bundle automorphism. Similarly, when $E$ is an $N$-dimensional real/complex vector bundle equipped with a fixed Riemannian/Hermitean fiber metric and $G = \mathrm{O}(N)\,/\,G = \mathrm{U}(N)$, $P$ will be the corresponding orthonormal frame bundle of $E$ and a bundle automorphism of $E$ will be a $G$-bundle automorphism if and only if it is fiberwise linear and isometric, that is, an isometric vector bundle automorphism. In general, we postulate that when $E$ is a $G$-bundle, symmetries of Lagrangians on $JE$ should be $G$-bundle automorphisms. In other words, we require the $G$-structure of $E$ to be fixed under all automorphisms considered as candidates for symmetries of physically interesting Lagrangians.

The aforementioned reinterpretation of a Riemannian/Hermitean fiber metric on an $N$-dimensional real/complex vector bundle as a $G$-structure with $G = \mathrm{O}(N)\,/\,G = \mathrm{U}(N)$ is easily extended from the positive definite to the indefinite case and can be further generalized by introducing the notion of a *G-invariant fiber metric*. Given a real/complex $G$-vector bundle $E = P \times_G E_0$ over $M$ associated to a principal $G$-bundle $P$ over $M$ via a representation of $G$ on a real / complex vector space $E_0$,[11] it is easy to see that any $G$-invariant non-degenerate scalar product $h_0^E$ on $E_0$ induces a fiber metric $h^E$ on $E$ which assigns to every point $x$ in $M$ a non-degenerate scalar product $h_x^E$ on the fiber $E_x$ of $E$ at $x$ of the same type as $h_0^E$ and depending smoothly on the base point $x$; it is $G$-invariant in the sense of being invariant under all $G$-bundle automorphisms. Explicitly, for $p$ in $P$ such that $\rho(p) = x$ and $v_1, v_2 \in E_0$,

$$h_x^E([p, v_1], [p, v_2]) = h_0^E(v_1, v_2) \tag{132}$$

where $[p, v_i] \in P \times_G E_0$ denotes the equivalence class of $(p, v_i) \in P \times E_0$. (Thus $G$ is now an extension, by the kernel of its representation on $E_0$, of a subgroup of the pertinent pseudo-orthogonal / pseudo-unitary group.) In any local trivialization of $E$ induced by a local section of $P$ and a choice of basis in $E_0$, the matrix elements of $h_x^E$ are independent of $x$ and equal to those of $h_0^E$. This construction admits a natural extension to the nonlinear situation. Namely, given a $G$-fiber bundle $E = P \times_G Q$ over $M$ associated to a principal $G$-bundle $P$ over $M$ via an action of $G$ on a manifold $Q$, it is easy to see that any $G$-invariant pseudo-Riemannian metric $h^Q$ on $Q$ induces a pseudo-Riemannian fiber metric $h^E$ on $E$ which assigns to every point $e$ in $E$ a non-degenerate scalar product $h_e^E$ on the vertical subspace $V_e E$ of the tangent space $T_e E$ at $e$ of the same type as $h^Q$ and depending smoothly on the base point $e$; it is $G$-invariant in the sense of being invariant under all $G$-bundle automorphisms. Explicitly, for $p$ in $P$, $q$ in $Q$ and $u_1, u_2 \in T_q Q$,

$$h_{[p,q]}^E([p, u_1], [p, u_2]) = h_q^Q(u_1, u_2) \tag{133}$$

where $[p, u_i] \in P \times_G TQ$ denotes the equivalence class of $(p, u_i) \in P \times TQ$ with respect to the induced action of $G$ on the tangent bundle $TQ$ of $Q$ and we have made use of the following canonical isomorphism of vector bundles over $P \times_G Q$:

$$V(P \times_G Q) \cong P \times_G TQ \,. \tag{134}$$

Thus we have an explicit example of the situation mentioned at the end of the previous subsection: if we think of a $G$-invariant fiber metric as a type of external field, this must be kept fixed under all variations – those used to derive the equations of motion as well as those generated by candidates for infinitesimal symmetries of the Lagrangian – and can therefore be made constant by restricting to special coordinates, namely fiber coordinates corresponding to local trivializations that are compatible with the given $G$-structure.

For what follows, it will be important to observe that all fiber metrics appearing in the construction of the standard Lagrangians of field theory are fixed external fields in this sense, with one possible exception: the metric tensor of space-time itself may be either external, as in special relativity, or dynamical, as in general relativity.

---

[11]For a vector bundle $E$ over $M$, we denote its typical fiber by $E_0$, rather than $Q$ as before.

Turning to connections, we recall first of all that a general connection in a general fiber bundle $E$ over $M$ can be defined in various equivalent ways. Two of these are:

- A connection in $E$ is given by the choice of a *horizontal bundle* $HE$, which is a vector subbundle of the tangent bundle $TE$ of the total space $E$ complementary to the vertical bundle $VE$:

$$TE \;=\; VE \oplus HE \;. \tag{135}$$

  Given a connection and a vector field $X_E$ on $E$, we write $(X_E)_V$ for its vertical part and $(X_E)_H$ for its horizontal part.

- A connection in $E$ is given by the choice of a *horizontal lifting map* $\Gamma_E$, which is simply a section $\Gamma_E : E \to JE$ of $JE$ over $E$. Equivalently, it can be viewed as a map $\Gamma_E : \pi^*(TM) \to TE$ of vector bundles over $E$ whose composition with the tangent map $T\pi$ to the projection $\pi : E \to M$, viewed as a map $T\pi : TE \to \pi^*(TM)$ of vector bundles over $E$, gives the identity on $\pi^*(TM)$:

$$T\pi \circ \Gamma_E \;=\; \mathrm{id}_{\pi^*(TM)} \;. \tag{136}$$

  Given a connection and a vector field $X_M$ on $M$, we write $\Gamma_E(X_M)$ for the vector field on $E$ obtained as its horizontal lift:

$$\Gamma_E(X_M)(e) \;=\; \Gamma_E(e) \cdot X_M(\pi(e)) \qquad \text{for } e \in E \;. \tag{137}$$

The equivalence between these two definitions is established by observing that for any point $e$ in $E$ with base point $x = \pi(e)$ in $M$, the horizontal space $H_e E$ at $e$ is the image of $T_x M$ under $\Gamma_E(e)$ and conversely, $\Gamma_E(e)$ is the inverse of the restriction of $T_e \pi$ to $H_e E$, which is a linear isomorphism. Given such a connection in $E$, we obtain a *covariant derivative* $D$ that maps every section $\varphi$ of $E$ to a 1-form $D\varphi$ on $M$ with values in the pull-back $\varphi^*(VE)$ of the vertical bundle $VE$ of $E$ via $\varphi$ and is defined as the difference

$$D\varphi \;=\; \partial \varphi - \Gamma_E(\varphi) \tag{138}$$

where $\partial \varphi$ is of course the ordinary derivative or 1-jet of $\varphi$ and $\Gamma_E(\varphi)$ the composition of the maps $\varphi : M \to E$ and $\Gamma_E : E \to JE$. We also define the *curvature form* of the connection to be the horizontal 2-form $\Omega_E$ on $E$ with values in the vertical bundle $VE$ of $E$ such that for any two vector fields $X_E$ and $Y_E$ on $E$, $\Omega_E(X_E, Y_E)$ is the vertical part of the Lie bracket between the horizontal parts of $X_E$ and $Y_E$:

$$\Omega_E(X_E, Y_E) \;=\; [\,(X_E)_H, (Y_E)_H\,]_V \;. \tag{139}$$

Thus $\Omega_E$ measures the extent to which the horizontal bundle fails to be involutive. In adapted local coordinates $(x^\mu, q^i)$ for $E$ derived from local coordinates $x^\mu$ for $M$, local coordinates $q^i$ for $Q$ and a local trivialization of $E$ over $M$, as well as the induced local coordinates $(x^\mu, q^i, q^i_\mu)$ for $JE$, we can write

$$\Gamma_E(\partial_\mu) \;=\; \partial_\mu + \Gamma^i_\mu \, \partial_i \;, \tag{140}$$

and

$$\Omega_E(\Gamma_E(\partial_\mu), \Gamma_E(\partial_\nu)) \;=\; \Omega^i_{\mu\nu} \, \partial_i \;, \tag{141}$$

to obtain

$$D_\mu \varphi^i \;=\; \partial_\mu \varphi^i - \Gamma^i_\mu(\varphi) \;, \tag{142}$$

and

$$\Omega^i_{\mu\nu} \;=\; \partial_\mu \Gamma^i_\nu - \partial_\nu \Gamma^i_\mu + \Gamma^j_\mu \, \partial_j \Gamma^i_\nu - \Gamma^j_\nu \, \partial_j \Gamma^i_\mu \;. \tag{143}$$

In the presence of a $G$-structure, it is appropriate to introduce the concept of a $G$-connection which, roughly speaking, is a connection compatible with the given $G$-structure. Again, this can be made more specific by first defining a principal connection in a principal $G$-bundle $P$ over $M$ to be a connection in $P$ as a fiber bundle over $M$ that is invariant under the right action of $G$ on $P$, which means that the corresponding horizontal bundle should be invariant under this action and the corresponding horizontal lifting map $\Gamma_P$ should be equivariant with respect to this action on $P$ and the induced action on $JP$. It is easy to see that such an equivariant connection $\Gamma_P$ induces a connection $\Gamma_{P \times_G Q}$ on any associated $G$-bundle $P \times_G Q$. Conversely, a connection $\Gamma_E$ of a $G$-bundle $E$ is called a $G$-connection if it can be obtained in this way. For example, when $E$ is an $N$-dimensional vector bundle and $G = \mathrm{GL}(N)$, $P$ will be the corresponding linear frame bundle of $E$ and a connection in $E$ will be a $G$-connection if and only if it is a linear connection. In this case, all relevant structures are simplified due to their linear behavior along the fibers of $E$. In particular, the pull-back $\varphi^*(VE)$ of the vertical bundle $VE$ of $E$ via any section $\varphi$ can be identified with the bundle $E$ itself, so that the covariant derivative becomes an operator taking sections $\varphi$ of $E$ to 1-forms $D\varphi$ on $M$ with values in $E$, whereas the curvature becomes a 2-form on $M$ with values in the bundle $L(E)$ of linear transformations of $E$ into itself. This linear behavior along the fibers is most easily seen in adapted local coordinates as above (where the $q^i$ should now be linear coordinates on the typical fiber), since

$$\Gamma^i_\mu(x, q) \;=\; -\, A_{\mu\,j}^{\;\;i}(x) \, q^j \;, \tag{144}$$

and

$$\Omega^i_{\mu\nu}(x, q) \;=\; -\, F_{\mu\nu\,j}^{\;\;\;i}(x) \, q^j \;, \tag{145}$$

so that

$$D_\mu \varphi^i \;=\; \partial_\mu \varphi^i + A_{\mu\,j}^{\;\;i} \, \varphi^j \;, \tag{146}$$

and

$$F_{\mu\nu\,j}^{\;\;\;i} \;=\; \partial_\mu A_{\nu\,j}^{\;\;i} - \partial_\nu A_{\mu\,j}^{\;\;i} + A_{\mu\,k}^{\;\;i} A_{\nu\,j}^{\;\;k} - A_{\nu\,k}^{\;\;i} A_{\mu\,j}^{\;\;k} \;. \tag{147}$$

Similarly, when $E$ is an $N$-dimensional real/complex vector bundle equipped with a fixed Riemannian/Hermitean fiber metric and $G = \mathrm{O}(N)/G = \mathrm{U}(N)$, $P$ will be the corresponding orthonormal frame bundle of $E$ and a connection in $E$ will be a $G$-connection if and only if it is a linear connection leaving the given fiber metric invariant.

Returning to the general case, we have seen that, by definition, $G$-connections in $G$-bundles are derived from principal connections in principal $G$-bundles, for which there is another widely used definition, based on the fact that the vertical bundle of a principal $G$-bundle $P$ over $M$ is globally trivial and admits a distinguished $G$-equivariant global trivialization

$$\begin{aligned} \tau_{VP}: \ P \times \mathfrak{g} &\longrightarrow\ VP \\ (p,X) &\longmapsto\ X_P(p) \end{aligned} \qquad (148)$$

given by the fundamental vector fields associated with the right action (129) of $G$ on $P$:

$$X_P(p) \ = \ \frac{d}{dt}\Big(p \cdot \exp(tX)\Big)\bigg|_{t=0} . \qquad (149)$$

Indeed:

- A principal connection in $P$ is given by the choice of a *connection form* $A$, which is a $G$-equivariant 1-form on $P$ with values in the Lie algebra $\mathfrak{g}$ of $G$ and such that, when viewed as a map $A: TP \to P \times \mathfrak{g}$ of vector bundles over $P$, its restriction to the vertical bundle $VP$ of $P$, composed with the isomorphism (148), gives the identity on $VP$:

$$\tau_{VP} \circ A\big|_{VP} \ = \ \mathrm{id}_{VP} . \qquad (150)$$

The equivalence with the previous two definitions can be established by postulating that the horizontal bundle is precisely the kernel of $A$:

$$HP \ = \ \ker A . \qquad (151)$$

This relation can indeed be read in both directions, since it not only defines the horizontal bundle in terms of the connection form but also yields the connection form in terms of the horizontal bundle. This statement, together with the relation between all three definitions, becomes even clearer if we express all objects in terms of the vertical projection from $TP$ onto $VP$ (with kernel $HP$) and the horizontal projection from $TP$ onto $HP$ (with kernel $VP$) associated with the direct decomposition (135), with $E$ replaced by $P$. In fact, the composition on the lhs of eq. (150) above is the vertical projection, whereas the composition of $\Gamma_P$, when viewed as a map $\Gamma_P: \rho^*(TM) \to TP$ of vector bundles over $P$, with the tangent map $T\rho$ to the bundle projection $\rho: P \to M$ (in the opposite order to that used in eq. (136)) is the horizontal projection. Thus using the obvious fact that these two projections must add up to the identity on $TP$, we can write

$$\tau_{VP} \circ A \ + \ \Gamma_P \circ T\rho \ = \ \mathrm{id}_{TP} . \qquad (152)$$

The *curvature form* associated with such a connection form is defined to be the $G$-equivariant horizontal 2-form $F$ on $P$ with values in the Lie algebra $\mathfrak{g}$ of $G$ defined as the composition of $\Omega_P$ with the inverse of the trivialization $\tau_{VP}$, except for a sign:

$$\tau_{VP} \circ F + \Omega_P = 0 \ . \tag{153}$$

Then

$$F = dA + \tfrac{1}{2} [A \wedge A] \ . \tag{154}$$

Note that $A$ is zero on the horizontal bundle while $F$ is zero on the vertical bundle. It is also clear that $G$-bundle automorphisms transform $G$-connections into $G$-connections, since automorphisms of principal $G$-bundles transform principal connections into principal connections. (For example, looking at the right action (129) of $G$ on $P$, it is clear that a $G$-equivariant automorphism will not only leave the $G$-invariant vertical bundle invariant but also transform a $G$-invariant horizontal bundle into another $G$-invariant horizontal bundle.) In particular, this is true for strict $G$-bundle automorphisms, known in physics as *gauge transformations*, which can also be interpreted as $G$-equivariant functions on $P$ with values in $G$, or equivalently, as sections of a certain bundle of Lie groups over $M$ constructed from $P$, namely the associated bundle $P \times_G G$, with respect to the action of $G$ on $G$ itself by conjugation. Similarly, infinitesimal strict $G$-bundle automorphisms, known in physics as *infinitesimal gauge transformations*, can also be interpreted as $G$-equivariant functions on $P$ with values in $\mathfrak{g}$, or equivalently, as sections of a certain bundle of Lie algebras over $M$ constructed from $P$, namely the associated bundle $P \times_G \mathfrak{g}$, with respect to the adjoint representation of $G$ on $\mathfrak{g}$. This interpretation is useful, for example, for writing down the transformation law of connection forms $A$ and curvature forms $F$ under gauge transformations $g$,

$$\begin{aligned} A &\to g \cdot A = g A g^{-1} - dg\, g^{-1} \ , \\ F &\to g \cdot F = g F g^{-1} \ , \end{aligned} \tag{155}$$

and for the variation of connection forms $A$ and curvature forms $F$ under infinitesimal gauge transformations $\xi$,

$$\begin{aligned} \delta_\xi A &= [\xi, A] - d\xi = -D\xi \ , \\ \delta_\xi F &= [\xi, F] \ . \end{aligned} \tag{156}$$

For later use, we note that the curvature form can also be viewed as a 2-form $F$ on $M$ with coefficients in the associated Lie algebra bundle $P \times_G \mathfrak{g}$. For the connection form $A$, it is impossible to give an analogous interpretation as a section of a vector bundle over $M$, but it can be viewed as a section of an affine bundle over $M$.

Concluding this subsection, we wish to emphasize that, in view of the enormous importance of connections both in physics and in mathematics, it seems adequate to consider them as a special class of fields. It must be emphasized, however, that the only connections of interest in physics are the $G$-connections. The reason is that fields are functions on space-time or, in the general geometric setting discussed in the previous subsection, sections of bundles over space-time, while general connections on a general fiber bundle show an arbitrary dependence on the fiber coordinates: they are not fields! But $G$-connections are, precisely because they are derived from principal connections in principal $G$-bundles whose dependence along the fibers is completely fixed by the requirement of $G$-equivariance.

## 3.6 Construction of invariant Lagrangians

As has already been mentioned at the very beginning of this section, there are essentially three different classes of fields that appear in concrete models.

- *Linear matter fields* which are sections of $G$-vector bundles over space-time,

- *Nonlinear matter fields* which are sections of general $G$-fiber bundles over space-time, also called sigma model type fields,

- *Gauge fields* which are $G$-connections in $G$-bundles over space-time.
  We emphasize once again that general connections in general fiber bundles without any additional structure are not physical fields.

A $G$-connection is the essential ingredient for defining covariant derivatives of matter fields, both linear and nonlinear, and such covariant derivatives are always sections of $G$-vector bundles over space-time. Similarly, the curvature of a $G$-connection is also a section of a certain $G$-vector bundle over space-time. It is interesting to note that in all known Lagrangians of physical significance in field theory, $G$-connections enter only through covariant derivatives of matter fields or through curvature terms.[12]

In order to explain how to construct the standard invariant Lagrangians of field theory, let us first take a brief look at various types of fields of particular importance.

a) Tensor fields on the space-time manifold $M$ will almost always be part of the construction – partly because they appear, explicitly or tacitly, as soon as one takes space-time derivatives of fields. They are sections of the tensor bundles $T^r_s M$ of $M$ which can be constructed canonically from the tangent bundle $TM$ and its dual, the cotangent bundle $T^*M$, by taking tensor products. Derivatives of tensor fields are covariant derivatives with respect to some linear space-time connection $\Gamma$ and as such are again tensor fields. (A notable exception occurs for differential forms, where one can use Cartan's operator $d$ of exterior differentiation, which makes no reference to the choice of a space-time connection. Its definition is usually written using ordinary partial derivatives but in fact the same formula holds if these are replaced by covariant derivatives with respect to an arbitrary torsion-free linear connection on $M$.) Another special feature of tensor fields is that diffeomorphisms of $M$ can be lifted canonically to automorphisms of the tensor bundles $T_s^r M$ and hence act naturally on tensor fields. It should be noted that the lifting $T_s^r \phi_M$ to $T_s^r M$ of a diffeomorphism $\phi_M$ of $M$ is never strict (except when both are the identity). Strict automorphisms of the tensor bundles also play a role: they are the proper geometric concept for describing local frame transformations.

---

[12]There is one notable exception of considerable theoretical interest but whose direct physical significance is unclear, namely the Chern-Simons Lagrangian, which differs from all other Lagrangians containing gauge fields in that it is only gauge invariant up to a total divergence and describes a topological field theory, without true dynamics.

A specific tensor field that plays an outstanding role is the metric tensor $g$. A pseudo-Riemannian metric on $M$ is a rank 2 symmetric tensor field which is nowhere degenerate and has fixed signature $(p, q)$ (where $p + q = n$), the default case for classical field theory being of course that of Lorentz signature ($p = 1$ and $q = n - 1$ or $p = n - 1$ and $q = 1$, depending on the conventions used). Apart from providing a fiber metric on each of the tensor bundles $T_s^r M$, such a metric tensor induces useful additional structures. One of these is the Levi-Civita connection which is a family of linear connections, all commonly denoted by $\nabla$, on the tensor bundles $T_s^r M$ that commutes with the natural operations between tensor bundles (tensor products and contractions) and is uniquely characterized by two properties:

$$\nabla_X g \;=\; 0 \qquad \text{(metricity)}, \tag{157}$$

$$\nabla_X Y - \nabla_Y X - [X, Y] \;=\; 0 \qquad \text{(vanishing torsion)}. \tag{158}$$

Supposing $M$ to be oriented, as always, the other is the pseudo-Riemannian volume form $\tau$ which, just like the metric tensor $g$ itself, is covariant constant:

$$\nabla_X \tau \;=\; 0 . \tag{159}$$

In terms of local coordinates $x^\mu$ on $M$, we can write

$$\nabla_\mu \varphi^{\nu_1 \ldots \nu_s}_{\mu_1 \ldots \mu_r} \;=\; \partial_\mu \varphi^{\nu_1 \ldots \nu_s}_{\mu_1 \ldots \mu_r} + \sum_{l=1}^s \Gamma^{\nu_l}_{\mu \lambda} \varphi^{\nu_1 \ldots \nu_{l-1} \lambda \nu_{l+1} \ldots \nu_s}_{\mu_1 \ldots \mu_r} - \sum_{k=1}^r \Gamma^\kappa_{\mu \mu_k} \varphi^{\nu_1 \ldots \nu_s}_{\mu_1 \ldots \mu_{k-1} \kappa \mu_{k+1} \ldots \mu_r} \tag{160}$$

where

$$\Gamma^\kappa_{\mu \lambda} \;=\; \tfrac{1}{2} g^{\kappa \nu} \left( \partial_\mu g_{\lambda \nu} + \partial_\lambda g_{\mu \nu} - \partial_\nu g_{\mu \lambda} \right), \tag{161}$$

while

$$\tau \;=\; \sqrt{|\det g|} \; d^n x . \tag{162}$$

This allows us to rewrite the Lagrangian $\hat{\mathcal{L}}$ in terms of a Lagrangian function $L$, rather than a Lagrangian density $\mathcal{L}$ as in eq. (79):

$$\hat{\mathcal{L}} \;=\; L \, \tau . \tag{163}$$

Of course,
$$\mathcal{L} = \sqrt{|\det g|}\, L \,. \tag{164}$$

b) Spinor fields on the space-time manifold $M$ are closely related to tensor fields but somewhat more difficult to handle since their mere definition presupposes the choice of a metric tensor. Namely, they are sections of a complex vector bundle $SM$ over $M$ called the spinor bundle which, very loosely speaking, is a square root of the (complexified) tangent bundle.[13] This interpretation can be taken almost literally in the case of Riemann surfaces, but in general a more precise statement is that there is a canonical homomorphism

$$\gamma: \ T^c M \ \longrightarrow \ L(SM) \equiv S^* M \otimes SM \tag{165}$$

of vector bundles over $M$ which by taking Clifford products extends to a canonical isomorphism

$$\mathrm{Cliff}(T^c M) \ \cong \ L(SM) \ \equiv \ S^* M \otimes SM \tag{166}$$

of algebra bundles over $M$ (the superscript $c$ indicates complexification). Moreover, the spinor bundle $SM$ of $M$ carries a natural fiber metric which associates to any two sections $\psi$ and $\chi$ of $SM$ a function on $M$ denoted by $\bar\psi\chi$: it is non-degenerate but in general not positive definite and such that $\gamma$ restricted to $TM$ takes values in the bundle of pseudo-Hermitean linear transformations of $SM$. Finally, the Levi-Civita connection can be uniquely lifted from tensor fields to spinor fields in such a way that the property of metricity carries over from the metric tensor $g$ to the homomorphism $\gamma$:

$$\nabla_X \gamma = 0 \,. \tag{167}$$

In terms of local coordinates $x^\mu$ on $M$, $\gamma$ is defined by requiring it to take the coordinate vector fields $\partial_\mu$ to the Dirac $\gamma$-matrices $\gamma_\mu$, which must satisfy the Clifford algebra relations

$$\gamma_\mu \gamma_\nu + \gamma_\nu \gamma_\mu = 2\, g_{\mu\nu} \tag{168}$$

and we can write

$$\nabla_\mu \psi^\alpha = \partial_\mu \psi^\alpha + \Gamma^\alpha_{\mu\beta} \psi^\beta \tag{169}$$

where the $\Gamma^\alpha_{\mu\beta}$ are known as the spinor connection coefficients. An explicit formula for them can only be given in terms of orthonormal frame fields; this is deferred to Sect. 4.3 where it will be needed. Here, we just mention that the spin connection preserves the aforementioned natural fiber metric on the spinor bundle. Moreover, we note that each spinor field $\psi$ gives rise to the composite tensor fields

$$\bar\psi\, \gamma_{\mu_1} \ldots \gamma_{\mu_r} \psi \ dx^{\mu_1} \otimes \ldots \otimes dx^{\mu_r}$$

---

[13] The spinors considered here are Dirac spinors.

whose covariant derivatives can be calculated in terms of covariant derivatives of
the underlying spinor field by means of a Leibniz rule:

$$\nabla_X \left( \bar\psi \, \gamma_{\mu_1} \ldots \gamma_{\mu_r} \psi \right) \;=\; (\nabla_X \bar\psi) \, \gamma_{\mu_1} \ldots \gamma_{\mu_r} \psi \;+\; \bar\psi \, \gamma_{\mu_1} \ldots \gamma_{\mu_r} (\nabla_X \psi) \,. \qquad (170)$$

c) Sections of other $G$-vector bundles $V$ and, more generally, of $G$-fiber bundles $F$ over the space-time manifold $M$ appear in theories containing matter fields with internal symmetries that are realized linearly or, more generally, nonlinearly. The covariant derivative $D_X \varphi$ of such a section $\varphi$ along any vector field $X$ on $M$, with respect to a given $G$-connection $A$, is in the first case again a section of $V$ and in the second case a section of the pull-back $\varphi^*(VF)$ of the vertical bundle $VF$ of $F$ to $M$ via $\varphi$ itself. In both cases, all bundles involved are associated to some given principal $G$-bundle $P$ over $M$, namely $V = P \times_G V_0$ with respect to some given representation of $G$ on the vector space $V_0$ which is the typical fiber of $V$, $F = P \times_G Q$ with respect to some given action of $G$ on the manifold $Q$ which is the typical fiber of $F$ and $VF = P \times_G TQ$ with respect to the induced action of $G$ on the tangent bundle $TQ$ of $Q$; then $A$ is a principal connection in $P$. In terms of local coordinates $x^\mu$ on $M$, a basis of generators $T_a$ in the underlying Lie algebra $\mathfrak{g}$ and a local section of $P$, we can write

$$D_\mu \varphi^i \;=\; \partial_\mu \varphi^i \;+\; A_\mu^a \, (T_a)^i_j \, \varphi^j \qquad (171)$$

with respect to a basis in $V_0$ in the first case, where the $(T_a)^i_j$ are the matrix elements of the linear transformations on $V_0$ corresponding to the generators $T_a$ under the representation of $\mathfrak{g}$ induced from that of $G$, and

$$D_\mu \varphi^i \;=\; \partial_\mu \varphi^i \;+\; A_\mu^a \, T_a^i(\varphi) \qquad (172)$$

with respect to local coordinates $q^i$ on $Q$ in the second case, where the $T_a^i$ are the components of the fundamental vector fields on $Q$ corresponding to the generators $T_a$ under the action of $G$ on $Q$.

d) Gauge fields are described by principal connections in a principal $G$-bundle $P$ over the space-time manifold $M$, more precisely in terms of their connection form $A$ and curvature form $F$, as explained in the previous subsection. The latter can be viewed as a 2-form on $M$ with coefficients in the corresponding associated Lie algebra bundle $P \times_G \mathfrak{g}$.

More general fields are obtained as sections of bundles constructed from the building blocks mentioned under a)-d) above by taking duals, direct sums (or in the nonlinear case, fiber products) and tensor products. Similarly, their covariant derivatives are obtained using the corresponding constructions for connections. For instance, tensor fields or spinor fields with coefficients in a $G$-vector bundle $V$ over $M$ are sections of $T_s^r M \otimes V$ and of $SM \otimes V$, respectively, and their covariant derivatives must be taken using both the $G$-connection $A$ and the space-time connection $\Gamma$:

$$
\begin{aligned}
D_\mu \varphi^{\nu_1...\nu_s, i}_{\mu_1...\mu_r} &= \partial_\mu \varphi^{\nu_1...\nu_s, i}_{\mu_1...\mu_r} + A^a_\mu (T_a)^i_j \varphi^{\nu_1...\nu_s, j}_{\mu_1...\mu_r} \\
&\quad + \sum_{l=1}^{s} \Gamma^{\nu_l}_{\mu\lambda} \varphi^{\nu_1...\nu_{l-1}\lambda\nu_{l+1}...\nu_s, i}_{\mu_1...\mu_r} - \sum_{k=1}^{r} \Gamma^{\kappa}_{\mu\mu_k} \varphi^{\nu_1...\nu_s, i}_{\mu_1...\mu_{k-1}\kappa\mu_{k+1}...\mu_r} ,
\end{aligned} \qquad (173)
$$

$$
D_\mu \psi^{\alpha, k} = \partial_\mu \psi^{\alpha, k} + A^a_\mu (T_a)^k_l \psi^{\alpha, l} + \Gamma^{\alpha}_{\mu\beta} \psi^{\beta, k} . \qquad (174)
$$

Apart from this input, the construction of invariant Lagrangians hinges on the choice of a metric tensor $g$ and, in the presence of fields carrying internal symmetries, of other fiber metrics $h$ on the bundles involved; these will also be denoted by brackets of the form $(.\,,.)$ in the real case and by brackets of the form $\langle .\,,.\rangle$ in the complex case. The most important examples are the following.

- For scalar fields with linearly realized internal symmetry, that is, sections $\varphi$ of $V = P \times_G V_0$,

$$
L_{\text{RSC}} = \tfrac{1}{2} g^{\mu\nu} (D_\mu \varphi , D_\nu \varphi) - U(\varphi) = \tfrac{1}{2} g^{\mu\nu} h_{ij} D_\mu \varphi^i D_\nu \varphi^j - U(\varphi) \qquad (175)
$$

in the real case and

$$
L_{\text{CSC}} = g^{\mu\nu} \langle D_\mu \varphi , D_\nu \varphi \rangle - U(\varphi) = g^{\mu\nu} h_{ij} D_\mu \bar\varphi^i D_\nu \varphi^j - U(\varphi) \qquad (176)
$$

in the complex case, where $h$ is a $G$-invariant scalar product on $V_0$, giving rise to a $G$-invariant fiber metric on $V$, and $U$ is a potential describing self-interactions; it is simply a $G$-invariant function on $V_0$, giving rise to a $G$-invariant function on $V$. For $\dim V_0 = 1$, for example, choosing $U(\varphi) = \tfrac{1}{2} m^2 \varphi^2$ in the real case and $U(\bar\varphi, \varphi) = m^2 |\varphi|^2$ in the complex case gives the free scalar field of mass $m$, including a term of the form $|\varphi|^4$ leads to the $\varphi^4$-theory, etc..

- For scalar fields with nonlinearly realized internal symmetry, that is, sections $\varphi$ of $F = P \times_G Q$,

$$
L_{\text{GSM}} = \tfrac{1}{2} g^{\mu\nu} (D_\mu \varphi , D_\nu \varphi) - U(\varphi) = \tfrac{1}{2} g^{\mu\nu} h_{ij}(\varphi) D_\mu \varphi^i D_\nu \varphi^j - U(\varphi) \qquad (177)
$$

where $h$ is now a $G$-invariant pseudo-Riemannian metric on $Q$, giving rise to a $G$-invariant pseudo-Riemannian fiber metric on $F$, and $U$ is a potential describing self-interactions; it is simply a $G$-invariant function on $Q$, giving rise to a $G$-invariant function on $F$. This is the Lagrangian for a generalized sigma model which contains as a special case the ordinary sigma model with target space a Riemannian manifold $Q$ with metric $h$: it is obtained by choosing $P$ and $F$ to be trivial $G$-bundles, i.e., $P = M \times G$ and $F = M \times Q$, $A$ to be the trivial flat $G$-connection and $U$ to vanish. Indeed, sections of $F$ can then be identified with mappings from $M$ to $Q$ and the Lagrangian (177) reduces to the ordinary sigma model Lagrangian:

$$L_{\text{OSM}} \;=\; \tfrac{1}{2}\, g^{\mu\nu}\, (\partial_\mu \varphi\,,\partial_\nu \varphi) \;=\; \tfrac{1}{2}\, g^{\mu\nu}\, h_{ij}(\varphi)\, \partial_\mu \varphi^i\, \partial_\nu \varphi^j \qquad (178)$$

- For spinor fields with (necessarily linearly realized) internal symmetry, that is, sections $\psi$ of $SM \otimes W$ with $W = P \times_G W_0$,

$$L_{\text{DSP}} \;=\; \tfrac{i}{2}\, \bar{\psi} \overleftrightarrow{\slashed{D}} \psi \;-\; U(\bar{\psi},\psi) \;=\; \tfrac{i}{2}\, g^{\mu\nu}\, h_{kl}\, \bar{\psi}^k \gamma_\mu \overleftrightarrow{D}_\nu \psi^l \;-\; U(\bar{\psi},\psi) \qquad (179)$$

where, once again, $h$ is a $G$-invariant scalar product on $W_0$, giving rise to a $G$-invariant fiber metric on $W$, and $U$ is a potential describing self-interactions; usually, it is an invariant function of composite tensor fields of the form $\bar{\psi}^k \gamma_{\mu_1} \ldots \gamma_{\mu_r} \psi^l$ obtained by contracting all internal indices with $h$ and all space-time indices with $g$.[14] For $\dim W_0 = 1$, for example, choosing $U(\bar{\psi},\psi) = m\, \bar{\psi}\psi$ gives the free Dirac spinor field of mass $m$, including a term of the form $g^{\mu\nu}\, \bar{\psi}\gamma_\mu\psi\, \bar{\psi}\gamma_\nu\psi$ gives the Thirring model, including a term of the form $(\bar{\psi}\psi)^2$ or $(\bar{\psi}\psi)^2 - (\bar{\psi}\gamma_5\psi)^2$ gives the Gross-Neveu model and the chiral Gross-Neveu model, respectively, etc..

- For $G$-connections, that is, connection forms $A$ on $P$,

$$L_{\text{YM}} \;=\; -\,\tfrac{1}{4}\, g^{\mu\kappa} g^{\nu\lambda}\, (F_{\mu\nu}\,,F_{\kappa\lambda}) \;=\; -\,\tfrac{1}{4}\, g^{\mu\kappa} g^{\nu\lambda}\, h_{ab}\, F^a_{\mu\nu} F^b_{\kappa\lambda} \qquad (180)$$

where $F$ is the curvature form of $A$ and $h$ is now a $G$-invariant scalar product on the Lie algebra $\mathfrak{g}$, giving rise to a $G$-invariant fiber metric on $P \times_G \mathfrak{g}$. This is the well-known *Yang-Mills Lagrangian*.

- For the metric tensor $g$ of space-time, when regarded as a dynamical field,

$$L_{\text{EH}} \;=\; -\,\tfrac{1}{2}\,(R + 2\Lambda) \qquad (181)$$

where $R$ denotes the scalar curvature and $\Lambda$ is the cosmological constant.[15] This is the well-known *Einstein-Hilbert Lagrangian*.

The question whether the metric tensor $g$ of space-time is an external field or a dynamical field depends on the physical system to be described. Fixing the metric tensor $g$ as an external field means studying a non-gravity type field system on a given space-time background, thus taking into account the gravitational influence exerted on the system by the outside world but neglecting gravitational interactions within the system as well as the gravitational back reaction of the system on the outside world.

---

[14] Note also that, by definition, $A \overleftrightarrow{D}_\mu B = A\, D_\mu B - D_\mu A\, B$.

[15] For simplicity, we work in natural units where $c = 1$ and Newton's constant $\gamma = 1/8\pi$.

As is well known, this approach may lead to consistency problems whose consequences may however often be neglected from a practical point of view. But if the effects of gravity are to be fully included, the metric tensor $g$ must be treated as a dynamical field and the Einstein-Hilbert term should be incorporated into the full Lagrangian.

Apart from the coupling terms between matter fields and gauge fields that have already been taken into account above by adopting the usual prescription of minimal coupling (which states that ordinary partial derivatives should be replaced by covariant derivatives), there are of course other possibilities to introduce coupling terms between different fields. An important example is provided by cubic coupling terms between scalar fields $\varphi$ and spinor fields $\psi$, e.g.

$$L_{YC} \;=\; h_{i,kl}\, \varphi^i\, \bar{\psi}^k \psi^l \tag{182}$$

for real scalar fields or

$$L_{YC} \;=\; \mathrm{Re}\left(h_{i,kl}\, \varphi^i\, \bar{\psi}^k \psi^l\right) \tag{183}$$

for complex scalar fields, where $h$ is now a $G$-invariant tensor that provides an intertwining operator between the tensor product of the representation of $G$ on the internal space $W_0$ for the spinors with its dual and the representation of $G$ on the internal space $V_0$ for the scalars. This is the well-known *Yukawa coupling*.

From the example Lagrangians given above, one can construct Lagrangians for more complicated, composite systems by summing up the individual Lagrangians for the various different pieces. In particular, one can in this way obtain the complete Lagrangian for the standard model of elementary particle physics.

As a final example, we mention another Lagrangian for $G$-connections, that is, connection forms $A$ on $P$, which in many respects is entirely different from the Yang-Mills Lagrangian:

$$\hat{\mathcal{L}}_{\mathrm{CS}} \;=\; (A \curlywedge dA) \,+\, \tfrac{1}{3}\,(A \curlywedge [A \curlywedge A])\,. \tag{184}$$

This is the *Chern-Simons Lagrangian*, which of course makes sense only in $n=3$ space-time dimensions since $\hat{\mathcal{L}}_{\mathrm{CS}}$ as defined in this equation is a 3-form. Other peculiar features are that its definition does not require a space-time metric and that it fails to be gauge invariant, but on the other hand it is gauge invariant up to total divergences as well as space-time diffeomorphism invariant up to total divergences, or in other words, it changes by the exterior derivative of an appropriate 2-form when subjected to an arbitrary automorphism (strict or non-strict) of $P$. This means that the corresponding action functional, whose critical points are simply the flat connections, has an enormous symmetry group: it is invariant under all gauge transformations and all space-time diffeomorphisms satisfying appropriate boundary conditions at infinity. As a result, the Chern-Simons theory is a topological field theory, without any true dynamics.

# 4 Currents and Energy-Momentum Tensor

In this section, we present the general solution to the problem of finding the physically correct current in Lagrangian field theories with gauge invariance and the physically correct energy-momentum tensor in Lagrangian field theories with general covariance, or space-time diffeomorphism invariance. Both of these problems are quite similar in nature and we believe that their parallel treatment will help to further clarify the reasoning.

To begin with, let us briefly recall the central points of the "Noetherian approach" to the question as developed in Sect. 3, within a fully geometrized first order Lagrangian formalism. Starting from the canonical total Noether current associated with a given Lagrangian – an object that comprises a "current type piece" in the original sense of the word "current", referring to internal symmetries, as well as an "energy-momentum tensor type piece", referring to space-time symmetries, and that can be derived from the covariant momentum map of multisymplectic field theory [7–9] – we have first established a general framework for possible correction terms. Subsequently, we have formulated the "ultralocality principle", which requires the improved Noether current and the improved energy-momentum tensor to be ultralocal – or in mathematical terms, $\mathfrak{F}(M)$-linear – in the pertinent symmetry generators. However, the procedure is still incomplete because, for Lagrangians that admit a "sufficiently large" local symmetry group, the second Noether theorem forces all of these expressions to vanish "on shell", that is, when all fields in the theory satisfy their equations of motion. This problem, as we have argued, can only be overcome by appropriately splitting the fields in the theory into different sectors and analyzing the exchange of conserved quantities between them. What remains to be done is to perform this splitting concretely and then draw the consequences.

Before embarking on this program, we would like to mention two restrictions, both of which are important prerequisites for the construction of invariant model Lagrangians, as explained in Sect. 3.5. First, we shall consider only theories with an underlying $G$-structure, where $G$ is some given (finite-dimensional) Lie group. In other words, all fiber bundles occurring in the theory are assumed to be $G$-bundles, associated to some given principal $G$-bundle $P$ over $M$, and all pertinent symmetry groups are realized by $G$-bundle automorphisms, induced by automorphisms of $P$; this will enable us to make use of $G$-invariant fiber metrics which can be regarded as fixed external background fields and are needed to define interesting Lagrangians. Second, we shall assume the space-time manifold $M$ to be equipped with a metric tensor $g$ which in the discussion of the current for gauge theories will serve merely as another fixed external background field but in the discussion of the energy-momentum tensor for generally covariant field theories may acquire a dynamical character. Using the induced volume form, this allows us to express Lagrangians $\hat{\mathcal{L}}$ in terms of Lagrangian functions $L$ instead of Lagrangian densities $\mathcal{L}$ and to reinterpret $(n-1)$-forms, such as the various types of currents discussed in the previous section, as vector fields and $(n-2)$-forms as bivector fields.

It also allows us to rewrite various important formulas of Sect. 3.3 in a more covariant form, which will turn out to facilitate the following discussion. The main ingredient for this conversion is the standard fact that if, locally on $M$, the $Z^\mu$ are the components of a vector field, then the $\sqrt{|\det g|}\, Z^\mu$ are the components of an $(n-1)$-form, and

$$\partial_\mu \left( \sqrt{|\det g|}\, Z^\mu \right) = \sqrt{|\det g|}\, \nabla_\mu Z^\mu \,. \tag{185}$$

Now for example, the definition of the action $S_K$ over a compact subset $K$ of $M$ (cf. eqs (78) and (80)) reads

$$S_K[\varphi] = \int_K d^n x \, \sqrt{|\det g|}\, L(\varphi, \partial\varphi) \,. \tag{186}$$

while conversion of eq. (85) gives the definition of the variational derivative or Euler-Lagrange derivative of $L$, whose vanishing once again expresses the equations of motion. Explicitly, we have

$$\frac{\delta L}{\delta \varphi^i} = \frac{1}{\sqrt{|\det g|}} \left( \frac{\partial \left(\sqrt{|\det g|}\, L\right)}{\partial \varphi^i} - \partial_\mu \left( \frac{\partial \left(\sqrt{|\det g|}\, L\right)}{\partial\, \partial_\mu \varphi^i} \right) \right) \,. \tag{187}$$

Unfortunately, this formula is rarely useful, but simpler expressions in terms of fully covariant quantities can only be derived under additional assumptions on the nature of the field $\varphi$ that allow us to perform an appropriate change of variables, replacing the ordinary partial derivatives of $\varphi$ in the argument $(\varphi, \partial\varphi)$ of the Lagrangian by other, more natural variables. (A typical example is provided by linear matter fields which are sections of some tensor or spinor bundle of $M$ or its tensor product with some other "internal" vector bundle: it is then natural to replace $(\varphi, \partial\varphi)$ by $(\varphi, \nabla\varphi)$ or $(\varphi, D\varphi)$. Other examples are gauge fields, where one replaces $(A, \partial A)$ by $(A, F)$, and the metric tensor, where one replaces $(g, \partial g)$ by $(g, \Gamma)$; we shall deal with these two cases in more detail at the end of Sect. 4.1 and 4.2, respectively.) Next, conversion of eqs. (93) and (94) shows that the variation of $S_K$ under an infinitesimal $G$-bundle automorphism $X$ can be written in the form

$$\delta_X S_K[\varphi] = \int_K d^n x \, \sqrt{|\det g|} \left( \frac{\delta L}{\delta \varphi^i} \delta_X \varphi^i + \nabla_\mu \langle j_{\text{can}}^\mu, X \rangle(\varphi, \partial\varphi) \right) , \tag{188}$$

where now

$$\langle j_{\text{can}}^\mu, X \rangle(\varphi, \partial\varphi) = L X^\mu + \frac{\partial L}{\partial\, \partial_\mu \varphi^i} \delta_X \varphi^i \,. \tag{189}$$

Similarly, conversion of eq. (97) leads to the following version of the relation between the canonical and the improved total Noether current:

$$j_{\text{imp}}^\mu(X; \varphi, \partial\varphi) = \langle j_{\text{can}}^\mu, X \rangle(\varphi, \partial\varphi) + \nabla_\nu j_{\text{cor}}^{\mu\nu}(X; \varphi, \partial\varphi) \,. \tag{190}$$

It should be noted that the covariant divergence of the second term vanishes identically because, for any bivector field $\alpha$ on $M$,

$$2\,\nabla_\mu \nabla_\nu\,\alpha^{\mu\nu} \;=\; [\nabla_\mu,\nabla_\nu]\,\alpha^{\mu\nu} \;=\; R^\mu{}_{\kappa\mu\nu}\,\alpha^{\kappa\nu} + R^\nu{}_{\kappa\mu\nu}\,\alpha^{\mu\kappa} \;=\; 0\;.$$

Therefore, covariant conservation of the improved total Noether current is equivalent to covariant conservation of the canonical total Noether current, so that we may reformulate eq. (188) above in the form

$$\delta_X S_K[\varphi] \;=\; \int_K d^n x\,\sqrt{|\det g|}\,\left(\frac{\delta L}{\delta \varphi^i}\,\delta_X \varphi^i \;+\; \nabla_\mu\, j^\mu_{\mathrm{imp}}(X;\varphi,\partial\varphi)\right)\,. \qquad (191)$$

If $X$ is an infinitesimal symmetry of the Lagrangian, then this expression must vanish, and since $K$ was arbitrary, we obtain an equation for the integrand:

$$\frac{\delta L}{\delta \varphi^i}\,\delta_X \varphi^i \;+\; \nabla_\mu\, j^\mu_{\mathrm{imp}}(X;\varphi,\partial\varphi) \;=\; 0\;. \qquad (192)$$

In what follows, we shall perform a slight change of notation: rather than abbreviating infinitesimal $G$-bundle automorphisms of $E$ by a single letter such as $X$, we shall represent them explicitly as pairs $(X_M, X_E)$ of vector fields $X_M$ on $M$ and $X_E$ on $E$, which in turn are functionals of the pertinent symmetry generators, as already discussed in Sect. 3.4.

Finally, it should be noted that the principles of the above construction continue to prevail for higher order Lagrangian field theories: the first order hypothesis is not really essential. This can be seen by direct inspection of the preceding arguments and calculations in local coordinates: the partial integrations required to arrive at eq. (93) of Sect. 3.3 and at eq. (188) above can without any problem be carried over to this more general context, and the same goes for all the further partial integrations that may be needed to obtain the improved total Noether current from the canonical one, provided one uses an appropriately modified definition of the variational derivative and of the canonical total Noether current. The main example of practical importance are Lagrangians whose dependence on the metric tensor involves partial derivatives up to second order; the explicit formula for the variational derivative with respect to the metric tensor to be derived at the end of Sect. 4.2 will also cover this case. But the global formulation of the Noetherian approach presented in Sect. 3.3, from the identification of the proper geometric framework to the definition of the (canonical or improved) total Noether currents, is of course adapted to a first order formalism; it would be substantially more complicated for higher order Lagrangian field theories. (For example, to begin with, one would have to replace the first order jet bundle used in Sect. 3 by some higher order jet bundle.) However, there is almost nothing to be gained by such a generalization, because the final outcome (as formulated in Theorem 4.1 and Theorem 4.2 below) is exactly the same.

## 4.1 Currents

Continuing the discussion of field theories with gauge invariance begun in Sect. 3.4 under the same heading, we now make use of the fact that standard gauge theories normally contain a $G$-connection field $A$, which can be either external or dynamical: it occupies a special status as the mediator of gauge interactions with all the other fields appearing in the theory, collectively referred to as *matter fields* and now denoted by $\varphi$. Mathematically, $\varphi$ is a section of a $G$-bundle $E$ over $M$ which, as mentioned at the beginning of this section, is assumed to be associated to some given principal $G$-bundle $P$ over $M$ by means of a given action of $G$ on its typical fiber $Q$, whereas $A$ is a connection form on $P$, that is, it is a section of the connection bundle $CP$ of $P$. Therefore, the analysis carried out in Sects 3.2-3.4 continues to apply, with $E$ replaced by the fiber product $CP \times_M E$ and with $\varphi$ replaced by the pair $(A, \varphi)$.

The dynamics of the theory is governed by a total Lagrangian $L$ which, as already stated in eq. (2), is assumed to be the sum of two terms, a "pure gauge field" part $L_{\mathrm{g}}$ depending only on the connection form $A$ and its first order partial derivatives but not on the matter fields or their derivatives, and a "matter field part" $L_{\mathrm{m}}$ depending on the matter fields and their first order partial derivatives as well as on the connection form $A$ and its first order partial derivatives:[16]

$$L(A, \partial A, \varphi, \partial \varphi) \;=\; L_{\mathrm{g}}(A, \partial A) \;+\; L_{\mathrm{m}}(A, \partial A, \varphi, \partial \varphi) \;. \tag{193}$$

A standard additional hypothesis is that $L_{\mathrm{m}}$ depends on the connection form $A$ and its first order partial derivatives only through combinations constructed from the first order covariant derivatives of the matter fields, and possibly also on the curvature form $F$:

$$L_{\mathrm{m}}(A, \partial A, \varphi, \partial \varphi) \;=\; L_{\mathrm{m}}(\varphi, \nabla \varphi, F) \;. \tag{194}$$

Similarly, for the pure gauge field part, the usual hypothesis is that $L_{\mathrm{g}}$ depends on the connection form $A$ and its first order partial derivatives only through its curvature form $F$:

$$L_{\mathrm{g}}(A, \partial A) \;=\; L_{\mathrm{g}}(F) \;. \tag{195}$$

The standard example is of course the Yang-Mills Lagrangian (180), but more complicated polynomials in the curvature form also fit into this framework. (It should be noted tbat there is one important exception, namely the Chern-Simons Lagrangian (184).) In addition, the definition of these Lagrangians will depend on the choice of appropriate $G$-invariant fiber metrics which, being invariant under strict $G$-bundle automorphisms, can be regarded as fixed gauge invariant external background fields; among them is the metric tensor $g$ on space-time mentioned at the beginning of this section. (Again, there is one important exception, namely the Chern-Simons Lagrangian (184), whose definition does not require the choice of a metric on space-time. However, this will change as soon as the Chern-Simons gauge field is coupled to matter fields.)

---

[16]Thus the criterion for deciding whether a given term is to be included in $L_{\mathrm{g}}$ or in $L_{\mathrm{m}}$ is that all terms not depending on the matter fields nor on their derivatives should be incorporated into $L_{\mathrm{g}}$ and all others into $L_{\mathrm{m}}$.

In passing, we note that $L_{\mathrm{m}}$ and $L_{\mathrm{g}}$ are often not given from the very beginning. Rather, and this is the standard method for introducing gauge fields in particle physics, one starts out from a field theory for matter fields alone whose Lagrangian $L_{\mathrm{m}}^0$ shows invariance under some (compact connected) Lie group $G$ acting as an internal global symmetry group and then "gauges" this symmetry, i.e., extends this global symmetry to a local one by introducing the adequate gauge fields and substituting, in the definition of $L_{\mathrm{m}}^0$, the ordinary derivatives by gauge covariant derivatives to obtain $L_{\mathrm{m}}$; $L_{\mathrm{g}}$ is only added in the last step.

Having identified which part of the total Lagrangian is to be considered as the matter field Lagrangian, we proceed to study the consequences of its invariance under gauge transformations. Infinitesimal gauge transformations $\xi$ are sections of the vector bundle $P \times_G \mathfrak{g}^*$ over $M$; they are represented on $E$ and on $\hat{E} = CP \times_M E$ by infinitesimal strict $G$-bundle automorphisms $(0, \xi_E)$ and $(0, \xi_{\hat{E}})$, respectively. Applying eq. (192) with the abbreviation contained in eq. (119) (and suppressing the explicit indication of the field dependence in order to simplify the notation), we see that gauge invariance of the matter field Lagrangian implies

$$\frac{\delta L_{\mathrm{m}}}{\delta A_\mu^a} \, \delta_\xi A_\mu^a \;+\; \frac{\delta L_{\mathrm{m}}}{\delta \varphi^i} \, \delta_\xi \varphi^i \;+\; \nabla_\mu \, (j_a^\mu \, \xi^a) \;=\; 0 \; . \tag{196}$$

Assuming that the matter fields satisfy the equations of motion, this leads us to the following basic relation:

$$\boxed{\left\langle \frac{\delta L_{\mathrm{m}}}{\delta A_\mu} \, , \, \delta_\xi A_\mu \right\rangle \;+\; \nabla_\mu \, \langle j^\mu, \xi \rangle \;=\; 0} \tag{197}$$

To calculate $j$ explicitly, one would have to start out from the formula for $j_{\mathrm{m,can}}$ that follows directly from eq. (189),

$$\langle j_{\mathrm{m,can}}^\mu, \xi_{\hat{E}} \rangle \;=\; \frac{\partial L_{\mathrm{m}}}{\partial \, \partial_\mu A_\nu^a} \, \delta_\xi A_\nu^a \;+\; \frac{\partial L_{\mathrm{m}}}{\partial \, \partial_\mu \varphi^i} \, \delta_\xi \varphi^i \; , \tag{198}$$

and compute $j_{\mathrm{m,imp}}$ by partially integrating and discarding all resulting total divergences, until all terms containing partial derivatives of $\xi$ have disappeared. This can be a cumbersome procedure, and the net result cannot be cast into a simple general formula because the variation $\delta_\xi \varphi^i$ of the matter fields under infinitesimal gauge transformations (which also enters the definition of $\xi_E$ and hence of $\xi_{\hat{E}}$ in terms of $\xi$) depends on their specific nature; only the variation of the connection form is known in general (cf. eq. (156)):

$$\delta_\xi A_\mu \;=\; -\, D_\mu \xi \; . \tag{199}$$

Now there is one situation which is important in applications and is especially simple, namely when a) $L_\mathrm{m}$ does not contain explicit curvature terms and hence depends only on the connection form $A$ but not on its derivatives and b) $E$ is not only a fiber bundle but in fact a vector bundle associated to $P$, so that

$$\delta_\xi \varphi^i \;=\; \xi^a \, (T_a)^i_j \, \varphi^j \;. \tag{200}$$

In this case, there are no correction terms, and the improved current is equal to the canonical one. However, it is easy to construct models where this is no longer so, for example by including an explicit coupling between the curvature form $F$ and a spinor field $\psi$ of the form

$$(\bar\psi \, [\gamma^\mu, \gamma^\nu] \, \psi, F_{\mu\nu})$$

or by including nonlinear matter fields: a field $\varphi$ taking values in an associated affine bundle, rather than vector bundle, may already be sufficient to generate contributions to $\delta_\xi \varphi$ containing derivatives of $\xi$, as shown by the example of the connection form itself.

Fortunately, the details of the construction of $j$ following the strategy of improvement, as described in the previous paragraph, are largely irrelevant: all that we shall really need is eq. (197) above which, combined with the requirement that, as suggested by the notation, the expression $\langle j^\mu, \xi \rangle$ should be $\mathfrak{F}(M)$-linear in $\xi$, we shall call the *ultralocality condition* and which is easily shown to admit a unique solution. To see this, let us apply eq. (199) to rewrite eq. (197) in the form

$$\begin{aligned}
0 &= - \left\langle \frac{\delta L_\mathrm{m}}{\delta A_\mu}, D_\mu \xi \right\rangle + \nabla_\mu \langle j^\mu, \xi \rangle \\
&= \left\langle j^\mu - \frac{\delta L_\mathrm{m}}{\delta A_\mu}, D_\mu \xi \right\rangle + \langle D_\mu j^\mu, \xi \rangle \;,
\end{aligned} \tag{201}$$

where $D_\mu$ applied to $\xi$ denotes the gauge covariant derivative ($D_\mu = \partial_\mu + A_\mu$) and applied to $j^\mu$ denotes the gauge and space-time covariant derivative ($D_\mu = \partial_\mu + A_\mu + \Gamma_\mu$). Using the fact that the value of $\xi$ and of its covariant derivatives at each point of space-time can be chosen independently, we obtain two equations: one of them is eq. (3), which is the relation we really wanted to prove, and the other is the covariant conservation law[17]

$$D_\mu j^\mu \;=\; 0 \;. \tag{202}$$

---

[17] The term "covariant conservation law" traditionally used in this context is somewhat unfortunate since vanishing of the covariant divergence of $j^\mu$ does not describe the conservation of charge but rather the exchange of charge between matter and the gauge field. (The main exception is electrodynamics, where $G$ is Abelian and hence covariant conservation of the current is a conservation law in the traditional sense since the covariant divergence reduces to the ordinary one, in accordance with the fact that the electromagnetic field itself carries no charge and hence there is no exchange of charge between it and the matter fields.) In Yang-Mills theory, covariant conservation of the current is even raised to the status of a consistency condition for the Yang-Mills equation (see eq. (206) below) whose lhs has identically vanishing covariant divergence.

The first of these relations again establishes uniqueness. To show existence, we turn the argument around: using eq. (3) as a definition, we must prove eq. (202). But this is an immediate consequence of gauge invariance of the matter field action. Indeed, given an arbitrary infinitesimal gauge transformation $\xi$ with compact support, take $K$ to be any compact subset of space-time containing the support of $\xi$ and use eq. (188) or (191) to calculate the variation of the matter field action $S_{m,K}$ over $K$ with respect to $X$: this gives

$$\begin{aligned}\delta_\xi S_{m,K} &= \int_K d^n x \; \sqrt{|\det g|} \; \left(\left\langle \frac{\delta L_m}{\delta A_\mu}, \delta_\xi A_\mu \right\rangle + \frac{\delta L_m}{\delta \varphi^i} \delta_\xi \varphi^i \right) \\ &= -\int_K d^n x \; \sqrt{|\det g|} \; \langle j^\mu, D_\mu \xi \rangle \\ &= \int_K d^n x \; \sqrt{|\det g|} \; \langle D_\mu j^\mu, \xi \rangle \end{aligned}$$

where in the second step, we have used eqs (3) and (199) together with the equations of motion for the matter fields, and in the last step, we have discarded the surface term coming from the partial integration because it vanishes due to our support assumption. Since this expression vanishes and $X$ was arbitrary, eq. (202) follows.

We formulate this result as a theorem, continuing to use the term "on shell" to indicate the restriction that the respective fields must satisfy their equations of motion.

**Theorem 4.1** *In first order Lagrangian field theories with gauge invariance, and after splitting off the gauge field $A$ from the matter field(s) $\varphi$, the (field dependent) improved current $j$ derived from the matter field Lagrangian $L_m$ according to Def. 3.1 is the physical current of the theory. Assuming the matter fields to be "on shell", it is simply given by the variational derivative of $L_m$ with respect to the gauge field:*

$$j^\mu = \frac{\delta L_m}{\delta A_\mu} \; . \tag{203}$$

*Explicitly, this means that $j$ is the vector field on space-time $M$ with values in the dual $P \times_G \mathfrak{g}^*$ of the associated Lie algebra bundle $P \times_G \mathfrak{g}$ depending on the fields of the theory which satisfies*

$$\delta_A \int_K d^n x \; \sqrt{|\det g|} \; L_m = \int_K d^n x \; \sqrt{|\det g|} \; \langle j^\mu, \delta A_\mu \rangle \tag{204}$$

*for every compact subset $K$ of $M$ and every variation $\delta A_\mu$ of the connection form with support contained in $K$, where the integrand on the rhs of eq. (204) is understood to contain no derivatives of $\delta A_\mu$ (this requirement is, once again, the expression of an implicit ultralocality principle).*

This theorem entails all the standard properties of the physical current. First of all, it is clear that it depends only on the dynamical equivalence class of the matter field Lagrangian, that is, $j$ does not change when $L_\mathrm{m}$ is modified by the addition of a total divergence ($L_\mathrm{m} \to L_\mathrm{m} + \nabla_\mu C^\mu$) of an expression ($C^\mu$) which is a local function of the gauge field, simply because the addition of such a term does not affect the lhs of eq. (204). Second, gauge invariance of the matter field action forces the current to be covariantly conserved, as stated in eq. (202) and proved above, provided the matter fields satisfy the equations of motion. (On the other hand, this conclusion is completely independent of whether the gauge field satisfies any equations of motion.) Third, the current is the source of the gauge field in the sense that if, as already stated in eq. (2), the total Lagrangian $L$ is the sum of the given matter field Lagrangian $L_\mathrm{m}$ and a gauge field Lagrangian $L_\mathrm{g}$ depending only on the gauge field and its derivatives, then the equations of motion for the gauge field will be

$$\frac{\delta L_\mathrm{g}}{\delta A_\mu} = -j^\mu . \tag{205}$$

In particular, if $L_\mathrm{g}$ is the Yang-Mills Lagrangian (180), these will be the Yang-Mills field equations:

$$D_\nu F^{\nu\mu} = -j^\mu . \tag{206}$$

(Note the extra sign on the rhs, which is opposite to the usual sign convention for the current, e.g., in electrodynamics.) Of course, it seems tempting to also introduce a "gauge field current" $j_\mathrm{g}$, given by

$$j_\mathrm{g}^\mu = \frac{\delta L_\mathrm{g}}{\delta A_\mu} . \tag{207}$$

In particular, if $L_\mathrm{g}$ is the Yang-Mills Lagrangian (180), this expression reads

$$j_\mathrm{g}^\mu = D_\nu F^{\nu\mu} . \tag{208}$$

Then the equations of motion for the gauge field imply (and in fact become identical with) the statement that the total current $j_\mathrm{g} + j$ vanishes "on shell", as required by the second Noether theorem. It must however be pointed out that this interpretation is formal and has no deeper physical significance, since the expression in eq. (207) or (208) above does not represent a physical current for the gauge fields. (Consider electromagnetism, for example: it would seem weird to regard the divergence of the electric field as a charge density for the electromagnetic field, given the fact that the electromagnetic field carries no charge.)

Finally, it is instructive to give an explicit formula for the variational derivative appearing in eq. (203) for the general case; this can be derived directly from eq. (204). As mentioned before (see the comments following eq. (187)), the basic trick is to replace the pair of variables $(A, \partial A)$ by the pair of variables $(A, F)$, as indicated in eq. (194), and use the fact that for any given variation $\delta A_\mu$ of the connection form, we have

$$\delta F_{\mu\nu} = D_\mu \delta A_\nu - D_\nu \delta A_\mu \tag{209}$$

for the induced variation of the curvature form. Using these relations, we obtain for any given variation $\delta A_\mu$ of the connection form with support contained in some compact set $K$,

$$\delta_A \int_K d^n x \sqrt{|\det g|}\, L_\mathrm{m}$$

$$= \int_K d^n x \sqrt{|\det g|}\, \left( \left\langle \frac{\partial L_\mathrm{m}}{\partial A_\mu}, \delta A_\mu \right\rangle + \left\langle \frac{\partial L_\mathrm{m}}{\partial F_{\nu\mu}}, \delta F_{\nu\mu} \right\rangle \right)$$

$$= \int_K d^n x \sqrt{|\det g|}\, \left\langle \frac{\partial L_\mathrm{m}}{\partial A_\mu} - 2 D_\nu \frac{\partial L_\mathrm{m}}{\partial F_{\nu\mu}}, \delta A_\mu \right\rangle$$

where we have discarded the surface term coming from the partial integration because it vanishes due to our support assumption. As a result,

$$j^\mu \;=\; \frac{\delta L_\mathrm{m}}{\delta A_\mu} \;=\; \frac{\partial L_\mathrm{m}}{\partial A_\mu} \,-\, 2 D_\nu \frac{\partial L_\mathrm{m}}{\partial F_{\nu\mu}} \;. \qquad (210)$$

The simplest class of matter field Lagrangians is of course formed by those that depend only on the connection form $A$ itself but not on its derivatives; many important examples belong to this class. In this case, the variational derivative in eq. (210) reduces to an ordinary partial derivative:

$$j^\mu \;=\; \frac{\partial L_\mathrm{m}}{\partial A_\mu} \;. \qquad (211)$$

Moreover, choosing $L_\mathrm{g}$ to vanish will then force $A$ to be a non-dynamical external field, so that making $A$ dynamical will require a non-trivial gauge field Lagrangian.

## 4.2 Energy-momentum tensor

Continuing the discussion of field theories with general covariance, or space-time diffeomorphism invariance, begun in Sect. 3.4 under the same heading, we now make use of the fact that generally covariant field theories normally contain a metric tensor field $g$, which can be either external or dynamical: it occupies a special status as the mediator of gravitational interactions with all the other fields appearing in the theory, collectively referred to as *matter fields* and now denoted by $\varphi$. (Of course, both gauge fields and matter fields in the sense of the previous subsection are to be considered as matter fields in this new sense.) Mathematically, $\varphi$ is a section of a $G$-bundle $E$ over $M$ which, as mentioned at the beginning of this section, is assumed to be associated to some given principal $G$-bundle $P$ over $M$ by means of a given action of $G$ on its typical fiber $Q$, whereas $g$ is a section of the symmetric second order tensor bundle $\bigvee^2 T^*M$ of $M$. Therefore, the analysis carried out in Sects 3.2-3.4 continues to apply, with $E$ replaced by the fiber product $\bigvee^2 T^*M \times_M E$ and with $\varphi$ replaced by the pair $(g, \varphi)$.

The dynamics of the theory is governed by a total Lagrangian $L$ which, as already stated in eq. (4), is assumed to be the sum of two terms, a "purely gravitational" part $L_\mathrm{g}$ depending only on the metric tensor $g$ and its first and second order partial derivatives but not on the matter fields or their derivatives, and a "matter field part" $L_\mathrm{m}$ depending on the matter fields and their first order partial derivatives as well as on the metric tensor $g$ and its first and second order partial derivatives:[18]

$$L(g, \partial g, \partial^2 g, \varphi, \partial\varphi) \;=\; L_\mathrm{g}(g, \partial g, \partial^2 g) \;+\; L_\mathrm{m}(g, \partial g, \partial^2 g, \varphi, \partial\varphi) \,. \qquad (212)$$

A standard additional hypothesis is that $L_\mathrm{m}$ depends only on the metric tensor $g$ and its first order partial derivatives, and only through combinations constructed from $g$ itself and the first order covariant derivatives of the matter fields:

$$L_\mathrm{m}(g, \partial g, \varphi, \partial\varphi) \;=\; L_\mathrm{m}(g, \varphi, \nabla\varphi) \,. \qquad (213)$$

However, it will be important to also consider the more general case in which $L_\mathrm{m}$ is allowed to depend on the metric tensor $g$ and its first and second order partial derivatives through additional terms involving the Riemann curvature tensor $R$:

$$L_\mathrm{m}(g, \partial g, \partial^2 g, \varphi, \partial\varphi) \;=\; L_\mathrm{m}(g, R, \varphi, \nabla\varphi) \,. \qquad (214)$$

Similarly, for the purely gravitational part, the usual hypothesis is that $L_\mathrm{g}$ depends on the metric tensor $g$ and its first and second order partial derivatives only through combinations constructed from $g$ itself and the Riemann curvature tensor $R$:

$$L_\mathrm{g}(g, \partial g, \partial^2 g) \;=\; L_\mathrm{g}(g, R) \,. \qquad (215)$$

The standard example is of course the Einstein-Hilbert Lagrangian (181), but more complicated polynomials in the curvature tensor also fit into this framework. In addition, the definition of these Lagrangians will depend on the choice of appropriate $G$-invariant fiber metrics which, being invariant under $G$-bundle automorphisms – strict as well as non-strict – can be regarded as fixed gauge and space-time diffeomorphism invariant external background fields.

In passing, we note that $L_\mathrm{m}$ and $L_\mathrm{g}$ are often not given from the very beginning. Rather, and this is the standard method for handling matter fields when passing from special to general relativity, one starts out from a field theory for matter fields with a Lagrangian $L_\mathrm{m}^0$ on Minkowski space-time and "gauges" this metric structure by substituting, in the definition of $L_\mathrm{m}^0$, contractions with the flat Minkowski metric tensor by contractions with a general metric tensor and ordinary derivatives by space-time covariant derivatives (that is, covariant derivatives with respect to the corresponding Levi-Civita connection) to obtain $L_\mathrm{m}$; $L_\mathrm{g}$ is only added in the last step.

Another important point to be observed is that matter field Lagrangians of the form (214) are of course of second order and therefore not directly covered by a first order Lagrangian formalism, whereas matter field Lagrangians of the form (213) are. (Similarly, Lagrangians of the form (215) for the purely gravitational part, among them the Einstein-Hilbert Lagrangian, are also of second order.) But as has already been announced before, all arguments and calculations to be presented in what follows are equally valid in the more general case; the reader is invited to verify this, step by step, as we go along.

---

[18]See Footnote 16.

Having identified which part of the total Lagrangian is to be considered as the matter field Lagrangian, we proceed to study the consequences of its invariance under space-time diffeomorphisms. Infinitesimal space-time diffeomorphisms $X$ are simply vector fields on $M$; they are represented on $E$ and on $\hat{E} = \bigvee^2 T^*M \times_M E$ by infinitesimal $G$-bundle automorphisms $(X, X_E)$ and $(X, X_{\hat{E}})$, respectively, where some kind of lifting procedure at the infinitesimal level, as described in Sect. 3.4, has been invoked. Applying eq. (192) with the abbreviation contained in eq. (123) (and suppressing the explicit indication of the field dependence in order to simplify the notation), we see that general coordinate invariance of the matter field Lagrangian implies

$$\frac{\delta L_{\mathrm{m}}}{\delta g_{\mu\nu}} \, \delta_X g_{\mu\nu} \; + \; \frac{\delta L_{\mathrm{m}}}{\delta \varphi^i} \, \delta_X \varphi^i \; + \; \nabla_\mu \left( T^{\mu\nu} X_\nu \right) \;=\; 0 \; . \tag{216}$$

Assuming that the matter fields satisfy the equations of motion, this leads us to the following basic relation:

$$\boxed{\frac{\delta L_{\mathrm{m}}}{\delta g_{\mu\nu}} \, \delta_X g_{\mu\nu} \; + \; \nabla_\mu \left( T^{\mu\nu} X_\nu \right) \;=\; 0} \tag{217}$$

To calculate $T$ explicitly, one would have to start out from the formula for $j_{\mathrm{m,can}}$ that follows directly from eq. (189),

$$\langle j_{\mathrm{m,can}}^\mu, X_{\hat{E}} \rangle \;=\; L_{\mathrm{m}} \, X^\mu \;+\; \frac{\partial L_{\mathrm{m}}}{\partial \, \partial_\mu g_{\kappa\lambda}} \, \delta_X g_{\kappa\lambda} \;+\; \frac{\partial L_{\mathrm{m}}}{\partial \, \partial_\mu \varphi^i} \, \delta_X \varphi^i \; , \tag{218}$$

and compute $j_{\mathrm{m,imp}}$ by partially integrating and discarding all resulting total divergences, until all terms containing partial derivatives of $X$ have disappeared. This can be a cumbersome procedure, and the net result cannot be cast into a simple general formula because the variation $\delta_X \varphi^i$ of the matter fields under infinitesimal space-time diffeomorphisms (which also enters the definition of $X_E$ and hence of $X_{\hat{E}}$ in terms of $X$) depends on their specific nature; only the variation of the metric tensor is known in general:

$$\delta_X g_{\mu\nu} \;=\; \nabla_\mu X_\nu + \nabla_\nu X_\mu \; . \tag{219}$$

Now there is one situation which is important in applications and is especially simple, namely when all matter fields are scalars. In this case, $L_{\mathrm{m}}$ will not contain Christoffel symbols and hence depends only on the metric tensor $g$ but not on its derivatives, while according to eq. (64),

$$\delta_X \varphi^i \;=\; -\, X^\mu \, \partial_\mu \varphi^i \; . \tag{220}$$

Therefore, there are no correction terms, and the improved energy-momentum tensor is equal to the canonical one. However, in the majority of physically relevant models – beginning with standard electrodynamics – this is no longer so, simply because most of them contain tensor or spinor fields on which space-time diffeomorphisms act non-trivially so that the first term in eq. (64) no longer vanishes and contains derivatives of $X$ (corresponding to the Lie derivative of $\varphi^i$ along $X$) that must be removed by partial integration.

Fortunately, the details of the construction of $T$ following the strategy of improvement, as described in the previous paragraph, are largely irrelevant: all that we shall really need is eq. (217) above which, combined with the requirement that, as suggested by the notation, the expression $T^{\mu\nu} X_\nu$ should be $\mathfrak{F}(M)$-linear in $X$, we shall call the *ultralocality condition* and which is easily shown to admit a unique solution. To see this, let us apply eq. (219) to rewrite eq. (217) in the form

$$\begin{aligned} 0 &= 2\, \frac{\delta L_\mathrm{m}}{\delta g_{\mu\nu}}\, \nabla_\mu X_\nu + \nabla_\mu \left( T^{\mu\nu} X_\nu \right) \\ &= \left( T^{\mu\nu} + 2\, \frac{\delta L_\mathrm{m}}{\delta g_{\mu\nu}} \right) \nabla_\mu X_\nu + \left( \nabla_\mu T^{\mu\nu} \right) X_\nu \; . \end{aligned} \tag{221}$$

Using the fact that the value of $X$ and of its covariant derivatives at each point of space-time can be chosen independently, we obtain two equations: one of them is eq. (5) (the relation we wanted to prove originally) and the other is the covariant conservation law[19]

$$\nabla_\mu T^{\mu\nu} = 0 \; . \tag{222}$$

The first of these relations again establishes uniqueness. To show existence, we turn the argument around: using eq. (5) as a definition, we must prove eq. (222). But this is an immediate consequence of general coordinate invariance of the matter field action. Indeed, given an arbitrary vector field $X$ on $M$ with compact support, together with an infinitesimal $G$-bundle automorphism $X_E$ of $E$ covering $X$, take $K$ to be any compact subset of space-time containing the support of $X$ and use eq. (188) or (191) to calculate the variation of the matter field action $S_{\mathrm{m},K}$ over $K$ with respect to $X$: this gives

$$\begin{aligned} \delta_X S_{\mathrm{m},K} &= \int_K d^n x \, \sqrt{|\det g|} \, \left( \frac{\delta L_\mathrm{m}}{\delta g_{\mu\nu}}\, \delta_X g_{\mu\nu} + \frac{\delta L_\mathrm{m}}{\delta \varphi^i}\, \delta_X \varphi^i \right) \\ &= -\int_K d^n x \, \sqrt{|\det g|} \; T^{\mu\nu}\, \nabla_\mu X_\nu \\ &= \int_K d^n x \, \sqrt{|\det g|} \; \left( \nabla_\mu T^{\mu\nu} \right) X_\nu \end{aligned}$$

---

[19]The term "covariant conservation law" traditionally used in this context is somewhat unfortunate since vanishing of the covariant divergence of $T^{\mu\nu}$ does not describe the local conservation of energy and momentum but rather the exchange of energy and momentum between matter and the gravitational field. In general relativity, covariant conservation and symmetry of the energy-momentum tensor are even raised to the status of a consistency condition for the Einstein equation (see eq. (226) below) whose lhs is automatically symmetric and has identically vanishing covariant divergence.

where in the second step, we have used eqs (5) and (219) together with the equations of motion for the matter fields, and in the last step, we have discarded the surface term coming from the partial integration because it vanishes due to our support assumption.. Since this expression vanishes and $X$ was arbitrary, eq. (222) follows.

We formulate this result as a theorem, continuing to use the term "on shell" to indicate the restriction that the respective fields must satisfy their equations of motion.

**Theorem 4.2** *In generally covariant first order Lagrangian field theories, and after splitting off the metric tensor field $g$ from the matter field(s) $\varphi$, the (field dependent) improved energy-momentum tensor $T$ derived from the matter field Lagrangian $L_\mathrm{m}$ according to Def. 3.2 is the physical energy-momentum tensor of the theory. Assuming the matter fields to be "on shell", it is simply given by the variational derivative of $L_\mathrm{m}$ with respect to the metric tensor:*

$$T^{\mu\nu} = -2\,\frac{\delta L_\mathrm{m}}{\delta g_{\mu\nu}} \; . \qquad (223)$$

*Explicitly, this means that $T$ is the rank 2 tensor field on space-time $M$ depending on the fields of the theory which satisfies*

$$\delta_g \int_K d^n x \, \sqrt{|\det g|} \; L_\mathrm{m} = -\tfrac{1}{2} \int_K d^n x \, \sqrt{|\det g|} \; T^{\mu\nu} \, \delta g_{\mu\nu} \qquad (224)$$

*for every compact subset $K$ of $M$ and every variation $\delta g_{\mu\nu}$ of the metric tensor with support contained in $K$, where the integrand on the rhs of eq. (224) is understood to contain no derivatives of $\delta g_{\mu\nu}$ (this requirement is, once again, the expression of an implicit ultralocality principle). The same statement also holds if the matter field Lagrangian $L_\mathrm{m}$ is allowed to depend on second order partial derivatives of the metric tensor, e.g., through explicit curvature terms as in eq. (214) instead of eq. (213).*

This theorem entails all the standard properties of the physical energy-momentum tensor. First of all, it is clear that it depends only on the dynamical equivalence class of the matter field Lagrangian, that is, $T$ does not change when $L_\mathrm{m}$ is modified by the addition of a total divergence ($L_\mathrm{m} \to L_\mathrm{m} + \nabla_\mu C^\mu$) of an expression ($C^\mu$) which is a local function of the metric tensor, simply because the addition of such a term does not affect the lhs of eq. (224). Second, general coordinate invariance of the matter field action forces the energy-momentum tensor to be covariantly conserved, as stated in eq. (222) and proved above, provided the matter fields satisfy the equations of motion. (On the other hand, this conclusion is completely independent of whether the metric tensor satisfies any equations of motion.) Third, the current is the source of gravity in the sense that if, as already stated in eq. (4), the total Lagrangian $L$ is the sum of the given matter field Lagrangian $L_\mathrm{m}$ and a gravitational Lagrangian $L_\mathrm{g}$ depending only on the metric tensor and its derivatives, then the equations of motion for the metric tensor will be

$$2\,\frac{\delta L_\mathrm{g}}{\delta g_{\mu\nu}} = T^{\mu\nu} \; . \qquad (225)$$

In particular, if $L_g$ is the Einstein-Hilbert Lagrangian (181), these will be the Einstein field equations:

$$G_{\mu\nu} - \Lambda g_{\mu\nu} \equiv R_{\mu\nu} - \tfrac{1}{2} g_{\mu\nu} R - \Lambda g_{\mu\nu} = T_{\mu\nu} . \tag{226}$$

Of course, it seems tempting to also introduce a "gravitational energy-momentum tensor" $T_g$ by setting

$$T_g^{\mu\nu} = -2 \frac{\delta L_g}{\delta g_{\mu\nu}} . \tag{227}$$

In particular, if $L_g$ is the Einstein-Hilbert Lagrangian (181), this expression reads

$$T_g^{\mu\nu} = -G_{\mu\nu} + \Lambda g_{\mu\nu} \equiv -\left(R_{\mu\nu} - \tfrac{1}{2} g_{\mu\nu} R\right) + \Lambda g_{\mu\nu} . \tag{228}$$

Then the equations of motion for the metric tensor imply (and in fact become identical with) the statement that the total energy-momentum tensor $T_g + T$ vanishes "on shell", as required by the second Noether theorem. It must however be pointed out that this interpretation is formal and has no deeper physical significance, since the expression in eq. (227) or (228) above does not represent a physical energy-momentum tensor for the gravitational field: as is well known, such an object does not exist.

Finally, it is instructive to give an explicit formula for the variational derivative appearing in eq. (223) for the general case; this can be derived directly from eq. (224). As mentioned before (see the comments following eq. (187)), the basic trick is to replace the pair of variables $(g, \partial g)$ by the pair of variables $(g, \Gamma)$, as indicated in eq. (213), or more generally the triplet of variables $(g, \partial g, \partial^2 g)$ by the triplet of variables $(g, \Gamma, R)$, as indicated in eq. (214), and use the fact that for any given variation $\delta g_{\mu\nu}$ of the metric tensor, we have

$$\delta g^{\mu\nu} = -g^{\mu\kappa} g^{\nu\lambda} \delta g_{\kappa\lambda} \tag{229}$$

for the induced variation of the inverse metric tensor,

$$\delta \sqrt{|\det g|} = \tfrac{1}{2} \sqrt{|\det g|} \, g^{\mu\nu} \delta g_{\mu\nu} \tag{230}$$

for the induced variation of the metric determinant,

$$\delta \Gamma^{\kappa}_{\mu\lambda} = \tfrac{1}{2} g^{\kappa\nu} \left( \nabla_\mu \delta g_{\lambda\nu} + \nabla_\lambda \delta g_{\mu\nu} - \nabla_\nu \delta g_{\mu\lambda} \right) \tag{231}$$

for the induced variation of the Christoffel symbols and

$$\delta R^{\kappa}{}_{\lambda\mu\nu} = \nabla_\mu \delta \Gamma^{\kappa}_{\lambda\nu} - \nabla_\nu \delta \Gamma^{\kappa}_{\lambda\mu} \tag{232}$$

for the induced variation of the Riemann curvature tensor. Using these relations, we obtain for any given variation $\delta g_{\mu\nu}$ of the metric tensor with support contained in some compact set $K$,

$$\delta_g \int_K d^n x \sqrt{|\det g|}\, L_\mathrm{m}$$

$$= \int_K d^n x \sqrt{|\det g|}\, \left( \tfrac{1}{2} g^{\mu\nu} L_\mathrm{m}\, \delta g_{\mu\nu} + \frac{\partial L_\mathrm{m}}{\partial g_{\mu\nu}}\, \delta g_{\mu\nu} + \frac{\partial L_\mathrm{m}}{\partial \Gamma^\kappa_{\mu\nu}}\, \delta \Gamma^\kappa_{\mu\nu} + \frac{\partial L_\mathrm{m}}{\partial R^\kappa{}_{\lambda\mu\nu}}\, \delta R^\kappa{}_{\lambda\mu\nu} \right)$$

$$= \int_K d^n x \sqrt{|\det g|}\, \left( \tfrac{1}{2} g^{\mu\nu} L_\mathrm{m}\, \delta g_{\mu\nu} + \frac{\partial L_\mathrm{m}}{\partial g_{\mu\nu}}\, \delta g_{\mu\nu} \right.$$
$$+ \tfrac{1}{2} g^{\kappa\lambda} \left( \frac{\partial L_\mathrm{m}}{\partial \Gamma^\kappa_{\mu\nu}} + \nabla_\rho \frac{\partial L_\mathrm{m}}{\partial R^\kappa{}_{\mu\nu\rho}} + \nabla_\rho \frac{\partial L_\mathrm{m}}{\partial R^\kappa{}_{\nu\mu\rho}} \right)$$
$$\left. \times \left( \nabla_\mu \delta g_{\nu\lambda} + \nabla_\nu \delta g_{\mu\lambda} - \nabla_\lambda \delta g_{\mu\nu} \right) \right)$$

$$= \int_K d^n x \sqrt{|\det g|}\, \left( \tfrac{1}{2} g^{\mu\nu} L_\mathrm{m}\, \delta g_{\mu\nu} + \frac{\partial L_\mathrm{m}}{\partial g_{\mu\nu}}\, \delta g_{\mu\nu} \right.$$
$$- \tfrac{1}{2} \left[ g^{\kappa\mu} \nabla_\lambda \left( \frac{\partial L_\mathrm{m}}{\partial \Gamma^\kappa_{\lambda\nu}} + \nabla_\rho \frac{\partial L_\mathrm{m}}{\partial R^\kappa{}_{\lambda\nu\rho}} + \nabla_\rho \frac{\partial L_\mathrm{m}}{\partial R^\kappa{}_{\nu\lambda\rho}} \right) \right.$$
$$+ g^{\kappa\nu} \nabla_\lambda \left( \frac{\partial L_\mathrm{m}}{\partial \Gamma^\kappa_{\mu\lambda}} + \nabla_\rho \frac{\partial L_\mathrm{m}}{\partial R^\kappa{}_{\mu\lambda\rho}} + \nabla_\rho \frac{\partial L_\mathrm{m}}{\partial R^\kappa{}_{\lambda\mu\rho}} \right)$$
$$\left. \left. - g^{\kappa\lambda} \nabla_\lambda \left( \frac{\partial L_\mathrm{m}}{\partial \Gamma^\kappa_{\mu\nu}} + \nabla_\rho \frac{\partial L_\mathrm{m}}{\partial R^\kappa{}_{\mu\nu\rho}} + \nabla_\rho \frac{\partial L_\mathrm{m}}{\partial R^\kappa{}_{\nu\mu\rho}} \right) \right] \delta g_{\mu\nu} \right)$$

where we have discarded all surface terms coming from the partial integrations because they vanish due to our support assumption. As a result,

$$\begin{aligned}
T^{\mu\nu} =\; & -2\, \frac{\partial L_\mathrm{m}}{\partial g_{\mu\nu}} - g^{\mu\nu} L_\mathrm{m} \\
& + \nabla_\lambda \left( g^{\kappa\mu} \frac{\partial L_\mathrm{m}}{\partial \Gamma^\kappa_{\lambda\nu}} + g^{\kappa\nu} \frac{\partial L_\mathrm{m}}{\partial \Gamma^\kappa_{\lambda\mu}} - g^{\kappa\lambda} \frac{\partial L_\mathrm{m}}{\partial \Gamma^\kappa_{\mu\nu}} \right) \\
& + \nabla_\rho \nabla_\sigma \left( g^{\kappa\mu} \frac{\partial L_\mathrm{m}}{\partial R^\kappa{}_{\rho\nu\sigma}} + g^{\kappa\mu} \frac{\partial L_\mathrm{m}}{\partial R^\kappa{}_{\nu\rho\sigma}} \right. \\
& \qquad\qquad + g^{\kappa\nu} \frac{\partial L_\mathrm{m}}{\partial R^\kappa{}_{\rho\mu\sigma}} + g^{\kappa\nu} \frac{\partial L_\mathrm{m}}{\partial R^\kappa{}_{\mu\rho\sigma}} \\
& \qquad\qquad \left. - g^{\kappa\rho} \frac{\partial L_\mathrm{m}}{\partial R^\kappa{}_{\mu\nu\sigma}} - g^{\kappa\rho} \frac{\partial L_\mathrm{m}}{\partial R^\kappa{}_{\nu\mu\sigma}} \right)
\end{aligned} \qquad (233)$$

The simplest class of matter field Lagrangians is of course formed by those that depend only on the metric tensor $g$ itself but not on its derivatives; many important examples belong to this class. In this case, the variational derivative in eq. (223) reduces to an ordinary partial derivative, plus the contribution from the metric determinant:

$$T^{\mu\nu} \;=\; -\,2\,\frac{\partial L_{\mathrm{m}}}{\partial g_{\mu\nu}} \,-\, g^{\mu\nu} L_{\mathrm{m}} \;. \tag{234}$$

Moreover, choosing $L_{\mathrm{g}}$ to vanish will then force $g$ to be a non-dynamical external field, so that making $g$ dynamical will require a non-trivial gravitational Lagrangian.

## 4.3 Calculation of the energy-momentum tensor

In this final subsection, we want to present a series of examples for the calculation of the physical energy-momentum tensor, based on the result formulated in Theorem 4.2, devoting special attention to two cases that are physically relevant but do not fit into the standard framework: a) Lagrangians containing an explicit coupling between curvature and a scalar field and b) Dirac type Lagrangians for spinor fields. In both cases, the prescription of Theorem 4.2 does provide the correct answer, but some of the arguments required are far from straightforward and, to our knowledge, are scattered across the literature.

We begin with the case of a multiplet of scalar fields, with either linearly or non-linearly realized internal symmetry, whose Lagrangian is one of the expressions given in eqs (175)-(178). In all these cases, the variational derivative with respect to the metric tensor is simply given by eq. (234), so we can use the formula

$$\frac{\partial g^{\mu\nu}}{\partial g_{\kappa\lambda}} \;=\; -\,\tfrac{1}{2} \left( g^{\mu\kappa} g^{\nu\lambda} + g^{\mu\lambda} g^{\nu\kappa} \right) \tag{235}$$

(which follows from the fact that the matrix $(g^{\mu\nu})$ is defined to be the inverse of the matrix $(g_{\mu\nu})$) to rewrite eq. (234) in the equivalent form

$$T_{\mu\nu} \;=\; 2\,\frac{\partial L_{\mathrm{m}}}{\partial g^{\mu\nu}} \,-\, g_{\mu\nu} L_{\mathrm{m}} \;. \tag{236}$$

The same statement holds for gauge fields subject to the Yang-Mills Lagrangian (180), including as a special case that of the electromagnetic field. The (standard) results of these calculations are listed in Table 1.

As our second example, which will become important in the next section, we consider again the theory of a single real scalar field $\varphi$ but with a modified Lagrangian which is the sum

$$L_{\mathrm{MSC}} \;=\; L_{\mathrm{RSC}} + L_{\mathrm{CC}} \tag{237}$$

of the standard Lagrangian
$$L_{\text{RSC}} = \tfrac{1}{2} g^{\mu\nu} \partial_\mu \varphi \, \partial_\nu \varphi - U(\varphi) \tag{238}$$
for the real scalar field and a new term
$$L_{\text{CC}} = R f(\varphi) \tag{239}$$
describing an additional coupling of the scalar field to the Ricci scalar curvature $R$, where $f$ is some (unspecified) smooth function of its argument. The corresponding equation of motion reads
$$\Box \varphi + U'(\varphi) - R f'(\varphi) = 0 \tag{240}$$
where $\Box$ is the pertinent covariant wave operator:
$$\Box = g^{\mu\nu} \nabla_\mu \nabla_\nu \, . \tag{241}$$
To calculate the energy-momentum tensor for this Lagrangian, application of eq. (233) is less convenient than invoking a direct approach which follows the strategy used for deriving the Einstein field equations from the Einstein-Hilbert Lagrangian. To deal with the most difficult term first, note that for any given variation $\delta g_{\mu\nu}$ of the metric tensor with support contained in some compact set $K$, eqs (229)-(232) imply that the trace of the induced variation of the Ricci tensor is given by
$$g^{\mu\nu} \delta R_{\mu\nu} = \left( g_{\mu\nu} \Box - \nabla_\mu \nabla_\nu \right) \delta g^{\mu\nu} \, . \tag{242}$$
Therefore, the term $R f(\varphi)$ gives the following contribution to the lhs of eq. (224)
$$\delta_g \int_K d^n x \, \sqrt{|\det g|} \, R f(\varphi)$$
$$= \int_K d^n x \, \sqrt{|\det g|} \, \left( \tfrac{1}{2} g^{\mu\nu} \delta g_{\mu\nu} \, R f(\varphi) + \delta g^{\mu\nu} R_{\mu\nu} f(\varphi) + g^{\mu\nu} \delta R_{\mu\nu} f(\varphi) \right)$$
$$= \int_K d^n x \, \sqrt{|\det g|} \, \left( \left( g_{\mu\nu} \Box - \nabla_\mu \nabla_\nu \right) f(\varphi) + \left( R_{\mu\nu} - \tfrac{1}{2} g_{\mu\nu} R \right) f(\varphi) \right) \delta g^{\mu\nu}$$
where we have discarded the surface terms coming from the partial integrations because they vanish due to our support assumptions. Thus the energy-momentum tensor for this theory is the sum
$$T^{\text{MSC}}_{\mu\nu} = T^{\text{RSF}}_{\mu\nu} + T^{\text{CC}}_{\mu\nu} \tag{243}$$
of the standard energy-momentum tensor
$$T^{\text{RSC}}_{\mu\nu} = \partial_\mu \varphi \, \partial_\nu \varphi - \tfrac{1}{2} g_{\mu\nu} g^{\kappa\lambda} \partial_\kappa \varphi \, \partial_\lambda \varphi + g_{\mu\nu} U(\varphi) \tag{244}$$
for the real scalar field and a new term
$$T^{\text{CC}}_{\mu\nu} = 2 \left( g_{\mu\nu} \Box - \nabla_\mu \nabla_\nu \right) f(\varphi) + 2 \left( R_{\mu\nu} - \tfrac{1}{2} g_{\mu\nu} R \right) f(\varphi) \tag{245}$$
coming from the additional coupling of the scalar field to the curvature; this result is also included in Table 1.

Our third and most complicated example is the theory of Dirac spinors. Here, we face a new conceptual problem concerning the interpretation of the prescription of Theorem 4.2 for computing the energy-momentum tensor, according to eqs (223) and (224). As we have seen, these relations may be regarded as defining the energy-momentum tensor to be the entity describing the reaction of the matter field Lagrangian to variations of the metric, where it is tacitly understood that variations of the metric should be variations of the metric only, that is, the matter fields should be held fixed in this process.[20] Of course, this is an unambiguous rule as long as all the matter fields are tensor fields, but it is not clear "a priori" how to interpret it for spinor fields, whose very definition presupposes the choice of a metric. In fact, spinor fields are sections of a spinor bundle, obtained as an associated vector bundle from the bundle of spin frames (also called a spin structure) which in turn is a double covering of the bundle of (oriented and time-oriented) orthonormal frames: all these bundles depend explicitly on the metric. Generally speaking, we may state that fixing fields which are sections of bundles over space-time means fixing their components with respect to some (and hence any) fixed reference frame. For tensor fields, we may use a holonomic frame, induced by a coordinate system, but for spinor fields, we do not have this option: components of spinor fields refer to spin frames which cover (oriented and time-oriented) orthonormal frames, and these cannot be held fixed when we vary the metric.

In this situation, the adequate procedure is to describe the metric tensor in terms of linear frames, by specifying which of these are orthonormal, and similarly, to consider variations of the metric tensor (such as those needed for the application of Theorem 4.2) as being induced by corresponding variations of linear frames. Of course, it must then be checked that physical quantities such as the energy-momentum tensor are invariant under the new type of gauge transformations introduced by this method, namely the local frame rotations that leave the metric tensor invariant. In local coordinates $x^\mu$ on $M$, the metric tensor is represented by a symmetric matrix of functions $g_{\mu\nu}$ (together with the inverse matrix of functions $g^{\mu\nu}$) as before, while a linear frame is represented by a matrix of functions $e^a_\mu$ providing the coefficients of the expansion of the $\mu$-th coordinate basis vector in terms of the basis vectors of the frame (together with the inverse matrix of functions $e^\mu_a$ providing the coefficients of the expansion of the $a$-th basis vector of the frame in terms of the coordinate basis vectors $\partial_\mu$). Thus we have

$$e^a_\mu \, e^\nu_a \;=\; \delta^\nu_\mu \quad,\quad e^a_\mu \, e^\mu_b \;=\; \delta^a_b \tag{246}$$

and, for orthonormal frames,

$$g_{\mu\nu} \;=\; \eta_{ab} \, e^a_\mu \, e^b_\nu \quad,\quad g^{\mu\nu} \;=\; \eta^{ab} \, e^\mu_a \, e^\nu_b \tag{247}$$

---

[20]In the definition of partial derivatives of functions on $\mathbb{R}^n$ given in standard calculus courses, the necessity of specifying what variables are to be held fixed while varying the others is usually swept under the rug, but it becomes an important issue when ambiguities may arise. For example, in the definition of thermodynamical quantities through partial derivatives, it is important to specify what are the other variables (apart from the one with respect to which the derivative is being taken) that are supposed to be held fixed.

where $\eta$ is the standard flat space-time metric of the same signature as $g$. Next, we use the explicit expression (161) for the Christoffel symbols, which are the connection coefficients of the Levi-Civita connection on the tangent bundle with respect to a coordinate frame, to derive an equally explicit expression for the connection coefficients of the Levi-Civita connection on the tangent bundle with respect to an orthonormal frame (or vector connection coefficients, for short):

$$\Gamma_{\mu,ab} \;=\; -\tfrac{1}{2}\, e_a^\kappa\, e_b^\lambda\left(\partial_\kappa g_{\lambda\mu} - \partial_\lambda g_{\kappa\mu}\right) \;-\; \tfrac{1}{2}\left(\eta_{ac}\, e_b^\nu - \eta_{bc}\, e_a^\nu\right)\partial_\mu e_\nu^c \;. \qquad (248)$$

(The corresponding formula for arbitrary linear frames would contain an additional term of the form $\tfrac{1}{2}\, e_a^\kappa\, e_b^\lambda\, \partial_\mu\!\left(g_{\kappa\lambda} - \eta_{cd}\, e_\kappa^c\, e_\lambda^d\right)$.) This is easily shown by applying the transformation law (155) for connection forms under gauge transformations,

$$\Gamma_{\mu b}^a \;=\; e_\kappa^a\, \Gamma_{\mu\lambda}^\kappa\, e_b^\lambda \;-\; \partial_\mu e_\nu^a\, e_b^\nu \;,$$

which upon writing $\Gamma_{\mu,ab} \equiv \eta_{ac}\Gamma^c_{\mu b}$ and $\Gamma_{\mu,\kappa\lambda} \equiv g_{\kappa\nu}\Gamma^\nu_{\mu\lambda}$ takes the form

$$\Gamma_{\mu,ab} \;=\; e_a^\kappa\, e_b^\lambda\, \Gamma_{\mu,\kappa\lambda} \;-\; \eta_{ac}\, \partial_\mu e_\nu^c\, e_b^\nu \;.$$

Inserting eq. (161) and applying eq. (247) to the first of its three terms, we arrive at eq. (248). The connection coefficients of the Levi-Civita connection on the spinor bundle (or spinor connection coefficients, for short) are then given by the simple algebraic formula

$$\Gamma_{\mu\,\beta}^{\;\;\alpha} \;=\; \tfrac{1}{8}\, \Gamma_{\mu,ab}\, [\gamma^a,\gamma^b]_{\;\beta}^{\alpha} \qquad (249)$$

where the $\gamma^a$ and similarly the $\gamma_a$ are the Dirac $\gamma$-matrices with respect to the pertinent orthonormal frame, related to those with respect to the pertinent coordinate frame in the obvious way:

$$\gamma_\mu \;=\; e_\mu^a\, \gamma_a \quad,\quad \gamma^\mu \;=\; e_a^\mu\, \gamma^a \;. \qquad (250)$$

Turning to variations, note first that differentiating eq. (246) gives

$$\delta e_\mu^a\, e_a^\nu \;=\; -\, e_\mu^a\, \delta e_a^\nu \quad,\quad \delta e_\mu^a\, e_b^\mu \;=\; -\, e_\mu^a\, \delta e_b^\mu \;, \qquad (251)$$

while differentiating eq. (247) gives the variation of the metric tensor induced by a variation of frames

$$\delta g_{\mu\nu} \;=\; \eta_{ab}\left(\delta e_\mu^a\, e_\nu^b + e_\mu^a\, \delta e_\nu^b\right) \quad,\quad \delta g^{\mu\nu} \;=\; \eta^{ab}\left(\delta e_a^\mu\, e_b^\nu + e_a^\mu\, \delta e_b^\nu\right) \;. \qquad (252)$$

The induced variation of the Christoffel symbols is then given by eq. (231), while that of the vector connection coefficients is given by

$$\delta\Gamma_{\mu,ab} \;=\; -\tfrac{1}{2}\, e_a^\kappa\, e_b^\lambda\left(\nabla_\kappa \delta g_{\lambda\mu} - \nabla_\lambda \delta g_{\kappa\mu}\right) \;-\; \tfrac{1}{2}\left(\eta_{ac}\, e_b^\nu - \eta_{bc}\, e_a^\nu\right)\nabla_\mu \delta e_\nu^c \;, \qquad (253)$$

where the covariant derivatives are defined as usual, namely by

$$\nabla_\mu \delta g_{\kappa\lambda} = \partial_\mu \delta g_{\kappa\lambda} - \Gamma^\nu_{\mu\kappa} \delta g_{\nu\lambda} - \Gamma^\nu_{\mu\lambda} \delta g_{\kappa\nu}, \qquad (254)$$

and

$$\nabla_\mu \delta e^a_\nu = \partial_\mu \delta e^a_\nu + \Gamma^a_{\mu b} \delta e^b_\nu - \Gamma^\kappa_{\mu\nu} \delta e^a_\kappa, \qquad (255)$$

respectively. (With the same definition, the covariant derivatives of the unvaried expressions, $\nabla_\mu g_{\kappa\lambda}$ and $\nabla_\mu e^a_\nu$, are easily seen to vanish.) The proof of eq. (253) requires a short calculation:

$$\begin{aligned}
\delta \Gamma_{\mu,ab} &= -\tfrac{1}{2} e^\kappa_a e^\lambda_b \left( \partial_\kappa \delta g_{\lambda\mu} - \partial_\lambda \delta g_{\kappa\mu} \right) \\
&\quad - \tfrac{1}{2} \left( \delta e^\kappa_a e^\lambda_b + e^\kappa_a \delta e^\lambda_b \right) \left( \partial_\kappa g_{\lambda\mu} - \partial_\lambda g_{\kappa\mu} \right) \\
&\quad - \tfrac{1}{2} \left( \eta_{ac} e^\nu_b - \eta_{bc} e^\nu_a \right) \partial_\mu \delta e^c_\nu \\
&\quad - \tfrac{1}{2} \left( \eta_{ac} \delta e^\nu_b - \eta_{bc} \delta e^\nu_a \right) \partial_\mu e^c_\nu \\
&= -\tfrac{1}{2} e^\kappa_a e^\lambda_b \left( \nabla_\kappa \delta g_{\lambda\mu} - \nabla_\lambda \delta g_{\kappa\mu} \right) \\
&\quad - \tfrac{1}{2} e^\kappa_a e^\lambda_b \left( \Gamma^\nu_{\kappa\lambda} \delta g_{\nu\mu} + \Gamma^\nu_{\kappa\mu} \delta g_{\lambda\nu} - \Gamma^\nu_{\lambda\kappa} \delta g_{\nu\mu} - \Gamma^\nu_{\lambda\mu} \delta g_{\kappa\nu} \right) \\
&\quad - \tfrac{1}{2} \left( \delta e^\kappa_a e^\lambda_b + e^\kappa_a \delta e^\lambda_b \right) \left( g_{\lambda\nu} \Gamma^\nu_{\kappa\mu} + g_{\mu\nu} \Gamma^\nu_{\kappa\lambda} - g_{\kappa\nu} \Gamma^\nu_{\lambda\mu} - g_{\mu\nu} \Gamma^\nu_{\lambda\kappa} \right) \\
&\quad - \tfrac{1}{2} \left( \eta_{ac} e^\nu_b - \eta_{bc} e^\nu_a \right) \nabla_\mu \delta e^c_\nu \\
&\quad + \tfrac{1}{2} \left( \eta_{ac} e^\nu_b - \eta_{bc} e^\nu_a \right) \left( \Gamma^c_{\mu d} \delta e^d_\nu - \Gamma^\lambda_{\mu\nu} \delta e^c_\lambda \right) \\
&\quad + \tfrac{1}{2} \left( \eta_{ac} e^\kappa_d \delta e^d_\nu e^\nu_b - \eta_{bc} e^\kappa_d \delta e^d_\nu e^\nu_a \right) \partial_\mu e^c_\kappa \\
&= -\tfrac{1}{2} e^\kappa_a e^\lambda_b \left( \nabla_\kappa \delta g_{\lambda\mu} - \nabla_\lambda \delta g_{\kappa\mu} \right) - \tfrac{1}{2} \left( \eta_{ac} e^\nu_b - \eta_{bc} e^\nu_a \right) \nabla_\mu \delta e^c_\nu \\
&\quad - \tfrac{1}{2} \eta_{cd} e^\kappa_a e^\lambda_b \left( \Gamma^\nu_{\kappa\mu} \left( \delta e^c_\lambda e^d_\nu + e^c_\lambda \delta e^d_\nu \right) - \Gamma^\nu_{\lambda\mu} \left( \delta e^c_\kappa e^d_\nu + e^c_\kappa \delta e^d_\nu \right) \right) \\
&\quad - \tfrac{1}{2} \eta_{cd} \left( \delta e^\kappa_a e^\lambda_b + e^\kappa_a \delta e^\lambda_b \right) \left( e^c_\lambda e^d_\nu \Gamma^\nu_{\kappa\mu} - e^c_\kappa e^d_\nu \Gamma^\nu_{\lambda\mu} \right) \\
&\quad + \tfrac{1}{2} \left( \eta_{ac} e^\nu_b - \eta_{bc} e^\nu_a \right) \left( e^c_\kappa \Gamma^\kappa_{\mu\lambda} e^\lambda_d \delta e^d_\nu - \partial_\mu e^c_\kappa e^\kappa_d \delta e^d_\nu - \Gamma^\lambda_{\mu\nu} \delta e^c_\lambda \right) \\
&\quad + \tfrac{1}{2} \left( \eta_{ac} e^\nu_b - \eta_{bc} e^\nu_a \right) e^d_\kappa \delta e^d_\nu \partial_\mu e^c_\kappa
\end{aligned}$$

$$
\begin{aligned}
= &-\tfrac{1}{2} e_a^\kappa e_b^\lambda \left( \nabla_\kappa \delta g_{\lambda\mu} - \nabla_\lambda \delta g_{\kappa\mu} \right) - \tfrac{1}{2} \left( \eta_{ac} e_b^\nu - \eta_{bc} e_a^\nu \right) \nabla_\mu \delta e_\nu^c \\
&- \underbrace{\tfrac{1}{2} \eta_{cd} e_a^\kappa e_b^\lambda e_\nu^d \Gamma^\nu_{\kappa\mu} \delta e_\lambda^c}_{1} - \underbrace{\tfrac{1}{2} \eta_{bd} e_a^\kappa \Gamma^\nu_{\kappa\mu} \delta e_\nu^d}_{3} \\
&+ \underbrace{\tfrac{1}{2} \eta_{cd} e_a^\kappa e_b^\lambda e_\nu^d \Gamma^\nu_{\lambda\mu} \delta e_\kappa^c}_{2} + \underbrace{\tfrac{1}{2} \eta_{ad} e_b^\lambda \Gamma^\nu_{\lambda\mu} \delta e_\nu^d}_{3} \\
&- \underbrace{\tfrac{1}{2} \eta_{bd} e_\nu^d \Gamma^\nu_{\kappa\mu} \delta e_a^\kappa}_{4} - \underbrace{\tfrac{1}{2} \eta_{cd} e_a^\kappa e_\lambda^c e_\nu^d \Gamma^\nu_{\kappa\mu} \delta e_b^\lambda}_{1} \\
&+ \underbrace{\tfrac{1}{2} \eta_{cd} e_b^\lambda e_\kappa^c e_\nu^d \Gamma^\nu_{\lambda\mu} \delta e_a^\kappa}_{2} + \underbrace{\tfrac{1}{2} \eta_{ad} e_\nu^d \Gamma^\nu_{\lambda\mu} \delta e_b^\lambda}_{4} \\
&+ \tfrac{1}{2} \left( \eta_{ac} e_b^\nu - \eta_{bc} e_a^\nu \right) \Big( \underbrace{e_\kappa^c \Gamma^\kappa_{\mu\lambda} e_d^\lambda \delta e_\nu^d}_{4} - \underbrace{\Gamma^\lambda_{\mu\nu} \delta e_\lambda^c}_{3} \Big)
\end{aligned}
$$

Next, we note that a general frame variation $\delta e_\mu^a$ may be naturally decomposed into two parts: a "symmetric part" $\delta^+ e_\mu^a$ that by itself produces the induced variation of the metric and an "antisymmetric part" $\delta^- e_\mu^a$ that preserves the metric:

$$\delta e_\mu^a \;=\; \delta^+ e_\mu^a + \delta^- e_\mu^a \quad,\quad \delta^\pm e_\mu^a \;=\; \tfrac{1}{2} \left( \delta e_\mu^a \mp \eta^{ab} g_{\mu\nu} \delta e_b^\nu \right). \tag{256}$$

Indeed, this gives

$$
\begin{aligned}
\delta^\pm g_{\mu\nu} &= \eta_{ab} \left( \delta^\pm e_\mu^a \, e_\nu^b + e_\mu^a \, \delta^\pm e_\nu^b \right) \\
&= \tfrac{1}{2} \eta_{ab} \left( \delta e_\mu^a \, e_\nu^b \mp \eta^{ac} g_{\mu\kappa} \delta e_c^\kappa \, e_\nu^b + e_\mu^a \, \delta e_\nu^b \mp \eta^{bd} g_{\nu\lambda} e_\mu^a \, \delta e_d^\lambda \right) \\
&= \tfrac{1}{2} \eta_{ab} \, \delta e_\mu^a \, e_\nu^b \pm \tfrac{1}{2} g_{\mu\kappa} e_b^\kappa \, \delta e_\nu^b + \tfrac{1}{2} \eta_{ab} e_\mu^a \, \delta e_\nu^b \pm \tfrac{1}{2} g_{\nu\lambda} \delta e_\mu^a \, e_a^\lambda,
\end{aligned}
$$

that is,

$$\delta^+ g_{\mu\nu} \;=\; \delta g_{\mu\nu} \quad,\quad \delta^- g_{\mu\nu} \;=\; 0. \tag{257}$$

Similarly, we compute

$$
\begin{aligned}
\eta^{ab} e_b^\nu \, \delta g_{\mu\nu} &= \eta^{ab} e_b^\nu \eta_{cd} \left( \delta e_\mu^c \, e_\nu^d + e_\mu^c \, \delta e_\nu^d \right) \\
&= \delta e_\mu^a + \eta^{ab} \eta_{cd} e_\mu^c e_b^\nu \, \delta e_\nu^d \\
&= \delta e_\mu^a - \eta^{ab} \eta_{cd} e_\mu^c e_\nu^d \, \delta e_b^\nu,
\end{aligned}
$$

showing that the symmetric part of the frame variation is completely determined by the variation of the metric:

$$\delta^+ e_\mu^a \;=\; \tfrac{1}{2} \eta^{ab} e_b^\nu \, \delta g_{\mu\nu}. \tag{258}$$

On the other hand, defining $\Lambda_b^a = e_b^\mu \, \delta^- e_\mu^a$ or more explicitly,

$$\Lambda_b^a \;=\; \tfrac{1}{2} \left( e_b^\mu \, \delta e_\mu^a - \eta^{ac} \eta_{bd} e_c^\mu \, \delta e_\mu^d \right), \tag{259}$$

and $\Lambda_{ab} \equiv \eta_{ac}\Lambda^c_b$ or more explicitly,

$$\Lambda_{ab} = \tfrac{1}{2}\left(\eta_{ac}\,e^\mu_b - \eta^{bc}\,e^\mu_a\right)\delta e^c_\mu ,\qquad(260)$$

we obtain

$$\delta^- e^a_\mu = \Lambda^a_b\,e^\mu_b ,\qquad(261)$$

showing that the antisymmetric part of the frame variation corresponds to a local frame rotation. We also note that the induced variation of connection coefficients for the Levi-Civita connection admits the same kind of decomposition. For the Christoffel symbols, the variation induced by a frame variation is obviously of the purely symmetric type ($\delta^+\Gamma^\kappa_{\mu\lambda} = \delta\Gamma^\kappa_{\mu\lambda}$, $\delta^-\Gamma^\kappa_{\mu\lambda} = 0$) while for the vector connection coefficients, the decomposition is precisely the one given by the two terms in eq. (253), since in this equation, the second term is easily seen to vanish when $\delta^- e^a_\mu = 0$ whereas the first term vanishes according to eq. (257) when $\delta^+ e^a_\mu = 0$. Thus

$$\delta^+\Gamma_{\mu,ab} = -\tfrac{1}{2}\,e^\kappa_a\,e^\lambda_b\left(\nabla_\kappa\,\delta g_{\lambda\mu} - \nabla_\lambda\,\delta g_{\kappa\mu}\right) ,\qquad(262)$$

$$\delta^-\Gamma_{\mu,ab} = -\nabla_\mu\Lambda_{ab} = -\tfrac{1}{2}\left(\eta_{ac}\,e^\nu_b - \eta_{bc}\,e^\nu_a\right)\nabla_\mu\,\delta e^c_\nu .\qquad(263)$$

In passing, we note that the decomposition described above has a global, coordinate and frame independent meaning, which can be understood by observing that locally, i.e., over appropriate open subsets of a manifold $M$, a linear frame field $e$ for $M$ is a section of the linear frame bundle $LM$ of $M$ and its variations $\delta e$ are sections of the pull back $e^*(LM)$ of the vertical bundle $V(LM)$ of $LM$ to $M$ by $e$. But $LM$ is a principal bundle with structure group $\mathrm{GL}(n,\mathbb{R})$ and hence, as a vector bundle over $LM$, $V(LM)$ is canonically isomorphic to the trivial vector bundle $LM \times \mathfrak{gl}(n,\mathbb{R})$ – a property that is passed on to all of its pull backs. Therefore, the symmetric pair type decomposition

$$\mathfrak{gl}(n,\mathbb{R}) = \mathfrak{so}(p,q) \oplus \mathfrak{so}^\perp(p,q)$$

of the Lie algebra $\mathfrak{gl}(n,\mathbb{R})$ into the subalgebra $\mathfrak{so}(p,q)$ of antisymmetric matrices and the complementary subspace $\mathfrak{so}^\perp(p,q)$ of symmetric matrices with respect to the standard symmetric bilinear form $\eta$ of signature $(p,q)$,[21] generated by the involution

$$\begin{aligned}\theta:\ \mathfrak{gl}(n,\mathbb{R}) &\longrightarrow \mathfrak{gl}(n,\mathbb{R})\\ X &\longmapsto -\eta X^T\eta\end{aligned},$$

can be transferred to the fibers of any of these vector bundles and gives rise to the decomposition of frame variations introduced above. (The argument can be extended, from the infinitesimal to the global level, so as to provide an Iwasawa type decomposition for strict automorphisms of the linear frame bundle; we leave out the details since they will not be needed here.)

---

[21]The notion of orthogonality used for this complement is that induced by the non-degenerate symmetric bilinear form on $\mathfrak{gl}(n,\mathbb{R})$ given by $(X,Y) \mapsto \mathrm{tr}\left(\eta X^T\eta Y\right)$.

Making use of this decomposition, we now fix the transformation law of purely spinorial quantities, without ambiguity, by requiring them to transform according to the pertinent standard transformation law under the local frame rotation provided by the antisymmetric part alone. In other words, the symmetric part is to be discarded. In order to implement this prescription explicitly, we first define the matrix $\Lambda$ representing this local frame rotation in spinor space:

$$\Lambda^\alpha_\beta \;=\; \tfrac{1}{8}\,\Lambda_{ab}\,[\gamma^a,\gamma^b]^\alpha_\beta\;. \tag{264}$$

Then for Dirac spinor fields,

$$\delta^+\psi^\alpha \;=\; 0 \quad,\quad \delta^-\psi^\alpha \;=\; \Lambda^\alpha_\beta\,\psi^\beta \tag{265}$$

Correspondingly, the variation of the orthonormal frame Dirac $\gamma$-matrices is given by

$$\begin{aligned}
\delta^+\gamma_a &= 0 \quad,\quad \delta^-\gamma_a = -\,\Lambda^b_a\,\gamma_b + [\Lambda,\gamma_a]\;, \\
\delta^+\gamma^a &= 0 \quad,\quad \delta^-\gamma^a = +\,\Lambda^a_b\,\gamma^b + [\Lambda,\gamma^a]\;,
\end{aligned} \tag{266}$$

whereas the variation of the coordinate frame Dirac $\gamma$-matrices is given by

$$\begin{aligned}
\delta^+\gamma^\mu &= \tfrac{1}{2}\,\delta g^{\mu\nu}\,\gamma_\nu \quad,\quad \delta^-\gamma_\mu = [\Lambda,\gamma_\mu]\;, \\
\delta^+\gamma_\mu &= \tfrac{1}{2}\,\delta g_{\mu\nu}\,\gamma^\nu \quad,\quad \delta^-\gamma^\mu = [\Lambda,\gamma^\mu]\;,
\end{aligned} \tag{267}$$

These equations are consistent with the relation (250) between the two types of $\gamma$-matrices and also with the basic Clifford algebra relations (cf. eq. (168)):

$$\begin{aligned}
\delta\{\gamma_a,\gamma_b\} &= \{\delta\gamma_a,\gamma_b\} + \{\gamma_a,\delta\gamma_b\} \\
&= -\,\Lambda^c_a\{\gamma_c,\gamma_b\} - \Lambda^c_b\{\gamma_a,\gamma_c\} + \{[\Lambda,\gamma_a],\gamma_b\} + \{\gamma_a,[\Lambda,\gamma_b]\} \\
&= -\,\Lambda_{ba} - \Lambda_{ab} + \Lambda\gamma_a\gamma_b - \gamma_a\Lambda\gamma_b + \gamma_b\Lambda\gamma_a - \gamma_b\gamma_a\Lambda \\
&\quad + \gamma_a\Lambda\gamma_b - \gamma_a\gamma_b\Lambda + \Lambda\gamma_b\gamma_a - \gamma_b\Lambda\gamma_a \\
&= 0 \\
&= 2\,\delta\eta_{ab}\;,
\end{aligned}$$

$$\begin{aligned}
\delta\{\gamma_\mu,\gamma_\nu\} &= \{\delta\gamma_\mu,\gamma_\nu\} + \{\gamma_\mu,\delta\gamma_\nu\} \\
&= \tfrac{1}{2}\,\delta g_{\mu\kappa}\{\gamma^\kappa,\gamma_\nu\} + \tfrac{1}{2}\,\delta g_{\nu\kappa}\{\gamma_\mu,\gamma^\kappa\} + \{[\Lambda,\gamma_\mu],\gamma_\nu\} + \{\gamma_\mu,[\Lambda,\gamma_\nu]\} \\
&= \delta g_{\mu\kappa}\,\delta^\kappa_\nu + \delta g_{\nu\kappa}\,\delta^\kappa_\mu + \Lambda\gamma_\mu\gamma_\nu - \gamma_\mu\Lambda\gamma_\nu + \gamma_\nu\Lambda\gamma_\mu - \gamma_\nu\gamma_\mu\Lambda \\
&\quad + \gamma_\mu\Lambda\gamma_\nu - \gamma_\mu\gamma_\nu\Lambda + \Lambda\gamma_\nu\gamma_\mu - \gamma_\nu\Lambda\gamma_\mu \\
&= 2\,\delta g_{\mu\nu}\;.
\end{aligned}$$

It is somewhat surprising that the variation of the commutators of the orthonormal frame Dirac $\gamma$-matrices also vanish:

$$\delta\,[\gamma_a,\gamma_b] \;=\; 0 \quad,\quad \delta\,[\gamma^a,\gamma^b] \;=\; 0\;. \tag{268}$$

Indeed,

$$\begin{aligned}
\delta\,[\gamma^a,\gamma^b] &= [\delta\gamma^a,\gamma^b]\,+\,[\gamma^a,\delta\gamma^b] \\
&= \Lambda^a_c\,[\gamma^c,\gamma^b]\,+\,\Lambda^b_d\,[\gamma^a,\gamma^d]\,+\,[[\Lambda,\gamma^a],\gamma^b]\,+\,[\gamma^a,[\Lambda,\gamma^b]] \\
&= \Lambda_{cd}\left(-\eta^{ac}\,[\gamma^b,\gamma^d]\,+\,\eta^{bc}\,[\gamma^a,\gamma^d]\right)\,+\,[\Lambda,[\gamma^a,\gamma^b]] \\
&= \tfrac{1}{2}\,\Lambda_{cd}\left(-\eta^{ac}\,[\gamma^b,\gamma^d]\,+\,\eta^{ad}\,[\gamma^b,\gamma^c]\,+\,\eta^{bc}\,[\gamma^a,\gamma^d]\,-\,\eta^{bd}\,[\gamma^a,\gamma^c]\right) \\
&\quad -\,\tfrac{1}{8}\,\Lambda_{cd}\,[[\gamma^a,\gamma^b],[\gamma^c,\gamma^d]] \\
&= 0\,,
\end{aligned}$$

where we have made use of the identity expressing the fact that the commutators of the orthonormal frame Dirac $\gamma$-matrices are, except for a factor $\tfrac{1}{4}$, the generators of the Lie algebra $\mathfrak{so}(p,q)$ in the spinor representation:

$$[[\gamma_a,\gamma_b],[\gamma_c,\gamma_d]] \;=\; -\,4\,\eta_{ac}\,[\gamma_b,\gamma_d]\,+\,4\,\eta_{ad}\,[\gamma_b,\gamma_c]\,+\,4\,\eta_{bc}\,[\gamma_a,\gamma_d]\,-\,4\,\eta_{bd}\,[\gamma_a,\gamma_c]\,. \tag{269}$$

This implies that the variation of the spinor connection coefficients is essentially the same as that of the vector connection coefficients given by eqs (253), (262) and (263):

$$\delta^+\Gamma^\alpha_{\mu\beta} \;=\; -\,\tfrac{1}{16}\left(\nabla_\kappa\,\delta g_{\lambda\mu}\,-\,\nabla_\lambda\,\delta g_{\kappa\mu}\right)[\gamma^\kappa,\gamma^\lambda]^\alpha_\beta\,, \tag{270}$$

$$\delta^-\Gamma^\alpha_{\mu\beta} \;=\; -\,\nabla_\mu\Lambda^\alpha_\beta \;=\; -\,\tfrac{1}{16}\left(\eta_{ac}\,e^\nu_b\,-\,\eta_{bc}\,e^\nu_a\right)\nabla_\mu\delta e^c_\nu\,[\gamma^a,\gamma^b]^\alpha_\beta\,. \tag{271}$$

Now we can compute the variation of covariant derivatives of spinor fields: using that $\nabla_\mu\Lambda^\alpha_\beta = \partial_\mu\Lambda^\alpha_\beta + \Gamma^\alpha_{\mu\gamma}\Lambda^\gamma_\beta - \Gamma^\gamma_{\mu\beta}\Lambda^\alpha_\gamma$, we get

$$\begin{aligned}
\delta\left(\nabla_\mu\psi^\alpha\right) &= \delta\,\partial_\mu\psi^\alpha\,+\,\delta\Gamma^\alpha_{\mu\beta}\,\psi^\beta\,+\,\Gamma^\alpha_{\mu\gamma}\,\delta\psi^\gamma \\
&= \partial_\mu\,\delta^-\psi^\alpha\,+\,\delta^+\Gamma^\alpha_{\mu\beta}\,\psi^\beta\,+\,\delta^-\Gamma^\alpha_{\mu\beta}\,\psi^\beta\,+\,\Gamma^\alpha_{\mu\gamma}\,\delta^-\psi^\gamma \\
&= \underline{\partial_\mu\Lambda^\alpha_\beta\,\psi^\alpha}\,+\,\Lambda^\alpha_\beta\,\partial_\mu\psi^\beta\,+\,\delta^+\Gamma^\alpha_{\mu\beta}\,\psi^\beta\,-\,\underline{\nabla_\mu\Lambda^\alpha_\beta\,\psi^\beta}\,+\,\underline{\Gamma^\alpha_{\mu\gamma}\,\Lambda^\gamma_\beta\,\psi^\beta} \\
&= \Lambda^\alpha_\beta\,\partial_\mu\psi^\beta\,+\,\underline{\Gamma^\gamma_{\mu\beta}\,\Lambda^\alpha_\gamma\,\psi^\beta}\,+\,\delta^+\Gamma^\alpha_{\mu\beta}\,\psi^\beta \\
&= \Lambda^\alpha_\beta\,\nabla_\mu\psi^\beta\,+\,\delta^+\Gamma^\alpha_{\mu\beta}\,\psi^\beta
\end{aligned}$$

and hence

$$\delta^+\left(\nabla_\mu\psi^\alpha\right) \;=\; \delta^+\Gamma^\alpha_{\mu\beta}\,\psi^\beta\,,\qquad \delta^-\left(\nabla_\mu\psi^\alpha\right) \;=\; \Lambda^\alpha_\beta\,\nabla_\mu\psi^\beta\,. \tag{272}$$

Therefore, the variation of a typical contribution to the kinetic term in the Dirac Lagrangian (179) takes the form

$$\delta\left(\bar{\psi}\overset{\leftrightarrow}{\nabla}\chi\right) = \delta\left(\bar{\psi}\,\gamma^\mu\,\nabla_\mu\chi\right) - \delta\left(\overline{\nabla_\mu\psi}\,\gamma^\mu\,\chi\right)$$

$$= +\,\overline{\delta\psi}\,\gamma^\mu\,\nabla_\mu\chi + \bar{\psi}\,\delta\gamma^\mu\,\nabla_\mu\chi + \bar{\psi}\,\gamma^\mu\,\delta\left(\nabla_\mu\chi\right)$$
$$-\,\overline{\delta\left(\nabla_\mu\psi\right)}\,\gamma^\mu\,\chi - \overline{\nabla_\mu\psi}\,\delta\gamma^\mu\,\chi - \overline{\nabla_\mu\psi}\,\gamma^\mu\,\delta\chi$$

$$= -\,\bar{\psi}\,\Lambda\,\gamma^\mu\,\nabla_\mu\chi + \tfrac{1}{2}\,\bar{\psi}\,\gamma_\nu\,\nabla_\mu\chi\,\delta g^{\mu\nu} + \bar{\psi}\,[\Lambda,\gamma^\mu]\,\nabla_\mu\chi$$
$$+\,\bar{\psi}\,\gamma^\mu\,\delta^+\Gamma_\mu\,\chi + \bar{\psi}\,\gamma^\mu\,\Lambda\,\nabla_\mu\chi$$
$$+\,\bar{\psi}\,\delta^+\Gamma_\mu\,\gamma^\mu\,\chi + \overline{\nabla_\mu\psi}\,\Lambda\,\gamma^\mu\,\chi - \tfrac{1}{2}\,\overline{\nabla_\mu\psi}\,\gamma_\nu\,\chi\,\delta g^{\mu\nu} - \overline{\nabla_\mu\psi}\,[\Lambda,\gamma^\mu]\,\chi$$
$$-\,\overline{\nabla_\mu\psi}\,\gamma^\mu\,\Lambda\,\chi$$

$$= \tfrac{1}{2}\,\bar{\psi}\,\gamma_\nu\,\overset{\leftrightarrow}{\nabla}_\mu\chi\,\delta g^{\mu\nu}$$
$$-\,\tfrac{1}{16}\left(\nabla_\kappa\,\delta g_{\lambda\mu} - \nabla_\lambda\,\delta g_{\kappa\mu}\right)\bar{\psi}\left(\gamma^\mu\,[\gamma^\kappa,\gamma^\lambda] + [\gamma^\kappa,\gamma^\lambda]\,\gamma^\mu\right)\chi\;.$$

(For simplicity, we have omitted internal indices and suppressed the spinor indices; moreover, we have used the fact that the Dirac $\gamma$-matrices are hermitean and hence their commutators are antihermitean with respect to Dirac adjoint, so that, for example, $\bar{\Lambda} = -\Lambda$ and $\bar{\Gamma}_\mu = -\Gamma_\mu$.) Now the second term, which comes from the variation of the Christoffel symbols, is actually zero because the expression $\gamma^\mu\,[\gamma^\kappa,\gamma^\lambda] + [\gamma^\kappa,\gamma^\lambda]\,\gamma^\mu$ is totally antisymmetric. Indeed, it is obviously antisymmetric in $\kappa$ and $\lambda$ but is also antisymmetric in $\lambda$ and $\mu$ because

$$\gamma^\mu\,[\gamma^\kappa,\gamma^\lambda] + [\gamma^\kappa,\gamma^\lambda]\,\gamma^\mu$$
$$= \gamma^\mu\gamma^\kappa\gamma^\lambda - \gamma^\mu\gamma^\lambda\gamma^\kappa + \gamma^\kappa\gamma^\lambda\gamma^\mu - \gamma^\lambda\gamma^\kappa\gamma^\mu$$
$$= 2\,\gamma^\mu\gamma^\kappa\gamma^\lambda - \underline{2\,g^{\kappa\lambda}\gamma^\mu} + \underline{2\,g^{\kappa\lambda}\gamma^\mu} - 2\,\gamma^\lambda\gamma^\kappa\gamma^\mu\;,$$

and is antisymmetric in $\kappa$ and $\mu$ because

$$\gamma^\mu\,[\gamma^\kappa,\gamma^\lambda] + [\gamma^\kappa,\gamma^\lambda]\,\gamma^\mu$$
$$= \gamma^\mu\gamma^\kappa\gamma^\lambda - \gamma^\mu\gamma^\lambda\gamma^\kappa + \gamma^\kappa\gamma^\lambda\gamma^\mu - \gamma^\lambda\gamma^\kappa\gamma^\mu$$
$$= -\,2\,\gamma^\mu\gamma^\lambda\gamma^\kappa + \underline{2\,g^{\kappa\lambda}\gamma^\mu} + 2\,\gamma^\kappa\gamma^\lambda\gamma^\mu - \underline{2\,g^{\kappa\lambda}\gamma^\mu}\;.$$

Therefore, only the first term survives. Potential terms also give no contribution since the variation of composite tensor fields is easily seen to be the expected one, namely

$$\delta\left(\bar{\psi}\,\gamma_{\mu_1}\ldots\gamma_{\mu_r}\chi\right) = \tfrac{1}{2}\sum_{i=1}^{r}\bar{\psi}\,\gamma_{\mu_1}\ldots\gamma_{\mu_{i-1}}\gamma_\nu\gamma_{\mu_{i+1}}\ldots\gamma_{\mu_r}\chi\;\delta g^{\mu_i\nu}\;, \qquad (273)$$

and similarly,

$$\delta\left(\bar{\psi}\,\gamma_{a_1}\ldots\gamma_{a_r}\chi\right) = -\,\tfrac{1}{2}\sum_{i=1}^{r}\bar{\psi}\,\gamma_{a_1}\ldots\gamma_{a_{i-1}}\gamma_b\gamma_{a_{i+1}}\ldots\gamma_{a_r}\chi\;\Lambda^b{}_{a_i}\;, \qquad (274)$$

so that for example,

$$\begin{aligned}
\delta\left(\left(\bar{\psi}\,\gamma^{\mu}\chi\right)\left(\bar{\psi}\,\gamma_{\mu}\chi\right)\right) &= \delta\left(\bar{\psi}\,\gamma^{\mu}\chi\right)\left(\bar{\psi}\,\gamma_{\mu}\chi\right) + \left(\bar{\psi}\,\gamma^{\mu}\chi\right)\delta\left(\bar{\psi}\,\gamma_{\mu}\chi\right) \\
&= \tfrac{1}{2}\left(\bar{\psi}\,\gamma_{\nu}\chi\right)\left(\bar{\psi}\,\gamma_{\mu}\chi\right)\delta g^{\mu\nu} + \tfrac{1}{2}\left(\bar{\psi}\,\gamma^{\mu}\chi\right)\left(\bar{\psi}\,\gamma^{\nu}\chi\right)\delta g_{\mu\nu} \\
&= 0\,.
\end{aligned}$$

It is interesting to note that the transformation law of spinors under frame variations postulated in eq. (265) above is completely fixed by requiring the isomorphism of algebra bundles $\gamma$ in eq. (166) to be equivariant under arbitrary linear frame transformations: this means precisely that the transformation law of spinors should be fixed so as to guarantee that all composite tensor fields which are bilinear in spinors should transform as tensor fields do, namely according to eqs (273) and (274).

Having collected all the basic and nontrivial ingredients that are needed, computing the energy-momentum tensor for the Dirac Lagrangian (179) is now a simple exercise; the result can be found in Table 1.

Finally, we would like to mention an alternative approach to calculating the energy-momentum tensor in theories where orthonormal frames are used in the construction of the Lagrangian, provided that the resulting frame dependence of the Lagrangian is ultimately an implicit one that can be reduced entirely to its dependence on the underlying metric tensor; this will be the case if and only if the Lagrangian is gauge invariant under local frame rotations. In this case, we can exploit eq. (247) together with the chain rule for variational derivatives, which expresses the variational derivative of $L_{\mathrm{m}}$ with respect to orthonormal frames in terms of that with respect to the metric tensor through the formula

$$\frac{\delta L_{\mathrm{m}}}{\delta e^a_{\mu}} = \frac{\delta L_{\mathrm{m}}}{\delta g_{\rho\sigma}}\frac{\delta g_{\rho\sigma}}{\delta e^a_{\mu}} \tag{275}$$

where of course

$$\frac{\delta g_{\rho\sigma}}{\delta e^a_{\mu}} = \frac{\partial g_{\rho\sigma}}{\partial e^a_{\mu}} = \eta_{ab}\left(\delta^{\mu}_{\rho}\,e^b_{\sigma} + \delta^{\mu}_{\sigma}\,e^b_{\rho}\right). \tag{276}$$

A general proof of this chain rule can be given by arguing that it is the direct expression of the chain rule for the functional derivative of the corresponding action functional. An alternative, "purely finite-dimensional" proof can be given in the context of the first order Lagrangian formalism exposed in Sect. 3.2, in the form of a chain rule for the variational derivative under bundle homomorphisms $E \to F$ and their jet extensions $JE \to JF$; we leave this to the reader. In this way, the fundamental formula (5) defining the energy-momentum tensor can be rephrased as stating that the mixed coordinate frame / orthonormal frame components $T^{\mu}_a = \eta_{ab}T^{\mu\nu}e^b_{\nu}$ of the energy-momentum tensor are given by

$$T^{\mu}_a = -\frac{\delta L_{\mathrm{m}}}{\delta e^a_{\mu}}. \tag{277}$$

| Field type | Lagrangian and energy-momentum tensor |
|---|---|
| Real scalar field (cf. eq. (175)) | $L_{\text{RSC}} = \frac{1}{2} g^{\mu\nu} (D_\mu \varphi, D_\nu \varphi) - U(\varphi)$ <br> $T_{\mu\nu}^{\text{RSC}} = (D_\mu \varphi, D_\nu \varphi) - g_{\mu\nu} L_{\text{RSC}}$ |
| Complex scalar field (cf. eq. (176)) | $L_{\text{CSC}} = g^{\mu\nu} \langle D_\mu \varphi, D_\nu \varphi \rangle - U(\varphi)$ <br> $T_{\mu\nu}^{\text{CSC}} = \langle D_\mu \varphi, D_\nu \varphi \rangle + \langle D_\nu \varphi, D_\mu \varphi \rangle - g_{\mu\nu} L_{\text{CSC}}$ |
| Generalized sigma model (cf. eq. (177)) | $L_{\text{GSM}} = \frac{1}{2} g^{\mu\nu} (D_\mu \varphi, D_\nu \varphi) - U(\varphi)$ <br> $T_{\mu\nu}^{\text{GSM}} = (D_\mu \varphi, D_\nu \varphi) - g_{\mu\nu} L_{\text{GSM}}$ |
| Ordinary sigma model (cf. eq. (178)) | $L_{\text{OSM}} = \frac{1}{2} g^{\mu\nu} (\partial_\mu \varphi, \partial_\nu \varphi)$ <br> $T_{\mu\nu}^{\text{OSM}} = (\partial_\mu \varphi, \partial_\nu \varphi) - g_{\mu\nu} L_{\text{OSM}}$ |
| Dirac spinor field (cf. eq. (179)) | $L_{\text{DSP}} = \frac{i}{2} g^{\mu\nu} \bar{\psi} \gamma_\mu \vec{D}_\nu \psi - U(\bar{\psi}, \psi)$ <br> $T_{\mu\nu}^{\text{DSP}} = \frac{i}{4} \left( \bar{\psi} \gamma_\mu \overleftrightarrow{D}_\nu \psi + \bar{\psi} \gamma_\nu \overleftrightarrow{D}_\mu \psi \right) - g_{\mu\nu} L_{\text{DSP}}$ |
| Gauge field (cf. eq. (180)) | $L_{\text{YM}} = -\frac{1}{4} g^{\mu\kappa} g^{\nu\lambda} (F_{\mu\nu}, F_{\kappa\lambda})$ <br> $T_{\mu\nu}^{\text{YM}} = -g^{\kappa\lambda} (F_{\mu\kappa}, F_{\nu\lambda}) - g_{\mu\nu} L_{\text{YM}}$ |
| Modified scalar field with additional $Rf(\varphi)$ term (cf. eqs (237)–(239)) | $L_{\text{MSC}} = \frac{1}{2} g^{\mu\nu} \partial_\mu \varphi \, \partial_\nu \varphi - U(\varphi) + R f(\varphi)$ <br> $T_{\mu\nu}^{\text{MSC}} = \partial_\mu \varphi \, \partial_\nu \varphi - g_{\mu\nu} L_{\text{MSC}}$ <br> $\qquad + 2 \left( g_{\mu\nu} \Box - \nabla_\mu \nabla_\nu + R_{\mu\nu} \right) f(\varphi)$ |

Table 1: Lagrangians and energy-momentum tensors for various types of fields

# 5  Scale Invariance and the Trace of the Energy-Momentum Tensor

Over the last few decades, a great deal of attention has been devoted to two important special classes of field theories that can be characterized by properties of their energy-momentum tensor:

- **conformal field theories**, whose energy-momentum tensor is traceless,

- **topological field theories**, whose energy-momentum tensor vanishes.

Using this characterization as a definition of course raises two questions which are far from trivial, namely how to construct, from a given Lagrangian, the correct energy-momentum tensor to be used for deciding whether a concrete field theoretical model does or does not fall into one of these classes, and also how to relate that construction to symmetry properties of the Lagrangian, in the spirit of Noether's theorem.

In what follows, we shall discuss this question for the case of conformal field theories, elucidating the relationship that exists between "scale invariance", which is often but somewhat imprecisely also referred to as "conformal invariance", and tracelessness of the energy-momentum tensor. The analogous relationship between "diffeomorphism invariance" and vanishing of the energy-momentum tensor is much simpler and will therefore only be briefly commented at the end of the section.

The most pragmatic method for testing scale invariance of any given Lagrangian in field theory is based on assigning to every field a *scaling dimension* such that the contribution to the action coming from the standard kinetic term for that field becomes scale invariant: it must then be verified whether the contributions to the action coming from the various interaction terms are so as well. This will be the case if and only if all terms in the Lagrangian have the same (composite) scaling dimension. In flat space-time, this technique is well known from quantum field theory and there is no difficulty in adapting it to classical field theory. Scale transformations are in this case the dilatations $D(\lambda)$ ($\lambda > 0$) given by

$$D(\lambda): \begin{array}{rcl} \mathbb{R}^n & \longrightarrow & \mathbb{R}^n \\ x & \longmapsto & \lambda x \end{array} \tag{278}$$

under which components of fields $\varphi$ transform according to $\varphi \mapsto D(\lambda)\varphi$ with

$$(D(\lambda)\varphi)(x) \;=\; \lambda^{-d_\varphi}\, \varphi(\lambda^{-1}x) \tag{279}$$

where $d_\varphi$ is a real number called the *scaling dimension* of $\varphi$. (Cf. the discussion in Sect. 2; in particular, the corresponding transformation law at the infinitesimal level has been written down in eq. (20).) Note that the metric tensor, which in this framework is the flat background metric $\eta$ of special relativity, is understood to be invariant under dilatations and so can be viewed as having scaling dimension 0. It is also clear from eq. (279) that if a field $\varphi$ has scaling dimension $d_\varphi$, then all its partial derivatives $\partial_\mu \varphi$ will have scaling dimension $d_\varphi + 1$: this means that the operator $\partial_\mu$ itself must carry scaling dimension 1. The same value must be attributed to gauge fields ($G$-connections) in order to ensure that the covariant derivative $D_\mu$ becomes homogeneous and carries the same scaling dimension as the ordinary derivative $\partial_\mu$. Moreover, the scaling dimension of scalar fields and spinor fields is found by inspection of the kinetic terms in the Lagrangians (175)–(179):

- Scalar fields have scaling dimension $\frac{1}{2}(n-2)$ in the case of linearly realized internal symmetry and scaling dimension 0 in the case of nonlinearly realized internal symmetry, since there is then no reasonable nontrivial prescription for making dilatations act on them.

- Spinor fields have scaling dimension $\frac{1}{2}(n-1)$.

The assignments thus obtained are summarized in the column labelled "SD" of Table 2 below. Further inspection of the Lagrangians (175)–(184) then reveals the following possibilities for obtaining scale invariant actions:

- Scalar fields. In $n=2$ dimensions, the Lagrangians (175), (176), (177) and (178) all yield scale invariant actions provided we suppose the potential $U$ to vanish. In $n>2$ dimensions, the Lagrangians (175) and (176) yield scale invariant actions provided the potential $U$ is homogeneous of degree $2n/(n-2)$, the only solutions for which $U$ is a polynomial being $\varphi^6$ for $n=3$, $\varphi^4$ for $n=4$ and $\varphi^3$ for $n=6$.

- Spinor fields. The Lagrangian (179) yields a scale invariant action provided the potential $U$ is homogeneous of degree $2n/n-1$, the only solutions for which $U$ is a polynomial being $\eta^{\mu\nu} \bar\psi \gamma_\mu \psi \bar\psi \gamma_\nu \psi$, $(\bar\psi\psi)^2$ and $(\bar\psi\gamma_5\psi)^2$ in $n=2$ dimensions.

- Yukawa coupling. The only possible polynomial interaction terms between scalar fields and spinor fields that give rise to a scale invariant contribution to the action are trilinear couplings o the Yukawa type, such as in eqs (182) and (183), in $n=4$ dimensions.

- $G$-connections. The Yang-Mills Lagrangian (180) and Chern-Simons Lagrangian (184) yield scale invariant actions in $n=4$ and $n=3$ dimensions, respectively.

These results are summarized in the column labelled "SI = GWI" of Table 3 below.

If on the other hand one looks at the energy-momentum tensors derived from the Lagrangians (175)–(180) according to the prescription (5), it turns out that, even "on shell" (i.e., when the equations of motion for the matter fields are taken into account), scale invariance of the action does not always force the energy-momentum tensor to be traceless: discrepancies occur in the presence of scalar fields in $n>2$ dimensions, the simplest examples being the massless free field theory and, at the level of interacting field theories, the massless $\varphi^4$ theory in $n=4$ dimensions. This problem has been noted long ago and a solution was proposed in the beginning of the 1970's [3,4], based on a further "ad hoc" improvement of the canonical energy-momentum tensor which goes beyond the standard prescription of Belinfante and Rosenfeld but, as we have shown in Sect. 2, can be understood in a similar spirit. It may thus seem as if the prescription (5) for finding the energy-momentum tensor might still not be the correct one – a conclusion that is hardly acceptable in view of the wealth of arguments presented in the previous sections. So something else must have gone wrong.

To find the way out of this dilemma, it is necesary to extend the concept of scale invariance and of the scaling dimension of fields to general space-time manifolds. To begin with, it is clear that the definition of scale transformations given in eq. (278) above makes sense in a vector space but not in a general manifold. Moreover, from a geometric point of view by which we aim at freeing ourselves from the assumption of a pre-existing linear structure, dilatations on a pseudo-Euclidean vector space can and should be viewed as conformal isometries of a pseudo-Riemannian manifold since they preserve the standard Minkowski metric up to a scale factor. Generically, however, pseudo-Riemannian manifolds do not admit any conformal isometries at all. This means that there is no way to define a scale transformation as being some kind of diffeomorphism of space-time. Of course, the key to the solution of this problem is well known: it consists in switching from the active to the passive point of view. Scale transformations are not really active transformations that move points in space-time but rather passive transformations that change the scale of the metric by which we measure distances between points in space-time. To avoid confusion, we shall as usual refer to scale transformations in this sense as *Weyl rescalings*. The connection between the two points of view can then be established by requiring that in flat space-time, a Weyl rescaling of the metric and a dilatation, by the same constant factor $\lambda$, should lead to the same rescaling for the distance between points; this leads to the following transformation law for the metric under Weyl rescalings:

$$g_{\mu\nu} \;\to\; \lambda^2 \, g_{\mu\nu} \quad , \quad g^{\mu\nu} \;\to\; \lambda^{-2} \, g^{\mu\nu} \; . \tag{280}$$

More generally, components of fields $\varphi$ transform according to

$$\varphi \;\to\; \lambda^{-w_\varphi} \, \varphi \tag{281}$$

where $w_\varphi$ is a real number called the *Weyl dimension* of $\varphi$. Of course, such a transformation law makes sense for fields that are sections of some vector bundle over space-time, whereas sections of nonlinear fiber bundles over space-time can only be supposed to remain invariant under Weyl rescalings. In particular, eq. (280) states that the metric tensor $g_{\mu\nu}$ and the inverse metric tensor $g^{\mu\nu}$ have Weyl dimension $-2$ and 2, respectively.

It is also clear from eq. (281) that if a field $\varphi$ has Weyl dimension $w_\varphi$, then all its partial derivatives $\partial_\mu \varphi$ will have Weyl dimension $w_\varphi$ as well: this means that the operator $\partial_\mu$ itself must carry Weyl dimension 0. The same value must be attributed to gauge fields ($G$-connections) in order to ensure that the covariant derivative $D_\mu$ becomes homogeneous and carries the same Weyl dimension as the ordinary derivative $\partial_\mu$; note that this is in agreement with eq. (280) which forces the Christoffel symbols for the Levi-Civita connection to have Weyl dimension 0, and the same goes for the Riemann curvature tensor and the Ricci tensor. Finally, the Weyl dimension of scalar fields and spinor fields is found by inspection of the kinetic terms in the Lagrangians (175)–(179), leading to the conclusion that for scalar fields and spinor fields, the Weyl dimension equals the scaling dimension. The assignments thus obtained are summarized in the column labelled "WD" of Table 2. Further inspection of the Lagrangians (175)–(184) then reveals that the list of those that are invariant under Weyl rescalings is exactly the same as the list of those that are invariant under dilatations; they can be read from the column labelled "SI = GWI" of Table 3 below. Thus it seems as if the correct generalization of the concept of scale invariance from flat to curved space-time is really that of invariance under Weyl rescalings, and we shall adopt this point of view from now on.

| Field type | SD (flat space-time) | WD (curved space-time) |
|---|---|---|
| Scalar field | $\frac{1}{2}(n-2)$ | $\frac{1}{2}(n-2)$ |
| Sigma model | 0 for $n=2$<br>undefined for $n>2$ | 0 for $n=2$<br>undefined for $n>2$ |
| Dirac spinor field | $\frac{1}{2}(n-1)$ | $\frac{1}{2}(n-1)$ |
| Gauge field | 1 | 0 |
| Metric tensor | 0<br>($g=\eta$ only) | $g_{\mu\nu} : -2$<br>$g^{\mu\nu} : +2$ |

Table 2: Scaling dimension (SD) and Weyl dimension (WD) for various types of fields in $n$ space-time dimensions

There is however one aspect of this reinterpretation that has not been addressed so far and that will turn out to be crucial: the possibility of distinguishing between global and local Weyl invariance; by abuse of language, these are also often referred to as global and local scale invariance, respectively. The former refers to *global Weyl rescalings*, where the metric is rescaled by a positive numerical factor $\lambda$, whereas the latter refers to *local Weyl rescalings*, where the metric is rescaled by a positive function $\lambda$. In the discussion above, we have tacitly assumed the rescaling factor to be numerical, which is sufficient for defining the concept of the Weyl dimension of a field, but after all, nothing prevents us from allowing to rescale the metric by different factors at different space-time points. Fields that transform according to eq. (281) even when $\lambda$ is a positive function will in what follows be called *Weyl covariant*. With this terminology, it is clear that, in general, partial derivatives of Weyl covariant fields are no longer Weyl covariant and that this defect can only be cured by "gauging" the global symmetry under Weyl rescalings so that it becomes local. As is well known, this requires introducing a corresponding new gauge field and replacing ordinary partial derivatives by corresponding covariant derivatives, which we might call scaling covariant derivatives in order to distinguish them from the gauge covariant derivatives (referring to internal symmetries) and the space-time covariant derivatives considered before. Of course, this is precisely Weyl's original procedure that gave rise to gauge theories in the first place! However, Weyl himself discarded this theory soon after, namely when, as a result of discussions with Einstein, he realized that it did not describe electromagnetism, as he had originally hoped. All that has remained of it is the word "gauge" (or "Eichung", in German), which in its original sense refers to the arbitrariness in the choice of units, in this case for measuring distances in space-time (rods and clocks). For our present purposes, we shall not need it either, since as we shall see below, there are many important situations where invariance of the action under *local* Weyl rescalings can be established *without* the need for introducing a new gauge field – for which no clear evidence in nature has been found so far.

Before analyzing which of the globally Weyl invariant actions derived from the Lagrangians listed in Sect. 3.6 are even locally Weyl invariant, we formulate the main theorem of this section, which clearly shows the usefulness of this property.

**Theorem 5.1** *"On shell", that is, assuming the matter fields to satisfy their equations of motion, the matter field action is locally Weyl invariant if and only if the corresponding energy-momentum tensor is traceless.*

Proof: In completely general terms, the variation of the matter field action $S_{\mathrm{m},K}$ over any given compact subset $K$ of $M$ induced by an infinitesimal local Weyl rescaling, parametrized by some given function $\omega$ on $M$, is given by

$$\delta_\omega S_{\mathrm{m},K} \;=\; \int_K d^n x \, \sqrt{|\det g|} \, \left( \frac{\delta L_{\mathrm{m}}}{\delta g_{\mu\nu}} \, \delta_\omega g_{\mu\nu} \,+\, \frac{\delta L_{\mathrm{m}}}{\delta \varphi^i} \, \delta_\omega \varphi^i \right) .$$

Now if we assume the matter fields $\varphi$ to satisfy their equations of motion, the second term drops out, so the only non-vanishing contribution comes from the induced variation of the metric. Hence using the definition (5) of the energy-momentum tensor, together with the formula for the induced variation of the metric,

$$\delta_\omega g_{\mu\nu} \;=\; 2\,\omega\,g_{\mu\nu} \quad, \quad \delta_\omega g^{\mu\nu} \;=\; -\,2\,\omega\,g^{\mu\nu} \;, \tag{282}$$

which is just the infinitesimal form of eq. (280), derived by taking $\omega = \ln\lambda$, we obtain

$$\delta_\omega S_{\mathrm{m},K} \;=\; -\int_K d^n x\,\sqrt{|\det g|}\,g_{\mu\nu}\,T^{\mu\nu}\,\omega\;,$$

which implies the equivalence stated in the theorem since $\omega$ was an arbitrary function.
$\square$

**Remark**: The simplest case is of course when the matter field Lagrangian $L_\mathrm{m}$ is locally Weyl invariant, but the above proof shows that in order to guarantee tracelessness of the energy-momentum tensor, much less is needed: under infinitesimal local Weyl rescalings, $L_\mathrm{m}$ may be allowed to pick up a total divergence, as well as terms that vanish upon insertion of the equations of motion for the matter fields.[22]

A simple example of a Lagrangian which is globally as well as locally Weyl invariant is the Yang-Mills Lagrangian in $n=4$ dimensions. A less trivial example is the covariant free Dirac Lagrangian (where the term "covariant" indicates that minimal coupling to gravity and also to gauge fields is allowed): here, local Weyl invariance can be shown to hold "on shell". To do so, we need the transformation law of spinor fields and their covariant derivatives under local Weyl rescalings. To derive these, note first that the transformation law (280) for the metric can be supplemented by the following simple transformation laws for orthonormal frames,

$$e^a_\mu \;\to\; \lambda\,e^a_\mu \quad, \quad e^\mu_a \;\to\; \lambda^{-1}\,e^\mu_a\;, \tag{283}$$

which is obviously consistent with the orthonormality condition (247), and for the coordinate frame Dirac $\gamma$-matrices,

$$\gamma_\mu \;\to\; \lambda\,\gamma_\mu \quad, \quad \gamma^\mu \;\to\; \lambda^{-1}\,\gamma^\mu\;, \tag{284}$$

which is obviously consistent with the Clifford algebra relations (168). At the infinitesimal level, we conclude that eq. (282) must be supplemented by

$$\delta_\omega e^a_\mu \;=\; \omega\,e^a_\mu \quad, \quad \delta_\omega e^\mu_a \;=\; -\,\omega\,e^\mu_a\;, \tag{285}$$

and

$$\delta_\omega \gamma_\mu \;=\; \omega\,\gamma_\mu \quad, \quad \delta_\omega \gamma^\mu \;=\; -\,\omega\,\gamma^\mu\;, \tag{286}$$

---
[22] In the remainder of this section, we omit the index "m", which stands for "matter fields".

respectively. In particular, it is worth noting that the frame variation generated by an infinitesimal Weyl rescaling is of purely symmetric type, so that according to the analysis carried out in Sect. 4.3, the induced variation of spinor fields (which we shall refer to as the indirect contribution to their total variation) and the induced variation of the orthonormal frame Dirac $\gamma$-matrices (cf. eqs (265) and (266)) vanish while the induced variation of the coordinate frame Dirac $\gamma$-matrices (cf. eq. (267)) is precisely the expression given above. Similarly, the induced variation of the spinor connection coefficients is given by inserting eq. (286) into eq. (270), with the result ,

$$\delta_\omega \Gamma^\alpha_{\mu\beta} \;=\; \tfrac{1}{16} \left( g_{\kappa\mu}\, \partial_\lambda \omega \;-\; g_{\lambda\mu}\, \partial_\kappa \omega \right) [\gamma^\kappa, \gamma^\lambda]^\alpha_\beta \;. \tag{287}$$

On the other hand, it must be emphasized that a Weyl rescaling is *not* merely a frame transformation (i.e., an automorphism of the frame bundle): this becomes obvious when one realizes that the scaling dimension of a field is not merely given by its tensor or spinor degree. (For example, frame transformations act trivially on scalar fields while Weyl rescalings act nontrivially, except when $n = 2$.) Accordingly, there is a further direct contribution to the variation of spinor fields under infinitesimal Weyl rescalings which is of course given by eq. (281) with the appropriate Weyl dimension for spinor fields, which is $\tfrac{1}{2}(n-1)$, so

$$\psi^\alpha \;\to\; \lambda^{-(n-1)/2} \, \psi^\alpha \;, \tag{288}$$

and hence, at the infinitesimal level, eq. (265) must be replaced by

$$\delta_\omega \psi^\alpha \;=\; -\tfrac{1}{2}\,(n-1)\,\omega\,\psi^\alpha \;. \tag{289}$$

Repeating the calculation leading to eq. (272), we thus get

$$\begin{aligned}
\delta_\omega \left( \nabla_\mu \psi^\alpha \right) &= \delta_\omega \, \partial_\mu \psi^\alpha \;+\; \delta_\omega \Gamma^\alpha_{\mu\beta}\, \psi^\beta \;+\; \Gamma^\alpha_{\mu\gamma}\, \delta_\omega \psi^\gamma \\
&= \partial_\mu \delta_\omega \psi^\alpha \;+\; \delta_\omega \Gamma^\alpha_{\mu\beta}\, \psi^\beta \;+\; \Gamma^\alpha_{\mu\gamma}\, \delta_\omega \psi^\gamma \\
&= -\tfrac{1}{2}(n-1)\, \nabla_\mu (\omega\,\psi^\alpha) \;+\; \delta_\omega \Gamma^\alpha_{\mu\beta}\, \psi^\beta \;,
\end{aligned}$$

that is,

$$\delta_\omega\!\left(\nabla_\mu \psi^\alpha\right) \;=\; -\tfrac{1}{2}(n-1)\,\omega\,\nabla_\mu \psi^\alpha \;-\; \tfrac{1}{2}(n-1)\,\partial_\mu \omega\, \psi^\alpha \;+\; \delta_\omega \Gamma^\alpha_{\mu\beta}\, \psi^\beta \;. \tag{290}$$

Therefore, the variation of the free Dirac Lagrangian becomes

$$\begin{aligned}
\delta_\omega \!\left( \bar\psi\, \overleftrightarrow{\nabla}\, \chi \right) &= \delta_\omega \left( \bar\psi\, \gamma^\mu\, \nabla_\mu \chi \right) \;-\; \delta_\omega \left( \overline{\nabla_\mu \psi}\, \gamma^\mu\, \chi \right) \\
&= +\, \overline{\delta_\omega \psi}\, \gamma^\mu\, \nabla_\mu \chi \;+\; \bar\psi\, \delta_\omega \gamma^\mu\, \nabla_\mu \chi \;+\; \bar\psi\, \gamma^\mu\, \delta_\omega\!\left( \nabla_\mu \chi \right) \\
&\quad -\; \overline{\delta_\omega\!\left(\nabla_\mu \psi\right)}\, \gamma^\mu\, \chi \;-\; \overline{\nabla_\mu \psi}\, \delta_\omega \gamma^\mu\, \chi \;-\; \overline{\nabla_\mu \psi}\, \gamma^\mu\, \delta_\omega \chi
\end{aligned}$$

$$\begin{aligned}
=\ &-\tfrac{1}{2}(n{-}1)\,\omega\,\bar\psi\,\gamma^\mu\,\nabla_\mu\chi\ -\ \omega\,\bar\psi\,\gamma^\mu\,\nabla_\mu\chi\\
&-\tfrac{1}{2}(n{-}1)\,\omega\,\bar\psi\,\gamma^\mu\,\nabla_\mu\chi\ -\ \tfrac{1}{2}(n{-}1)\,\partial_\mu\omega\,\bar\psi\,\gamma^\mu\,\chi\ +\ \bar\psi\,\gamma_\mu\,\delta_\omega\Gamma_\mu\,\chi\\
&+\tfrac{1}{2}(n{-}1)\,\omega\,\overline{\nabla_\mu\psi}\,\gamma^\mu\,\chi\ +\ \tfrac{1}{2}(n{-}1)\,\partial_\mu\omega\,\bar\psi\,\gamma^\mu\,\chi\ +\ \bar\psi\,\delta_\omega\Gamma_\mu\,\gamma^\mu\,\chi\\
&+\omega\,\overline{\nabla_\mu\psi}\,\gamma^\mu\,\chi\ +\ \tfrac{1}{2}(n{-}1)\,\omega\,\bar\psi\,\gamma^\mu\,\nabla_\mu\chi\ -\ \omega\,\overline{\nabla_\mu\psi}\,\gamma^\mu\,\chi\\
=\ &-n\,\omega\,\bar\psi\,\overset{\leftrightarrow}{\slashed\nabla}\chi\\
&+\tfrac{1}{16}\left(g_{\kappa\mu}\,\partial_\lambda\omega\ -\ g_{\lambda\mu}\,\partial_\kappa\omega\right)\bar\psi\left(\gamma^\mu\,[\gamma^\kappa,\gamma^\lambda]\ +\ [\gamma^\kappa,\gamma^\lambda]\,\gamma^\mu\right)\chi\,,
\end{aligned}$$

where we note that two terms containing partial derivatives of $\omega$ have cancelled and that the last term vanishes for symmetry reasons, as before. Thus as long as the covariant free Dirac Lagrangian is the only part of the total Lagrangian containing spinor fields, the corresponding equation of motion will be the covariant free Dirac equation, and whenever this is satisfied, the variation that we have just computed will vanish. Briefly, we may express this by saying that the variation of the covariant free Dirac Lagrangian under infinitesimal local scale transformations is proportional to the covariant free Dirac Lagrangian itself, and this vanishes "on shell".

Turning to scalar fields, it becomes clear from the above theorem that the source of the problem with the non-vanishing trace of the energy-momentum tensor in scale invariant field theories containing scalar fields lies in the fact that the corresponding action is globally but not locally Weyl invariant. No such problem arises in two space-time dimensions since, for $n=2$, scalar fields have Weyl dimension 0, which means that they are manifestly invariant under Weyl rescalings, global as well as local ones. Thus in this case, both types of Weyl invariance hinge on the same single hypothesis, namely vanishing of the potential $U$, and correspondingly, direct inspection shows that the energy-momentum tensors given in the first four entries of Table 1 are all traceless when the potential term is absent. But for $n>2$ and a potential $U$ which is homogeneous of degree $2n/(n{-}2)$, the Lagrangians (175) and (176) yield globally Weyl invariant actions which are not locally Weyl invariant: variation of the kinetic term in the Lagrangian under an infinitesimal local Weyl rescaling produces an additive contribution depending on the derivatives of the rescaling factor, and this lack of invariance reflects itself in a non-vanishing trace of the corresponding energy-momentum tensor.

As mentioned before, it has been observed long ago in the literature [3,4] that this problem can be overcome by a further "improvement" of the energy-momentum tensor, which also turns out to improve its behavior under renormalization. In the same context, it has also been observed that the necessary correction can be traced back to a modification of the pertinent Lagrangian, which amounts to adding a term of the form $R\varphi^2$. This means, of course, that the term "improvement" is not really adequate to describe the procedure since, after all, the definition of the energy-momentum tensor as the variational derivative of the Lagrangian with respect to the metric remains unaltered. Instead, what has been abandoned is the prescription of minimal coupling to gravity, by allowing for the possibility of introducing an additional contribution to the Lagrangian that vanishes in the flat space-time limit. But why exactly this term? What seems to be missing is to establish a motivation for the choice made, and this motivation can only come from studying the relation with Weyl invariance in more detail.

In order to gain further insight into the problem and its solution, let us concentrate on the simplest situation – that of a single real scalar field, with Lagrangian (237), equation of motion (240) and energy-momentum tensor given by eqs (243)-(245). The specific dependence of this Lagrangian on the Riemann curvature tensor (a function of the matter field multiplied by the Ricci scalar curvature) is dictated by the fact that this seems to be the only way to guarantee that the equations of motion for the metric can still be written in the form of Einstein's equations. Concerning the dependence on the matter field, we have already seen that global Weyl invariance requires the potential $U$ to be homogeneous of degree $2n/(n-2)$; similarly, it forces the function $f$ to be homogeneous of degree 2. Therefore, we put

$$L \;=\; \tfrac{1}{2} (\partial\varphi)^2 \;-\; a\,\varphi^{2n/(n-2)} \;+\; b\,R\,\varphi^2 \;, \tag{291}$$

with coefficients $a$ and $b$ that remain to be determined. The corresponding equation of motion reads

$$\Box\varphi \;+\; \frac{2n\,a}{n-2}\,\varphi^{(n+2)/(n-2)} \;-\; 2\,b\,R\,\varphi \;=\; 0 \;. \tag{292}$$

Next, we calculate the trace of the corresponding energy-momentum tensor:

$$g^{\mu\nu} T_{\mu\nu} \;=\; -\,\tfrac{1}{2}\,(n{-}2)\,(\partial\varphi)^2 \;+\; n\,a\,\varphi^{2n/(n-2)} \;+\; 2\,(n{-}1)\,b\,\Box\varphi^2 \;-\; (n{-}2)\,b\,R\,\varphi^2 \;.$$

Inserting eq. (292), multiplied by $\tfrac{1}{2}\,(n-2)\,\varphi$, this reduces to

$$g^{\mu\nu} T_{\mu\nu} \;=\; -\,\tfrac{1}{2}\,(n{-}2)\,(\partial\varphi)^2 \;-\; \tfrac{1}{2}\,(n{-}2)\,\varphi\,\Box\varphi \;+\; 2\,(n{-}1)\,b\,\Box\varphi^2 \;,$$

which vanishes if and only if we choose

$$b \;=\; \frac{1}{8}\,\frac{n-2}{n-1} \;,$$

whereas $a$ remains undetermined. Similarly, we can check directly that, with this choice, the resulting Lagrangian

$$L \;=\; \frac{1}{2}\,(\partial\varphi)^2 \;-\; a\,\varphi^{2n/(n-2)} \;+\; \frac{n-2}{8\,(n-1)}\,R\,\varphi^2 \tag{293}$$

is locally Weyl invariant; more precisely, this property holds for the Lagrangian

$$L \;=\; -\,\frac{1}{2}\,\varphi\,\Box\varphi \;-\; a\,\varphi^{2n/(n-2)} \;+\; \frac{n-2}{8\,(n-1)}\,R\,\varphi^2 \tag{294}$$

which differs from the previous one by a covariant total divergence (of the form $\frac{1}{2} g^{\mu\nu} \nabla_\mu(\varphi \, \partial_\nu \varphi)$, to be precise) and hence has the same equations of motion and the same energy-momentum tensor. To show this, we need the transformation law of the Ricci scalar curvature and of the covariant wave operator under local Weyl rescalings. Setting $\omega = \ln \lambda$, note first that the Christoffel symbols for the Levi-Civita connection are globally Weyl invariant but pick up an additive term under local Weyl rescalings:

$$\Gamma^\kappa_{\mu\nu} \quad \rightarrow \quad \Gamma^\kappa_{\mu\nu} + \left( \delta^\kappa_\nu \, \partial_\mu \omega + \delta^\kappa_\mu \, \partial_\nu \omega - g_{\mu\nu} \, g^{\kappa\lambda} \, \partial_\lambda \omega \right) \tag{295}$$

The same then goes for the Riemann curvature tensor: after some calculation, we get

$$\begin{aligned} R^\kappa{}_{\lambda\mu\nu} \quad \rightarrow \quad & R^\kappa{}_{\lambda\mu\nu} - \left( \delta^\kappa_\mu \, \nabla_\nu \partial_\lambda \omega - \delta^\kappa_\nu \, \nabla_\mu \partial_\lambda \omega \right) \\ & + \left( g_{\lambda\mu} \, g^{\kappa\rho} \, \nabla_\nu \partial_\rho \omega - g_{\lambda\nu} \, g^{\kappa\rho} \, \nabla_\mu \partial_\rho \omega \right) \\ & + \left( \delta^\kappa_\mu \, \partial_\nu \omega \, \partial_\lambda \omega - \delta^\kappa_\nu \, \partial_\mu \omega \, \partial_\lambda \omega \right) \\ & - \left( g_{\lambda\mu} \, g^{\kappa\rho} \, \partial_\nu \omega \, \partial_\rho \omega - g_{\lambda\nu} \, g^{\kappa\rho} \, \partial_\mu \omega \, \partial_\rho \omega \right) \\ & - \left( \delta^\kappa_\mu \, g_{\nu\lambda} - \delta^\kappa_\nu \, g_{\mu\lambda} \right) (\partial \omega)^2 \, . \end{aligned} \tag{296}$$

Hence for the Ricci tensor,

$$\begin{aligned} R_{\mu\nu} \quad \rightarrow \quad & R_{\mu\nu} - (n-2) \, \nabla_\nu \partial_\mu \omega - g_{\mu\nu} \, \Box \omega \\ & + (n-2) \, \partial_\nu \omega \, \partial_\mu \omega - (n-2) \, g_{\mu\nu} \, (\partial \omega)^2 \, , \end{aligned} \tag{297}$$

while the Ricci scalar curvature, which has Weyl dimension $-2$, transforms according to

$$R \quad \rightarrow \quad \exp(-2\omega) \left( R - 2 \, (n-1) \, \Box \omega - (n-1)(n-2) \, (\partial \, \omega)^2 \right) \, . \tag{298}$$

On the other hand, the transformation law (281), with the value $w_\varphi = \frac{1}{2}(n-2)$ already substituted, gives

$$\partial_\mu \varphi \quad \rightarrow \quad \exp(-\tfrac{1}{2} \, (n-2) \, \omega) \left( \partial_\mu \varphi - \tfrac{1}{2} \, (n-2) \, \partial_\mu \omega \, \varphi \right) \, ,$$

and after iteration

$$\partial_\mu \partial_\nu \varphi \quad \rightarrow \quad \exp(-\tfrac{1}{2} \, (n-2) \, \omega) \Big( \partial_\mu \partial_\nu \varphi - \tfrac{1}{2} \, (n-2) \, \partial_\mu \partial_\nu \omega \, \varphi - \tfrac{1}{2} \, (n-2) \, \partial_\nu \omega \, \partial_\mu \varphi \\ - \tfrac{1}{2} \, (n-2) \, \partial_\mu \omega \, \partial_\nu \varphi + \tfrac{1}{4} \, (n-2)^2 \, \partial_\mu \omega \, \partial_\nu \omega \, \varphi \Big) \, .$$

Thus

$$g^{\mu\nu} \, \partial_\mu \partial_\nu \varphi \quad \rightarrow \quad \exp(-\tfrac{1}{2} \, (n+2) \, \omega) \Big( g^{\mu\nu} \, \partial_\mu \partial_\nu \varphi - \tfrac{1}{2} \, (n-2) \, g^{\mu\nu} \, \partial_\mu \partial_\nu \omega \, \varphi \\ - (n-2) \, g^{\mu\nu} \, \partial_\mu \omega \, \partial_\nu \varphi + \tfrac{1}{4} \, (n-2)^2 \, (\partial \omega)^2 \, \varphi \Big) \, ,$$

whereas

$$g^{\mu\nu}\,\Gamma^{\kappa}_{\mu\nu}\,\partial_{\kappa}\varphi \;\to\; \exp(-\tfrac{1}{2}(n+2)\,\omega)\left(g^{\mu\nu}\,\Gamma^{\kappa}_{\mu\nu} - (n-2)\,g^{\kappa\lambda}\,\partial_{\lambda}\omega\right)$$
$$\times\left(\partial_{\kappa}\varphi - \tfrac{1}{2}(n-2)\,\partial_{\kappa}\omega\,\varphi\right).$$

Taking the difference gives

$$\Box\varphi \;\to\; \exp(-\tfrac{1}{2}(n+2)\,\omega)\left(\Box\varphi - \tfrac{1}{2}(n-2)\,\Box\omega\,\varphi - \tfrac{1}{4}(n-2)^2\,(\partial\omega)^2\,\varphi\right). \qquad(299)$$

Together, eqs (298) and (299) show that

$$\Box \;-\; \frac{n-2}{4\,(n-1)}\,R \qquad(300)$$

is the (locally) *scale covariant wave operator*, and inserting this result into eq. (294), we immediately obtain

$$L \;\to\; \exp(-\omega)\,L \qquad(301)$$

Note that the resulting local Weyl invariance of the action is valid even "off shell": at no point of the proof of the transformation law (301) did we use the equations of motion for $\varphi$. In fact, there is a very simple reason for this invariance, which comes to light if we use the scalar field $\varphi$ and the metric tensor $g$ to define a new metric tensor $\hat{g}$ which is manifestly Weyl invariant by its mere definition:

$$\hat{g}_{\mu\nu} \;=\; \varphi^{4/(n-2)}\,g_{\mu\nu}. \qquad(302)$$

Then the Lagrangian (291) is, up to a constant multiple, nothing but the Einstein-Hilbert Lagrangian for $\hat{g}$, with the potential term corresponding to the cosmological constant! Indeed, applying eq. (298) with $\lambda = \varphi^{2/(n-2)}$, we get

$$\sqrt{|\det\hat{g}|}\,\left(\hat{R} + 2\Lambda\right)$$
$$= \lambda^{n}\,\sqrt{|\det g|}\,\left(\lambda^{-2}\left(R - 2(n-1)\,\Box\ln\lambda - (n-1)(n-2)\,(\partial\ln\lambda)^2\right) + 2\Lambda\right)$$
$$= \sqrt{|\det g|}\,\left(\varphi^2\left(R - \frac{4\,(n-1)}{n-2}\,\Box\ln\varphi - \frac{4\,(n-1)}{n-2}\,(\partial\ln\varphi)^2\right) + 2\Lambda\,\varphi^{2n/(n-2)}\right)$$
$$= \sqrt{|\det g|}\,\left(\varphi^2\left(R - \frac{4\,(n-1)}{n-2}\,\left(g^{\mu\nu}\,\nabla_{\mu}(\varphi^{-1}\,\partial_{\nu}\varphi) + \varphi^{-2}\,(\partial\varphi)^2\right)\right) + 2\Lambda\,\varphi^{2n/(n-2)}\right)$$

Thus expressing the constant $a$ in eqs (291)-(294) in terms of the cosmological constant $\Lambda$, the Lagrangians (293) and (294) assume the form

$$L \;=\; \tfrac{1}{2}(\partial\varphi)^2 + \frac{n-2}{4\,(n-1)}\,\Lambda\,\varphi^{2n/(n-2)} + \frac{n-2}{8\,(n-1)}\,R\,\varphi^2 \qquad(303)$$

and

$$L \;=\; -\tfrac{1}{2}\,\varphi\,\Box\varphi + \frac{n-2}{4\,(n-1)}\,\Lambda\,\varphi^{2n/(n-2)} + \frac{n-2}{8\,(n-1)}\,R\,\varphi^2 \qquad(304)$$

respectively, with corresponding energy-momentum tensor

$$\begin{aligned}T_{\mu\nu} &= \partial_\mu\varphi\,\partial_\nu\varphi - \tfrac{1}{2}g_{\mu\nu}(\partial\varphi)^2 - \frac{n-2}{4(n-1)}\Lambda g_{\mu\nu}\varphi^{2n/(n-2)} \\ &+ \frac{n-2}{4(n-1)}\left(g_{\mu\nu}\Box - \nabla_\mu\nabla_\nu\right)\varphi^2 + \frac{n-2}{4(n-1)}\left(R_{\mu\nu} - \tfrac{1}{2}g_{\mu\nu}R\right)\varphi^2\end{aligned} \qquad (305)$$

which in flat space-time coincides with the result derived in Sect. 2, eq. (54).

In $n=4$ space-time dimensions, the construction can be extended to include a Dirac spinor field with a Yukawa coupling to the scalar field. In this case, the Lagrangian reads

$$L = \tfrac{1}{2}(\partial\varphi)^2 + \tfrac{i}{2}\bar{\psi}\overrightarrow{\slashed{\nabla}}\psi + \tfrac{1}{6}\Lambda\varphi^4 + \tfrac{1}{12}R\varphi^2 - \lambda_Y\varphi\bar{\psi}\psi\,, \qquad (306)$$

or equivalently

$$L = -\tfrac{1}{2}\varphi\Box\varphi + \tfrac{i}{2}\bar{\psi}\overrightarrow{\slashed{\nabla}}\psi + \tfrac{1}{6}\Lambda\varphi^4 + \tfrac{1}{12}R\varphi^2 - \lambda_Y\varphi\bar{\psi}\psi\,, \qquad (307)$$

The corresponding equations of motion read

$$\Box\varphi - \tfrac{2}{3}\Lambda\varphi^3 - \tfrac{1}{6}R\varphi + \lambda_Y\bar{\psi}\psi = 0\,, \qquad (308)$$

$$i\slashed{\nabla}\psi - \lambda_Y\varphi\psi = 0\,, \qquad (309)$$

while the energy-momentum tensor is

$$\begin{aligned}T_{\mu\nu} &= \partial_\mu\varphi\,\partial_\nu\varphi + \tfrac{i}{4}\left(\bar{\psi}\gamma_\mu\overrightarrow{\nabla}_\nu\psi + \bar{\psi}\gamma_\nu\overrightarrow{\nabla}_\mu\psi\right) \\ &- \tfrac{1}{2}g_{\mu\nu}(\partial\varphi)^2 - \tfrac{i}{2}g_{\mu\nu}\bar{\psi}\overrightarrow{\slashed{\nabla}}\psi - \tfrac{1}{6}\Lambda g_{\mu\nu}\varphi^4 + \lambda_Y g_{\mu\nu}\varphi\bar{\psi}\psi \\ &+ \tfrac{1}{6}\left(g_{\mu\nu}\Box - \nabla_\mu\nabla_\nu\right)\varphi^2 + \tfrac{1}{6}\left(R_{\mu\nu} - \tfrac{1}{2}g_{\mu\nu}R\right)\varphi^2\end{aligned} \qquad (310)$$

Its trace

$$g^{\mu\nu}T_{\mu\nu} = -(\partial\varphi)^2 - 3\tfrac{i}{2}\bar{\psi}\overrightarrow{\slashed{\nabla}}\psi - \tfrac{2}{3}\Lambda\varphi^4 + 4\lambda_Y\varphi\bar{\psi}\psi + \tfrac{1}{2}\Box\varphi^2 - \tfrac{1}{6}R\varphi^2$$

vanishes "on shell", as can be seen by multiplying the equation of motion (308) for $\varphi$ by $\varphi$ and the equation of motion (309) for $\psi$ by $\tfrac{3}{2}\bar{\psi}$ and adding. Similarly, we can check directly that the Lagrangian (306) or rather (307) is locally Weyl invariant "on shell" by combining our previous analysis of local Weyl invariance for the covariant free Dirac Lagrangian with that for the scalar field, the only modification being that it is now the sum of the covariant free Dirac Lagrangian and the Yukawa term that vanishes "on shell".

| Field type and Lagrangian | SI = GWI | LWI |
|---|---|---|
| Scalar field (cf. eqs (175) and (176)) | $U = 0$ for $n=2$ | yes |
| | $\left\{\begin{array}{l}\deg(U) = 6 \text{ for } n=3 \\ \deg(U) = 4 \text{ for } n=4 \\ \deg(U) = 3 \text{ for } n=6\end{array}\right\}$ | no |
| | $U = 0$ for other values of $n$ | no |
| Sigma model (cf. eqs (177) and (178)) | $U = 0$ ($n=2$ only) | yes |
| Dirac spinor field (cf. eq. (179)) | $\deg(U) = 4$ for $n=2$ $U = 0$ for $n>2$ | yes (on shell) |
| Gauge field Yang-Mills Lagrangian (180) | $n=4$ | yes |
| Metric tensor Einstein-Hilbert Lagrangian (181) | $n=2$ | no |
| Scalar + Dirac spinor field with Yukawa coupling (cf. eq. (182)) | $n=4$, $\deg(U_{\text{scalar}}) = 4$, $U_{\text{spinor}} = 0$ | no |
| Gauge field Chern-Simons Lagrangian (184) | $n=3$ | yes |
| Modified scalar field with an additional $R\varphi^2$ term (cf. eqs (303) or (304)) ($n>2$ only) | $\left\{\begin{array}{l}\deg(U) = 6 \text{ for } n=3 \\ \deg(U) = 4 \text{ for } n=4 \\ \deg(U) = 3 \text{ for } n=6\end{array}\right\}$ | yes |
| | $U = 0$ for other values of $n$ | yes |
| Modified scalar + Dirac spinor field with Yukawa coupling (cf. eq. (182)) and an additional $R\varphi^2$ term (cf. eqs (303) or (304)) | $n=4$, $\deg(U_{\text{scalar}}) = 4$, $U_{\text{spinor}} = 0$ | yes (on shell) |

Table 3: Conditions for scale invariance (SI) or global Weyl invariance (GWI) and additional conditions for local Weyl invariance (LWI) of standard Lagrangians with homogeneous self-interaction potential $U$ in $n$ space-time dimensions

The procedure of extending a given globally Weyl invariant Lagrangian to a locally Weyl invariant one by adding an appropriate $R\varphi^2$ term can easily be made to work for multicomponent scalar fields as well: all we need to do is replace the square of the field by the square of its modulus, defined in terms of the given fiber metric. The same goes, in $n = 4$ dimensions, for multicomponent scalar fields with Yukawa couplings to multicomponent spinor fields. This fact is of considerable physical interest since it applies directly to the standard model of particle physics. Indeed, what has been shown here is that and in what sense the scaling limit of the standard model (obtained simply by neglecting all explicit mass terms, such as bare quark and lepton masses) is a conformal field theory: what is required to achieve this is the inclusion of an appropriate $R|\varphi|^2$ term in the standard model Lagrangian, where $\varphi$ is the Higgs field.

Another interesting feature of Weyl invariance is that there is a simple way to construct lots of locally Weyl invariant Lagrangians, namely by inverting the philosophy underlying the transition made in eq. (302), as follows. Consider any Lagrangian field theory containing, apart from a bunch of matter fields $\hat{\varphi}^i$, a metric tensor $\hat{g}$, all of which are defined to be invariant under local Weyl rescalings. Now introduce a new Weyl covariant scalar field $\phi$ of Weyl dimension $\frac{1}{2}(n-2)$, together with new Weyl covariant matter fields $\varphi^i$ of Weyl dimension $w_i$ (where $w_i = 0$ for components of nonlinear matter fields of sigma model type) and a new metric tensor $g$ with standard Weyl dimension ($-2$ for $g_{\mu\nu}$, $+2$ for $g^{\mu\nu}$), defined by making the substitutions

$$\hat{g}_{\mu\nu} = \phi^{4/(n-2)} g_{\mu\nu} \quad , \quad \hat{g}^{\mu\nu} = \phi^{-4/(n-2)} g^{\mu\nu} ,$$
$$\hat{\varphi}^i = \phi^{-2w_i/(n-2)} \varphi^i , \tag{311}$$

together with the corresponding substitutions in the partial derivatives of these fields (including the Christoffel symbols that appear in the space-time covariant derivatives and in the components of the Riemann curvature tensor). Then it is obvious that the Lagrangian in the new fields $\varphi^i$, $g$ and $\phi$ obtained from the original Lagrangian in the old fields $\hat{\varphi}^i$ and $\hat{g}$ by this substitution is invariant under local Weyl rescalings

$$g_{\mu\nu} \to \lambda^2 g_{\mu\nu} \quad , \quad g^{\mu\nu} \to \lambda^{-2} g^{\mu\nu} ,$$
$$\varphi^i \to \lambda^{-w_i} \varphi^i \quad , \quad \phi \to \lambda^{-(n-2)/2} \phi . \tag{312}$$

Of course, the *dilaton field* $\phi$ thus introduced may be a new, additional field, but when the original Lagrangian already contains scalar fields, it may equally well be possible to identify it with one of these. Also, nothing guarantees "a priori" that the new Lagrangian contains derivatives of $\phi$; if not, the dilaton field will be a non-dynamical Lagrange multiplier.

We conclude our considerations with a few remarks on topological field theories. As mentioned at the beginning of this section, such theories can be defined to be field theories with vanishing energy-momentum tensor. But the latter represents the functional derivative of the matter field action with respect to the metric, so topological field theories are simply field theories whose formulation does not depend at all on the choice of a metric tensor on space-time. A notable example are gauge theories in $n = 3$ dimensions whose action is given by the Chern-Simons Lagrangian (184). Other examples can be obtained by starting out from a Lagrangian that does contain a metric tensor $g$ and eliminating it by "going to the stationary point". More specifically, this procedure works for Lagrangians depending on a metric tensor $g$ but not on its derivatives (that is, on the Christoffel symbols or the Riemann curvature tensor), so that the metric tensor is a Lagrange multiplier field with purely algebraic equations of motion that can be used to eliminate it, expressing it in terms of the other fields. As an example, consider the modified sigma model with fields that are maps from $n$-dimensional space-time $M$ to some target space $Q$ but with modified action

$$S_K^{\mathrm{MSM}} \;=\; \int_K d^n x \, \sqrt{|\det g|} \, L_{\mathrm{OSM}}^{n/2} \;, \qquad (313)$$

where $L_{\mathrm{OSM}}$ is the standard Lagrangian (178) for the ordinary sigma model and the extra power $\frac{n}{2}$ is fixed by requiring that the corresponding action should be (globally as well as locally) Weyl invariant, despite the fact that, of course, the Weyl dimension of the sigma type field $\varphi$ vanishes, independently of the value of $n$. This homogeneity of degree 0 is needed to guarantee the existence of non-trivial stationary points for $g$. Indeed, using eq. (236) to calculate the energy-momentum tensor corresponding to the action (313), we get

$$T_{\mu\nu}^{\mathrm{MSM}} \;=\; \frac{n}{2} \, L_{\mathrm{OSM}}^{(n-2)/2} \, h_{ij}(\varphi) \, \partial_\mu \varphi^i \, \partial_\nu \varphi^j \;-\; g_{\mu\nu} \, L_{\mathrm{OSM}}^{n/2} \;, \qquad (314)$$

which is easily verified to be traceless. Now the equations of motion for $g$ derived from the variational principle for the action (313) just state that this expression vanishes, or equivalently, that $g$ is proportional to the pull back $\varphi^* h$ of the metric $h$ on the target space $Q$ to space-time $M$ by the map $\varphi$:

$$g \;=\; \frac{n}{2 L_{\mathrm{OSM}}} \, \varphi^* h \;. \qquad (315)$$

Inserting this condition back into eq. (313), we obtain, up to an irrelevant numerical factor, the action for the theory of $n$-branes in $Q$, which simply measures the $n$-dimensional volume of the image of any compact subset $K$ of $M$ in $Q$ under $\varphi$ with respect to $h$:

$$S_K^{\mathrm{NBR}} \;=\; \int_K d^n x \, \sqrt{|\det \varphi^* h|} \;. \qquad (316)$$

Of course, the interpretation of this action in string theory ($n\!=\!1$) and and membrane theory ($n\!>\!1$) differs from that in field theory: now $Q$ is viewed as space-time, serving as a background for the motion of the string or membrane, while $M$ is just a parameter space, so the fact that this action corresponds to a topological field theory just translates into the postulate of reparametrization invariance of the string or membrane action.